\documentclass[11pt]{article}

\usepackage{amsmath}
\usepackage{widetext}
\usepackage[unicode]{hyperref}
\usepackage{caption}
\usepackage{color}
\usepackage{longtable}

\usepackage{graphicx,isolatin1}
\usepackage[squaren]{SIunits}

\usepackage{etoolbox}

\newcommand{\textroman}[1]{\mbox{\rm #1}}
\newcommand{\gfrac}[2]{\displaystyle\frac{#1}{#2}}

\newcommand{\dd}{\mbox{d}}
\newcommand{\Aeff}{A_{\footnotesize\textroman{eff}}}
\newcommand{\AQED}{A_{\footnotesize\textroman{QED}}}
\newcommand{\FermiLAT}{{\em Fermi}-LAT}

\newcommand{\LAT}{\footnotesize\textroman{\,LAT}}
\newcommand{\MC}{\footnotesize\textroman{MC}}
\newcommand{\ntrack}{n_{\footnotesize\textroman{track}}}
\newcommand{\E}{\textroman{E}}

\newcommand{\mine}{\textroman{\scriptsize min}}

\newcommand{\pitch}{\textroman{p}}
\newcommand{\Var}{\textroman{Var}}

\RequirePackage{xspace}

 \newcommand{\Red}[1]{\textcolor{red}{#1}}
 
 \newcommand{\Blue}[1]{\textcolor{blue}{#1}}
 
 \newcommand{\Magenta}[1]{\textcolor{magenta}{#1}}

\begin{document}

\title{MeV-GeV Polarimetry with $\gamma \to e^+e^-$:
 Asserting the Performance of Silicon Strip Detectors-Based Telescopes}

\author{D.~Bernard
 \\
LLR, Ecole Polytechnique, CNRS/IN2P3,
91128 Palaiseau, France}

\maketitle 

\begin{abstract}
The polarimetry of gamma rays converting to an $e^+e^-$ pair would
open a new window on the high-energy
gamma-ray sky by, among other things, providing
insight into the radiation mechanism in pulsars (curvature
or synchrotron) or deciphering the composition of the gamma-ray
emitting jets in blazars (leptonic or lepto-hadronic).

The performance of polarimeters based on homogeneous active targets
(gas detectors (MeV, HARPO) or emulsions (GeV, GRAINE) has been
studied both with simulation and by the analysis of data collected
with telescope prototypes on linearly-polarised gamma-ray beams, and
found to be excellent.
The present (\FermiLAT, AGILE) and future project (AMEGO, ASTROGAM)
gamma-ray missions, though, are using active targets based on silicon
strip detectors (SSD).
No demonstration of a non-zero effective polarisation asymmetry with
SSDs has been published to date, be it only with simulated data,
and sensitivity estimations were obtained from an assumed value of the
effective polarisation asymmetry.

I present a characterisation of the potential of SSD-based active
targets for polarimetry with gamma-ray conversions to pairs and the
development of various methods to improve on the sensitivity.
This work could pave the way to providing the polarimetry of the
brightest gamma-ray sources of the sky from the decade of data
collected by the \FermiLAT\ and by AGILE, and to guiding the design of
future missions.
\end{abstract}
 
{\em keywords }:
gamma rays, 
pair conversion, 
polarisation,
polarimeter,
silicon strip detector,
moments method

\section{High-Energy $\gamma$-Ray Polarimetry: Motivation}
\label{sec:Motivation}

The gamma-ray sky is dominated by sources emitting by non thermal
processes since a body hot enough to produce gamma-rays, that is, well
above Eddington luminosity, would blow its emitting layers off.
Several of these radiative processes such as synchrotron radiation and
curvature radiation involve a magnetic field and the emission is
polarised in a direction that depends on the direction of the field
and/or of the charged particle flow.
(Inverse) Compton scattering too can produce partially polarised
photons, as the differential cross section involves a polarised
component that has a specific angular distribution.
In addition, Compton scattering enables an efficient transfer of the
polarisation of an incident photon to the scattered one.
In contrast to the above, high-energy hadronic interactions of cosmic
rays with matter produce spin-zero neutral pions that eventually decay to
randomly polarised photons.
 The feasibility of gamma-ray polarimetry, the
measurement of the polarisation fraction and of the polarisation angle
of the high-energy radiation emitted by cosmic sources would be the
opening of a new window on the non-thermal Universe
\cite{Ilie:2019yvs}.

In particular the polarimetry of ${\cal O} (100\,\mega\electronvolt)$
gamma rays would sign the transition between synchrotron radiation and
curvature radiation in rotation-powered pulsars
\cite{Harding:2017tdm}.
``Since the pulsed radiation is emitted along the direction of
relativistic particle motion which is coupled to the magnetic field
direction, and the electric vector, parallel to the particle
acceleration, is different for curvature and synchrotron radiation,
the polarisation can be a very sensitive diagnostic''
\cite{Harding:2017ypk}.

Also gamma-ray polarimetry would enable the determination of the
composition of the gamma-ray-emitting high-energy charged particles in
blazar jets: an
electron-positron plasma is predicted to emit gamma rays with a low
polarisation fraction (leptonic model) while a jet consisting of
ionised matter (lepto-hadronic model) is expected to emit highly
polarised gamma rays \cite{Zhang:2013bna,Chakraborty:2014vma}.

Polarimetry is also a powerful tool to detect a possible
Lorentz-invariance violation (LIV) in the photon sector induced by
``new'' physics beyond the standard model: as the induced birefringence 
would have the polarisation direction of the radiation emitted by a
GRB rotated by a photon-energy-dependent angle during propagation,
any significant broad-band
observation of a non-zero linear polarisation of gamma rays from the GRB 
would constrain the amplitude of the violation: the larger the propagation
distance and the photon energies, the stronger the constraint (a recent
review is at \cite{Kislat:2019ovx}).
Polarimetry is an exercise that is very demanding in terms of photon
statistics though, as we shall see, and polarisation measurements of
GRBs with a sufficient precision, within the $e^+e^-$-pair creation energy
range, would need missions capable of sending hundreds of kilograms of
sensitive material to space \cite{Bernard:2013jea}, something that
might have to wait for some far-away future.

\section{High-Energy $\gamma$-Ray Polarimetry: Measurement}
\label{sec:Measurement}

The conversion of a photon to an $e^+e^-$ pair needs the interaction
with an extra charged particle here named the target, as the effective mass
of the produced pair is larger than $2 m c^2$ ($m$ the electron
mass), while the photon mass is zero and effective mass is bound to be
conserved upon conversion.
The final state then consists of three particles, the two leptons
of the pair (the electron and the positron) and the recoiling target.
The differential cross section is therefore five-dimensional
\cite{Bethe-Heitler};
its polarised version \cite{BerlinMadansky1950,May1951,jau} includes a
term that depends on the linear polarisation fraction and direction
of the incoming photon and on the azimuthal angles\footnote{The
 azimuthal angle of a particle is defined as the angle, in a plane
 perpendicular to the direction of the incident photon, between the
 direction of propagation of that particle and a fixed direction.}
of the final state particles.

The traditional way to perform the measurement is to use the
dependence of the singly-differential cross section, function of an
azimuthal angle, $\varphi$, that describes the orientation of the final state:
\begin{equation}
 \gfrac{ \dd \sigma}{ \dd \varphi} \propto
 \left(1 + A \times P \cos[2(\varphi-\varphi_0)] \right)
 ,
 \label{eq:1D}
\end{equation}
\begin{itemize}
\item
$A$, the polarisation asymmetry, depends on the energy, $E$, of the
 photon and varies from $\pi/4$ at threshold \cite{Gros:2016dmp} to
 $1/7$ at very high-energy \cite{Boldyshev:1972va}.
Over most of the MeV-GeV energy range which is presently accessible to
experimentalists, $A$ is close to $0.15 - 0.20$ (see e.g. Fig. 3 of 
\cite{Gros:2016dmp}).
 
\item $P$ is the linear polarisation fraction of the photon beam.

\item $A \times P$ is the modulation factor of the $\varphi$
 distribution.
\end{itemize} 

Only the linear (not the circular) polarisation of the photon takes
part in the differential cross section, at first order Born
approximation \cite{BerlinMadansky1950}.

After polarimetry with pair conversions was considered
\cite{BerlinMadansky1950,May1951,jau} it was realised that the
multiple scattering of the leptons in the detector material before
their azimuthal angle can be measured is a serious issue.
The blurring of the azimuthal information with an angular resolution,
$\sigma_{\varphi}$,
degrades the measurement of $P$ by a dilution
factor, the ratio of the polarisation asymmetry with and without
experimental effects,
$D \equiv \Aeff / \AQED$, of $D = e^{-2 \sigma_{\varphi}^2}$.
Calculations based on the small-polar-angle approximation and on
approximating the opening angle of the pair, $\theta_{+-}$, to the
most probable value of its distribution, $\hat\theta_{+-}$ \cite{Olsen:1963zz}, indicated
that the measurement of the azimuthal angle of the event must be
performed in the very few $10^{-3}$ radiation lengths ($X_0$)
downstream of the conversion vertex \cite{Kelner,Kotov,Mattox}.
$\sigma_{\varphi}$ increases in proportion to $\sqrt{t}$
during propagation in a material of path length $t \times X_0$
\cite{Patrignani:2016xqp}, 
so $D$ decreases exponentially with $t$;
for an angular resolution of several radians, there is simply no
meaningful angular measurement, and the effective asymmetry $\Aeff$ is
close to zero.
The telescope then has no sizeable sensitivity to the polarisation of
the incoming radiation.
Since the value of the most probable opening angle varies like the
inverse of the photon energy \cite{Olsen:1963zz},
since the value of the RMS multiple scattering angle of a
relativistic charged
particle varies like the inverse of the particle momentum $p$
\cite{Patrignani:2016xqp}, and
since the distribution of the momenta of the leptons scales
approximately like the inverse of the photon energy \cite{Bethe-Heitler},
the variation of the dilution with thickness turns out to be
independent of energy \cite{Kelner,Kotov,Mattox}.

In the first, to my knowledge, successful experimental attempt to
demonstrate the feasibility of polarimetry with pairs, the authors
used the conversion of the photons of a 1.5 -- 2.4\,GeV beam on a
100-micron-thick carbon-foil target ($0.5 \times 10^{-3} X_0$),
with silicon detectors 2 meters
 downstream of the target and located on both sides of the beam,
measuring the lepton track positions \cite{deJager:2007nf}.
Such a set-up, which is well tuned to a high-intensity beam
measurement (with an efficiency of 0.02\,\%),
is obviously not suitable for a space mission, for which one wishes 
 a high-effective-area high-acceptance detector.

 \begin{table} \small \caption{Properties of some elements. \label{tab:pdg} }
  \hfill
 \begin{tabular}{c|cc|cc|cccc} \hline \hline
 & & & & & \\ \noalign{\vskip-7pt}
 & specific & radiation & & & density \\
 & radiation & length & specific & \\
 & length & & & \\
 & $\rho X_0$ & $X_0$ & $\gfrac{1}{\rho} \left.\gfrac{\dd E}{\dd x}\right|_{\mine} $ & $\left.\gfrac{\dd E}{\dd x}\right|_{\mine}$ & $\rho$ \\
 & & & & & \\ \noalign{\vskip-7pt}
 & $\gram / \centi\meter^{2}$ & $\centi\meter$ & $\mega\electronvolt \, \centi\meter^{2} / \gram $ & $\mega\electronvolt / \centi\meter$ & $\gram / \centi\meter^{3}$ \\
 & \cite{Patrignani:2016xqp} & & \cite{Patrignani:2016xqp} & &\cite{Patrignani:2016xqp} \\
 \hline
 C & 42.7 & 19.3 & 1.742 & 3.85 & 2.21 \\
 Al & 24.0 & 8.90 & 1.62 & 4.36 & 2.70 \\
 Si & 21.8 & 9.37 & 1.66 & 3.88 & 2.33 \\
 W & 6.76 & 0.35 & 1.15 & 22.1 & 19.3 
\end{tabular}
  \hfill ~
\end{table}

It was then realised that, thanks to the large high-angle tail that
the pair-opening-angle distribution exhibits, the most-probable
approximation of \cite{Kelner,Kotov,Mattox} is actually quite
pessimistic at large thicknesses
(Fig. 17 of \cite{Bernard:2013jea}\footnote{An updated version,
 obtained with the optimal ``bisectrix'' event-azimuthal angle
 definition from \cite{Gros:2016dmp} can be found in Fig. 7 of
 \cite{Bernard:2019znc}.}).
Also instead of using the simple expression of the scattering angle
after traversing a slab of matter as an estimate of the track angle
RMS deflection, one can perform the reconstruction of the angle at the
production vertex in an optimal way using a Kalman-filter based
tracking \cite{Fruhwirth:1987fm}, in which case the angular precision
varies like $p^{-3/4}$ in the homogeneous-detector approximation
(eq. (4) of \cite{Bernard:2012uf}).
The dilution of the polarisation asymmetry then does not depend on the
converter thickness only, but also on the photon energy: there is
still some sizeable sensitivity
in the
lowest-energy part of the spectrum (Fig. 20 of \cite{Bernard:2013jea}).
This opportunity was explored by two recent projects using homogeneous
detectors
\begin{itemize}
\item The GRAINE experiment \cite{Takahashi:2015jza} developed a
 high-density, (sub-micron) hyper-high-resolution emulsion tracker
 with which they measured the polarisation fraction of a GeV test
 beam \cite{Ozaki:2016gvw}.

\item The HARPO experiment \cite{Bernard:2018jql} developed a
 low-density, mild-resolution gaseous time projection chamber (TPC),
 with which we measured the polarisation fraction of a MeV test beam
 \cite{Gros:2017wyj}.
\end{itemize} 
In both cases, the dilution of the polarisation asymmetry was found to
be excellent.

The present-day gamma-ray missions in orbit, though, the \FermiLAT\ 
\cite{Atwood:2009ez} and AGILE \cite{Tavani:2008sp} and the projects
AMEGO \cite{McEnery:2019tcm} and ASTROGAM \cite{DeAngelis:2016slk} use
telescopes in which the active target consists of a stack of
silicon strip detectors (SSDs) with thicknesses of 300 to 500 microns,
that is of (3.2 to 5.3)$\times 10^{-3} X_0$ (Tab. \ref{tab:pdg})
and that therefore in the present context can be described to be thick
wafers (Tab. \ref{tab:telescopes}), in particular when compared with
the state of the art \cite{Suljic:2021hfl}.
\begin{table}[t] 
% \small
% \footnotesize
 \scriptsize
 \caption{Properties of several past and future trackers for
 gamma-ray space missions. \label{tab:telescopes} }
  \hfill
\begin{tabular}{l|c|cccccc}
 & & ASTROGAM & AMEGO & LAT & AGILE \\ \hline
 Reading & & DSSSD & DSSSD & SSSSD & SSSSD \\
 Wafer thickness & $e$ & 500. & 500. & 400. & 410. & $\micro\meter$ \\
 Layer number & $N$ & 56 & 60 & 19 & 14 \\
 Distance & $\ell$ & 1. & 1. & 3. & 1.9 & cm \\
 Strip pitch & $\pitch$ & 240 & 500 & 228 & 242 & $\micro\meter$ \\
 Critical energy & $E_c$ & 67. & 32. & 211. & 126. & $\mega\electronvolt$ \\
 & & \cite{DeAngelis:2016slk} & \cite{McEnery:2019tcm} & \cite{Atwood:2009ez}& \cite{Tavani:2008sp} 
\end{tabular}
  \hfill ~
\end{table} 
Polarimetry with such detectors has never been demonstrated, to my
knowledge, be it by the analysis of the data collected from a detector
prototype exposed to a polarised gamma-ray beam nor from studies of
simulated data, which means that a non-zero, strictly positive value
of the effective polarisation asymmetry $\Aeff$ has never been obtained for a
SSD-based active-target polarimeter.
This is most likely what has lead the authors of prospective studies
to {\bf assume} a value of $\Aeff$ to issue a sensitivity
\cite{Tavani:2008sp,Giomi:2016brf}.
We can surmise that this inability to demonstrate a non-zero value of
$\Aeff$ may be due to a number of reasons or to a combination thereof:
\begin{itemize}
\item The dilution due to multiple scattering, as mentioned already.
Telescopes (AMEGO, ASTROGAM) that use double-sided silicon strip
detectors (DSSSDs) have a fraction of the conversion wafer,
downstream of the vertex, inflicting
multiple scattering on the tracks (i.e. a fraction of 
$\approx 4\times 10^{-3} X_0$) even before they exit from the conversion wafer;
in the case of single-sided SSDs (the LAT, AGILE), a pair of wafers is
needed with orthogonal strip orientation, and the thickness is even
larger.

\item The impossibility to reconstruct the azimuthal angle in the
 second layer when the two tracks ``fall'' in the same
 strip for the two transverse directions $x$ and $y$: the first track
 segments of the two tracks are in this case reconstructed as a single
 track, and there is no azimuthal information available.
Given the value of the most probable value $\hat\theta_{+-}$, of the
pair opening angle, $\theta_{+-}$ \cite{Olsen:1963zz}
\begin{equation}
 \hat\theta_{+-} = \gfrac{1.6\,\mega\electronvolt}{E} ~ \radian ,
\label{eq:theta:pm}
\end{equation}
and the limit pair opening angle that can be measured from the two
first layers, $\theta_c = \pitch /\ell$
(from pure geometry, no multiple scattering assumed)
where $\pitch$ is the strip pitch and $\ell$ the distance between
layers, it is possible to define a critical photon energy value
$E_c$ for which the most probable pair opening angle is equal to the
limit pair opening angle by
\begin{equation}
 E_c \equiv 1.6\,\mega\electronvolt \gfrac{\ell}{\pitch} .
 \label{eq:critical:energy}
\end{equation}
Since most cosmic sources have a flux that decreases strongly with photon
energy, $E_c$ gives the energy scale of the photon candidates that should
contribute mainly to the polarimetry measurement, would the telescope
effective area already plateau at that energy.
For the \FermiLAT, $E_c = 211\,\mega\electronvolt$ (Tab. \ref{tab:telescopes}).

\item For the \FermiLAT\ and for AGILE, for which a high value of the
 telescope effective area was obtained by interleaving the silicon
 detectors with high-$Z$ (tungsten) foils,
 the W foils must be located above the detectors (which is actually
 the case) and events with a conversion in the Si wafers should be
 identified (and those in W foils, rejected) with a high efficiency
 (something that was actually proven to be possible \cite{Giomi:2016brf}).
 
\item As a pad segmentation of the reading electrodes was out of reach
 due to power consumption limitations in orbit, two orthogonal series of strips
 are used, and in the case where the track hits produce separate clusters in both
the $x$ and $y$ strips, a two-fold ambiguity arises.
For homogeneous detectors, the issue is easily solved by matching the
violently varying profiles of the energy deposition along the tracks
(deposited-charge track matching, see Fig. 6 of \cite{Bernard:2012jy}).
For DSSSDs, deposited-charge track matching
can also be considered.
Even in case where no track-matching scheme is available, or if the
scheme used is inefficient, the azimuthal information is
still present in the data and, using the two possible configurations of
the event (each weighted by a factor of 1/2), yields, at worst, an
induced dilution factor of $D = 0.5$
(see Sec. \ref{sec:4tracks}).
\end{itemize}

In this paper I characterise the performance of telescopes using
DSSSD-based active targets for the polarimetry of linearly polarised
gamma-rays.

\begin{figure}[htbp!]
\hspace{4cm}
 \setlength{\unitlength}{1.375pt}
 \begin{picture}(300,300)(-210,0)
 \put(-210,0) {
% \hfill
 \includegraphics[width=7.73cm]{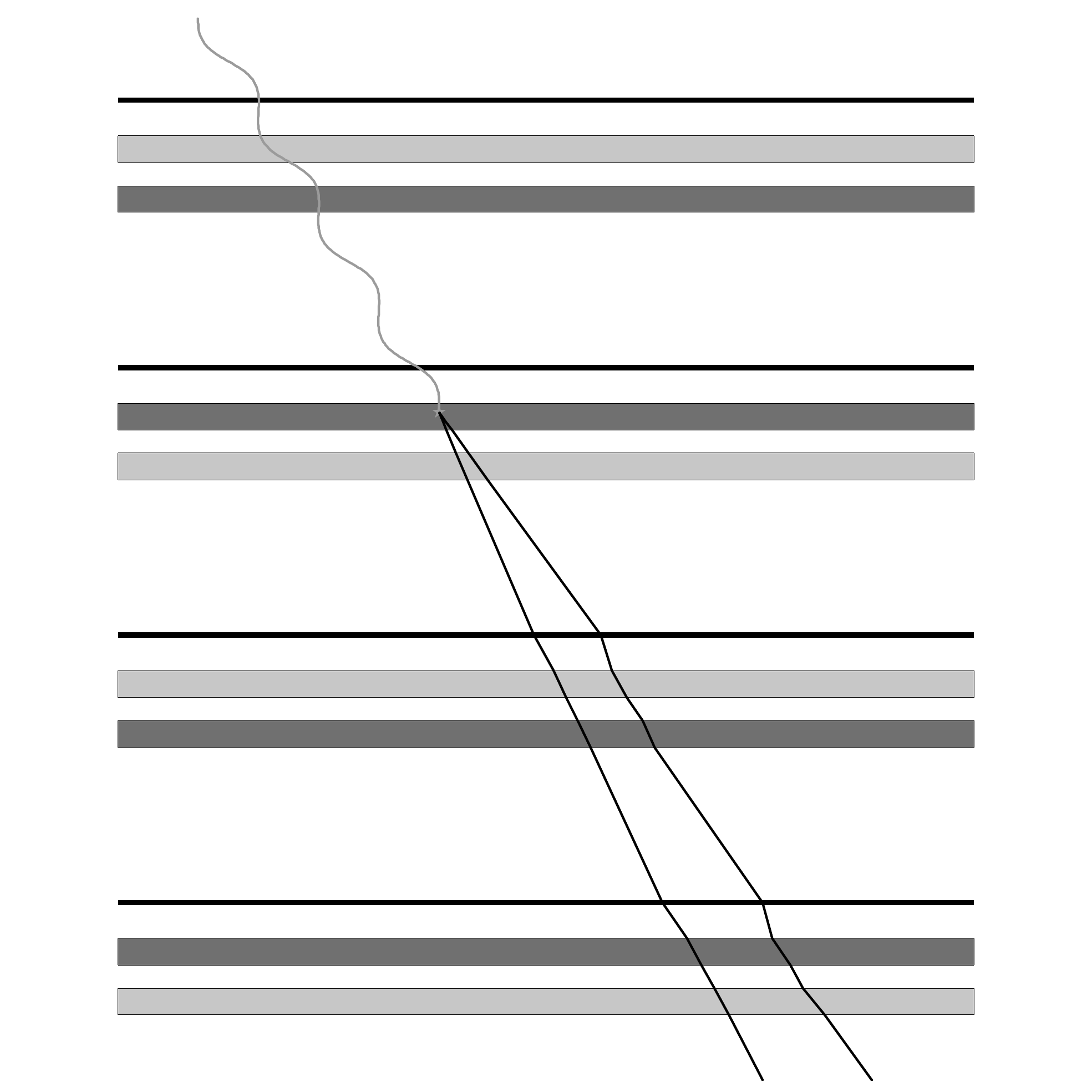}
% \hfill ~
 }
{
 \footnotesize
% \scriptsize
 %\tiny
 \put(-50,30){Tungsten}
 \put(-50,70){Tungsten}
 \put(-50,110){Tungsten}
 \put(-50,150){Tungsten}
 \put(-50,20){$x$ Wafer}
 \put(-50,60){$y$ Wafer}
 \put(-50,100){$x$ Wafer}
 \put(-50,140){$y$ Wafer}
 \put(-50,10){$y$ Wafer}
 \put(-50,50){$x$ Wafer}
 \put(-50,90){$y$ Wafer}
 \put(-50,130){$x$ Wafer}
 \put(-250,94){1st Layer}
 \put(-250,54){2nd Layer}
 \put(-220,4){ \resizebox{2cm}{!}{ \{ }}
 \put(-250,10){Rest of event}
 \put(-185,160){$\gamma$}
 \put(-105,0){$e^+$}
 \put(-75,0){$e^-$}
 }
 \end{picture}
\caption{Schema of the test detector model (not to scale).
Adapted from \cite{Atwood:2007ra}.
The thick low-density aluminium structural material is not shown.
\label{fig:schemaFermiLAT}
 }
\end{figure}
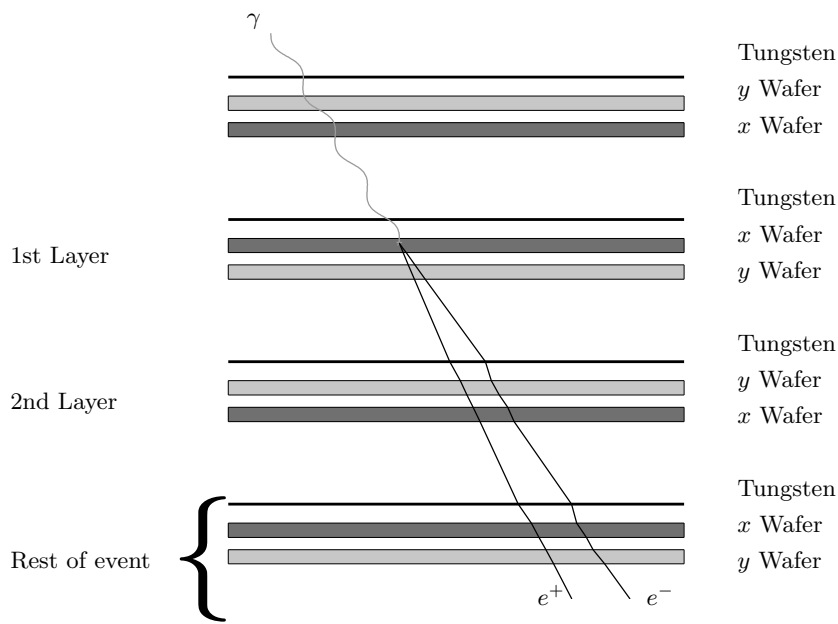

\section{Test detector model}
\label{sec:Model}

The test detector model used in this study is a simplified version of
the \FermiLAT\ active target (Fig. \ref{fig:schemaFermiLAT}): 
\begin{itemize}
\item
The longitudinal (vertical) structure is similar to that described in 
\cite{Atwood:2007ra}:
\begin{itemize}
\item 18 measurement layers at a distance of $3 \, \centi\meter$ from
 each other, each consisting of a pair of $400\,\micro\meter$-thick
 silicon wafers with a $2\,\milli\meter$ gap between them, one
 measuring the $x$, the other the $y$, transverse (horizontal)
 coordinate, with a pitch of $\pitch = 228\,\micro\meter$.
\item Each of the top 16 layers includes a tungsten foil, the
 first 12 planes of $95\,\micro\meter$ in thickness ($t = 2.7\,\%$),
 while the final four are $720 \,\micro\meter$ thick ($t = 18\,\%$).
\item Vented aluminium honeycomb panels, with an average density of
 $48 \,\kilo\gram/\meter^3$ for the heavy-converter and bottom trays and
 $16 \, \kilo\gram/\meter^3$ for the others, are filling the
 $3 \, \centi\meter$ space between layers to give to the structure
 its rigidity.
\end{itemize}
\item The horizontal structure, on the contrary, is different from
 that of the LAT: a unique tower-sized square wafer with 1536 strips
 in each direction,
 that is, $35\, \centi\meter$, with one of the central strips set to
 contain the conversion point.
\end{itemize}
The charge deposited by the passage of the leptons through the wafer
is assumed to be collected in the strip right below the creation
point, as charge sharing was shown to be small \cite{Brigida:2004ff}.
The discrimination threshold is set at $35\,\kilo\electronvolt$, which
corresponds to approximately 1/4 of the energy deposited by a
minimum-ionising particle crossing the wafer at a right angle.

\section{Assumptions}
\label{sec:Assumptions}

I simulate a point-like source and I assume that the combination of
the path of the LAT on its orbit, of the variation of the orbit and of
the attitude of the detector makes the exposure for that source
isotropic, in LAT coordinates.
I assume that the data have been analysed and that a list of
photons associated with that source has been obtained and therefore that 
\begin{itemize}
\item The direction of propagation of each incident photon is perfectly known
to be coming from the source;
\item An estimate of the photon energy is available.
\end{itemize}

\section{Simulation}
\label{sec:Simulation}

After an isotropic irradiation is generated, photons coming from above
the tracker are selected, and a $\cos{\theta_{\LAT}} > 0.25$ cut is
applied, which corresponds approximately to the angle of $75\degree$
for which the LAT acceptance reaches zero
\cite{Fermi:LAT:Performance}.

The conversion is generated with the five-dimensional, polarised event
generator documented in \cite{Bernard:2018hwf}, after which the
propagation of the leptons in the tracker is simulated with the EGS5
software, version 1.0.5 \cite{Hirayama:2005zm}.

\section{Tracking}
\label{sec:Tracking}

Before I engage in the description of the details of the
reconstruction, a couple of words about tracking are in order.
The Kalman-filter-based technology that is commonly used for the tracking of
the trajectories of charged particles in a tracker
is in principle able to make use optimally of all the information
available, taking into account in an appropriate way the correlations
induced by multiple scattering at each slice of the tracker
\cite{Fruhwirth:1987fm}.
The \FermiLAT\ documentation states that their tracking is able to
assign correctly hits to their original track with high certainty
\cite{Atwood:2009ez}
% in 2.4.1 pp 1082-3.
but the assignment performance in the second
layer, very close to the vertex and which is critical for the present
study, needs to be established.

Let us define ``the rest of the event'' to be the hits registered in the
tracker downstream with the exception of those in the first and the
second layer, and let's suppose that the hit-to-track assignment is
performed perfectly in the rest of the event. Let us fit them to a track.
The precision of track parameter measurements with an optimal tracking
in a segmented (i.e., not homogeneous) tracker was studied in
\cite{Frosini:2017ftq,Bernard:2019znc}.
The RMS angular resolution for a track impinging normally on the
tracker, $\sigma_{\theta}$,
is\footnote{Note that the specific energy loss of the leptons along
 their path in the detector, $\dd E / \dd x$, was neglected, so the
 momenta are assumed to be unchanged over the full trajectory.}
\begin{equation} 
 \sigma_{\theta} =
 \gfrac{\sigma}{\ell}
 \sqrt{
 \frac
 {2 \, x^3 \, \left(\sqrt{4 j - x^2} + \sqrt{- 4 j -x^2} \right)}
 { \left(\sqrt{4 j - x^2} + j x \right) \left(\sqrt{- 4 j - x^2} - j x \right)}
 }
 ,
 \label{eq:sigma:theta:track}
\end{equation}

where $x$ is the distance between wafers, $\ell$, normalised to the
detector scattering length $\lambda$ \cite{Innes:1992ge},
 $x\equiv \ell/\lambda = \sqrt{({\ell}/{\sigma}) ({p_0}/{p}) \sqrt{{\Delta}/{X_{0,W}}}}$,
$j$ is the imaginary unit, and other variables are defined in Table
 \ref{tab:geo}.

\begin{table}[h] \caption{Simple geometry for the tracking section.} \label{tab:geo} \small
 \begin{tabular}{llrlllllll} \hline \hline
distance between layers & $\ell$ & 3 & cm \\
single wafer precision & $\sigma =\pitch/\sqrt{12}$ & 66 & $\micro\meter$ \\
track momentum & $p$ \\
multiple scattering constant & $p_0$ \cite{Patrignani:2016xqp} & 13.6 & $\mega\electronvolt/c$ \\
tracker scattering momentum & $p_1$ see eq. (\ref{eq:p1}) & 0.367 & $\mega\electronvolt/c$ 
 \\ 
tracker scattering length & $\lambda$ see \cite{Innes:1992ge} \\
scatterer thickness & $\Delta$ & 95 & $\micro\meter$ \\
scatterer radiation length & $X_{0,W}$ & 0.35 & cm \\
\end{tabular}
\end{table}

\newtoggle{couleur}
\toggletrue{couleur}
%\togglefalse{couleur}

\begin{figure}[htbp!]
 \hfill
\iftoggle{couleur}{
 \includegraphics[width=12cm]{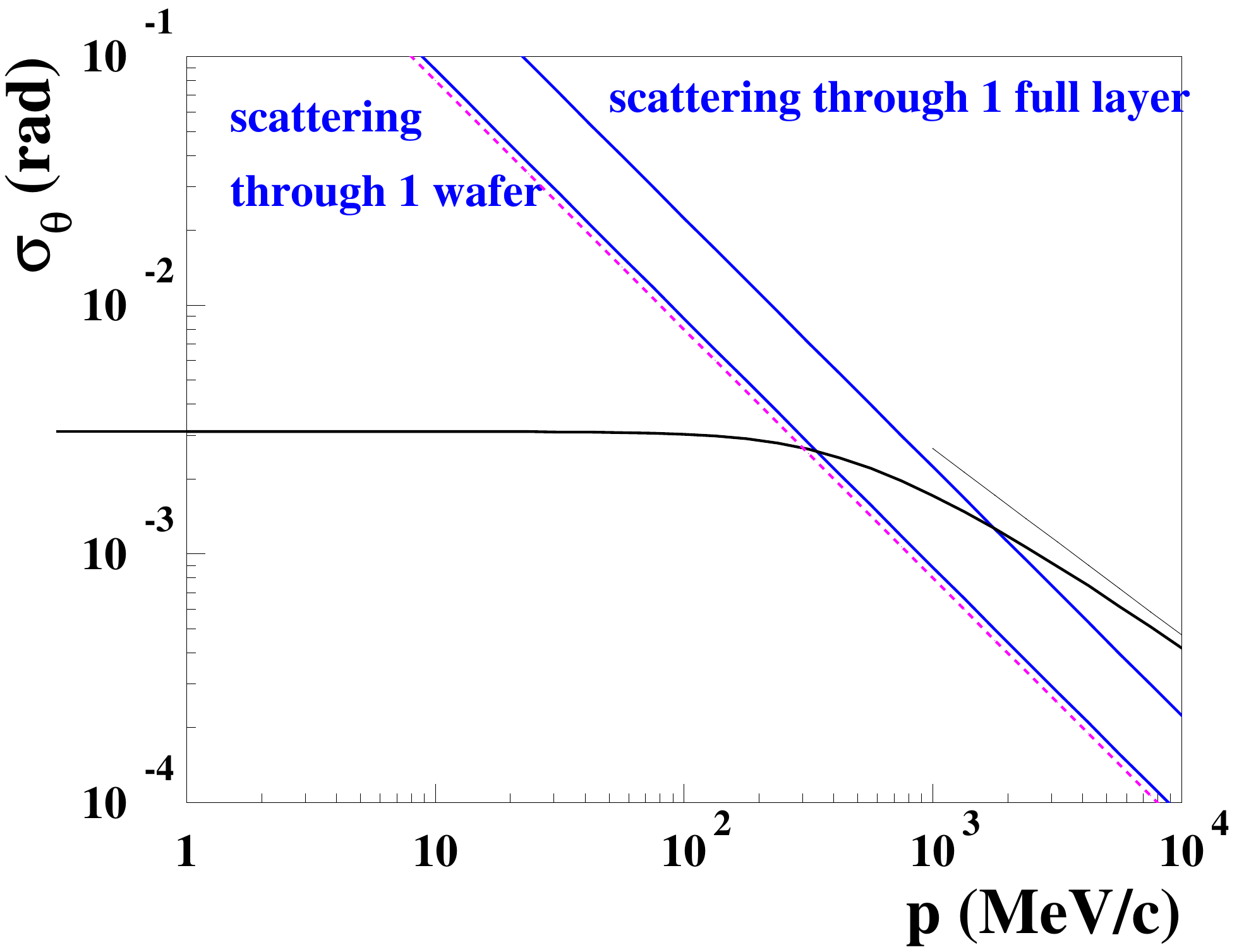}
 \put(-100,80){ \Magenta {$\hat{\theta}_{+-} $}}
}{
 \includegraphics[width=12cm]{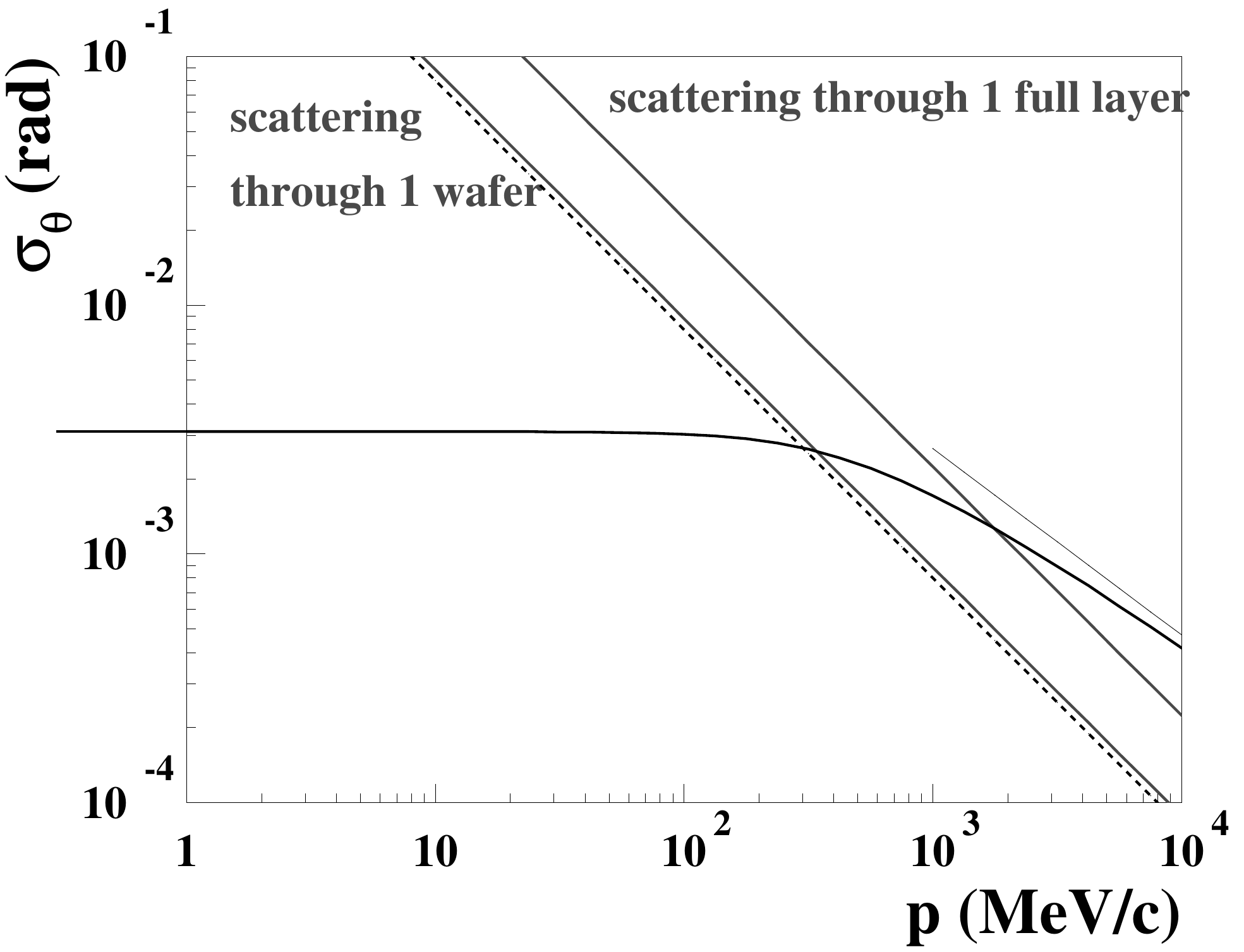}
 \put(-100,80){ $\hat{\theta}_{+-} $}
 }
 \put(-365,140){ $\sqrt{2} \sigma/\ell$}
 \put(-205,150){eq. (\ref{eq:sigma:theta:track})}
 \put(-70,135){ $({p}/{p_1})^{-3/4}$ }
 \hfill
 ~
\caption{Single-track RMS angular resolution of a segmented detector as a function of track momentum (eq. (\ref{eq:sigma:theta:track})).
 % The small-$x$ approximation, valid at high energies, is eq. (22) of \cite{Frosini:2017ftq},
 The thin high-$p$ line is the $\sigma_{\theta} \approx \left({p}/{p_1}\right)^{-3/4}$ ``homogeneous detector'' asymptote \cite{Bernard:2013jea}).
% The high-$x$ approximation, valid at low energies, is eq. (25) of \cite{Frosini:2017ftq}).
The low-$p$ asymptotic value is shown by the horizontal line at $\sigma_{\theta} \approx \sqrt{2} \sigma/\ell$.
The inclined continuous straight lines show the RMS multiple scattering;
the inclined dashed straight line show the pair opening angle
for equipartition ($E = 2p$), for comparison.
\label{fig:tracking}
 }
\end{figure}
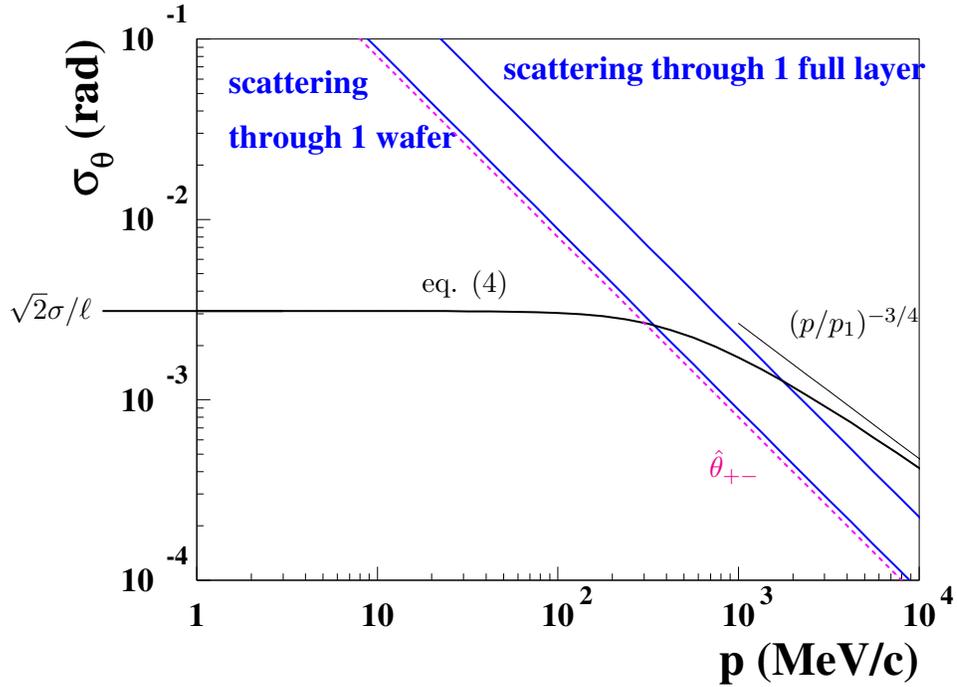

The variation of $\sigma_{\theta}$ with $p$ is shown in
Fig.~\ref{fig:tracking} for a track traversing the full 12 layers of
the ``front'' part of the tracker and a conversion at the very bottom
of the bottom wafer of a layer.
Figure ~\ref{fig:tracking} shows two regimes:
\begin{itemize}
\item at low momentum (large $x$, coarsely-segmented detector),
an optimal measurement can be obtained simply from the position
measurements in the two first wafers, no Kalman filter is needed:
asymptotically at very low momentum, the angular resolution is
obviously that of a segment between two measurement points, 
$\sigma_{\theta} \approx \sqrt{2} \sigma/\ell$.
 
\item at high momentum (small $x$), 
the thin straight line is the homogeneous-detector asymptote 
\cite{Bernard:2013jea},
$\sigma_{\theta} \approx \left({p}/{p_1}\right)^{-3/4}$, where the
momentum $p_1$ that characterises the
tracking-with-multiple-scattering properties of the homogeneous
detector is
\begin{equation}
 p_1 = p_0
 \left( \gfrac{2 \sigma}{\ell} \right)^{1/3}
 \left( \gfrac{\Delta}{X_{0,W}} \right)^{1/2} .
 \label{eq:p1}
\end{equation}
\end{itemize}

The limit between the high- and low-momentum ranges can be defined as
$x = 2$ \cite{Frosini:2017ftq}, which corresponds to
$p = 255\,\mega\electronvolt/c$ for the (front part of the) \FermiLAT.
The thicker curve represents the result
(eq. (\ref{eq:sigma:theta:track})) of the exact calculation
\cite{Bernard:2019znc}.
These expressions have been verified by comparison with the angular
resolution of an actual Kalman-filter-based tracking applied on a
Gaussian-generated tracker with Gaussian-distributed multiple
scattering distributions (Figure 8 of \cite{Frosini:2017ftq}).

In addition to the angular resolution discussed above for a conversion
just at the entrance of the 3\,cm gap, for a conversion inside a wafer
or in the tungsten foil, the multiple scattering inside that layer must
be taken into account.
The two parallel straight lines represent this contribution for
the crossing of a full layer and for the crossing of a full single
wafer, respectively.
\begin{itemize}
\item For a conversion in the tungsten, the multiple scattering in the first layer dominates the angular resolution below $1\,\giga\electronvolt/c$;
\item 
For a conversion in a bottom wafer, it dominates below
$0.3\,\giga\electronvolt/c$.
\end{itemize}

In addition, I have indicated with a dashed line the most probable
value of the opening angle \cite{Olsen:1963zz} for equipartition
($E \approx 2 p$):
wherever the conversion point (silicon, tungsten), the
projection of the opening angle on the second layer, that is, the
distance between the two tracks, is smaller than the resolution of the
position of each of the tracks as extrapolated from the tracking over the rest
of the event:
we can surmise that the contribution to track matching in the second
layer, from the transportation of the track matching in the rest of
the event even if assumed to be perfect, is most likely to be of little
help.

\section{Reconstruction}
\label{sec:Reconstruction}

The simple LAT model described above is exposed to an isotropic
gamma-ray irradiation from above with $\cos{\theta_{\LAT}} > 0.25$.
At this point the model is a simple stack of transversely infinite
planes of silicon, tungsten and low-density aluminium in vacuum.
The conversion point (CP) is recorded and a single tower is built with
the strips containing the CP at its centre.
It is checked that the transverse leakage from that ``tower'' is
 negligible.
Therefore the side effects induced by the dead zones at the wafer
borders and at the tower borders of the actual \FermiLAT\ are not
addressed in this study.
No dead channels, no wafer misalignments are simulated.
I require that three successive layers see a signal above threshold,
so as to emulate the main ingredient of the trigger of the \FermiLAT\ .

The conversion layer is defined as the first layer, starting
from above, that has a non zero signal in a silicon wafer.
After discrimination over threshold has been applied, the hit pattern
in the strips of the first and of the second layer is recorded.
Clusters are formed from groups of consecutive hit strips.
Track positions measured in that wafer are defined to be the position
of the geometrical centre of clusters.
The nature of the material in which the conversion took place is recorded: 
\begin{itemize}
 \item lower wafer;
 \item upper wafer;
 \item W front (a thin tungsten foil);
 \item W back (a thick tungsten foil).
\end{itemize}

We should keep in mind that this is a Monte Carlo information that is
used in the analysis, and that is not directly available in the
analysis of the data of the actual \FermiLAT.
\begin{itemize}
 \item 
Conversions in the lower wafer can be easily identified as no signal
is collected in the neighbouring upper wafer (up to sub-threshold
charge deposition, dead channel and border proximity analysis). 
 \item 
The identification of the conversions in the upper wafer has been
addressed for the actual \FermiLAT\ in \cite{Giomi:2016brf}.
\end{itemize}

The cluster configurations that will be of some use in the present
study are therefore:
\begin{itemize}
\item at least one cluster in the 1st layer;
\item 0 or 1 cluster in each wafer of the 1st layer;
\item 1 or 2 cluster(s) in each wafer of the 2nd layer;
\item 2 clusters in at least one wafer of the 2nd layer.
\end{itemize}

\begin{figure}[t]
 \hspace{-0.7cm}
 \begin{center}
 \includegraphics[trim=6 6 40 6 ,clip,width=3.0cm]{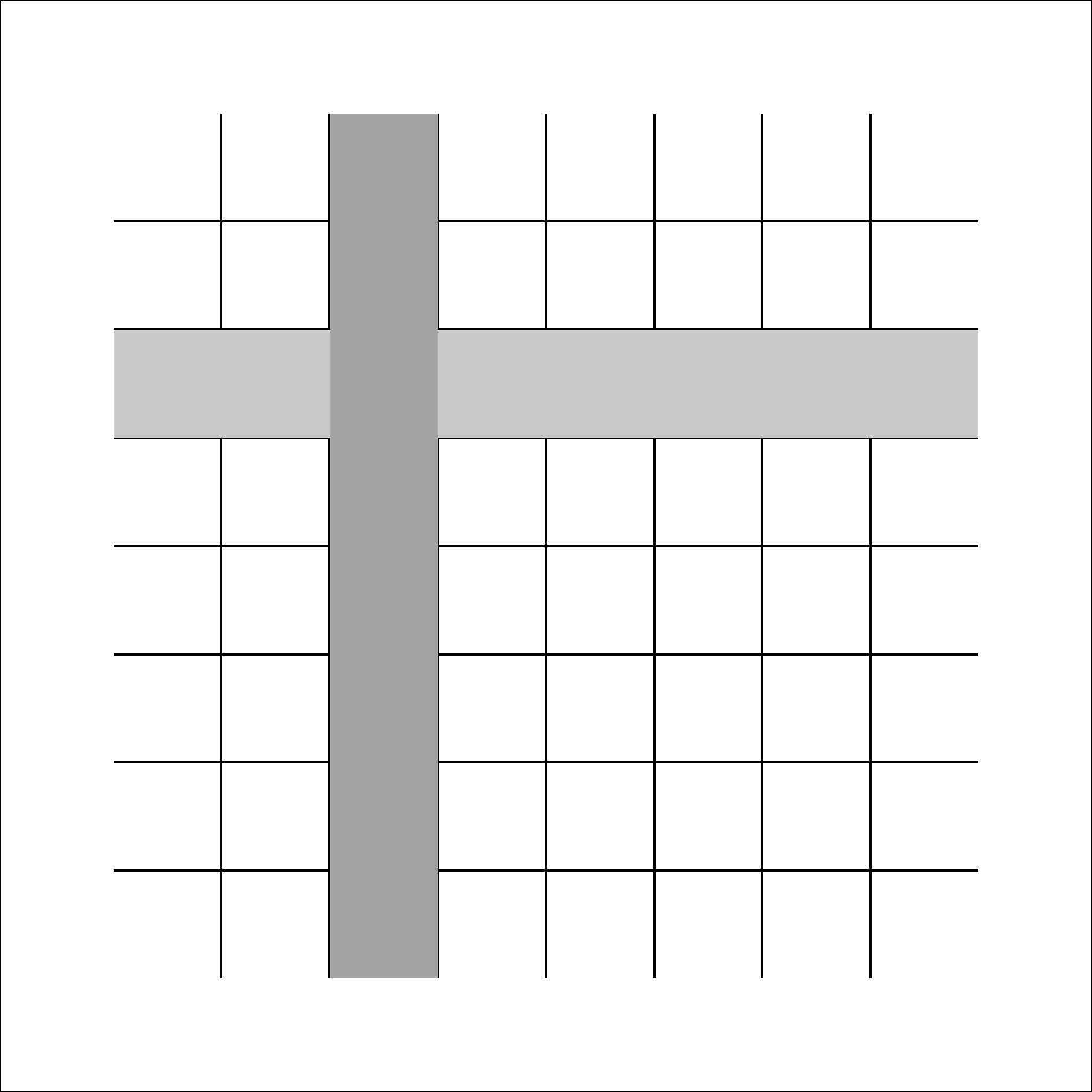}
 \put(-54,56){\Red{$\bullet$}}
 \put(-59,59){\Blue{$\bullet$}}
 \put(-50,0){$\ntrack =1$}
 ~
 ~
 \includegraphics[trim=40 6 40 6 ,clip,width=3.0cm]{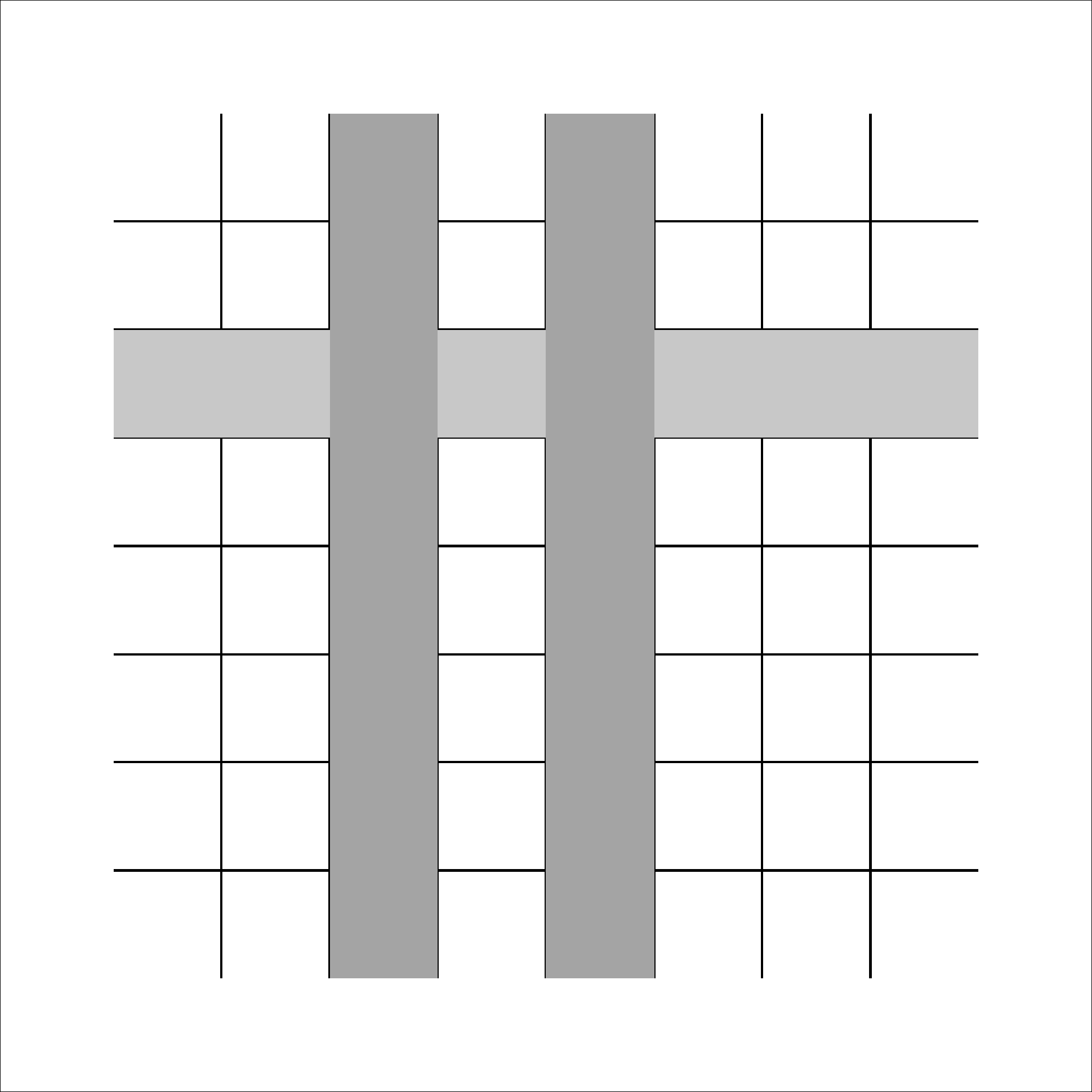}
 \put(-40,61){\Red{$\bullet$}}
 \put(-59,63){\Blue{$\bullet$}}
 \put(-50,0){$\ntrack =2$}
 ~
 ~
 \includegraphics[trim=40 6 6 6 ,clip,width=3.0cm]{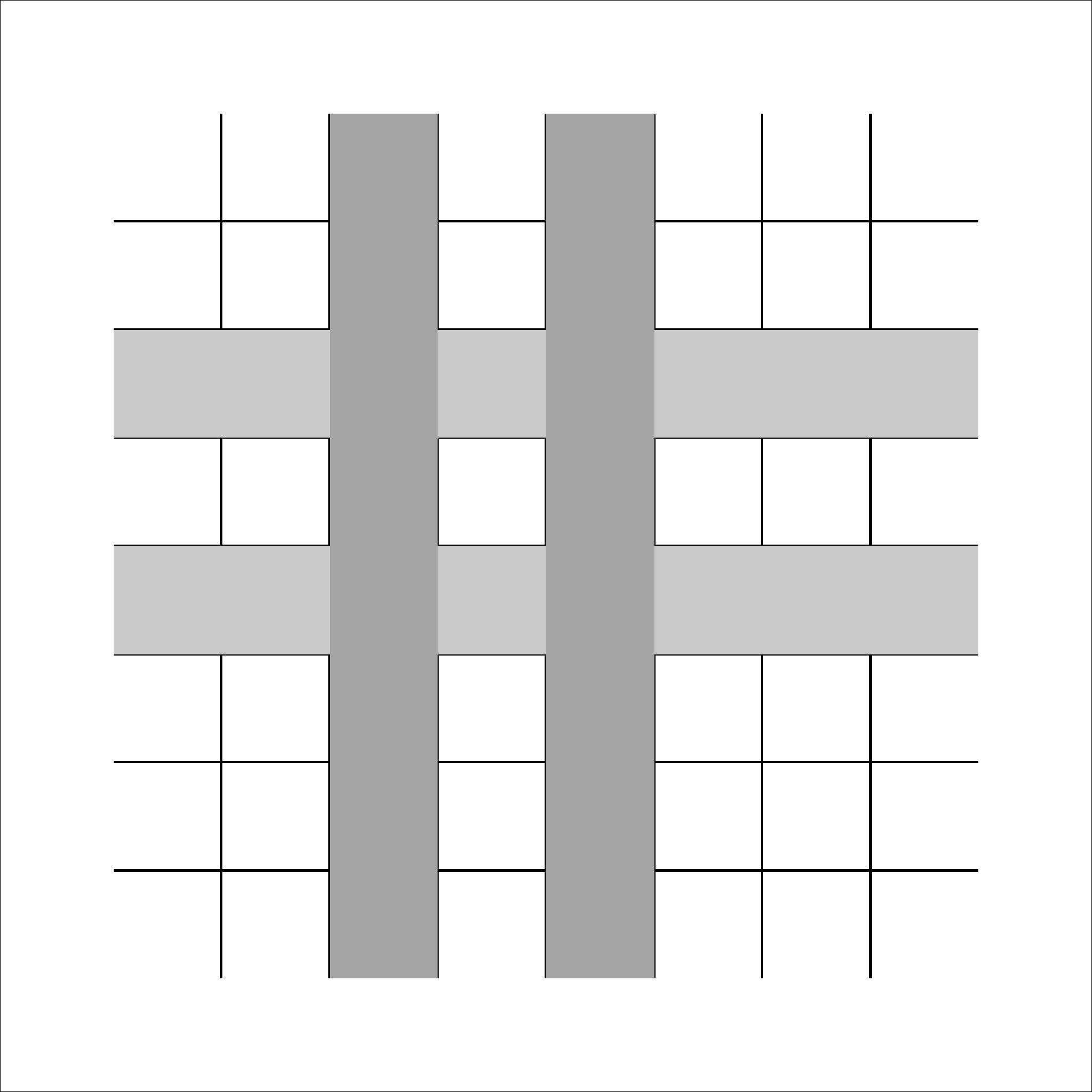}
 \put(-45,56){\Red{$\bullet$}}
 \put(-63,40){\Blue{$\bullet$}} 
 \put(-50,0){$\ntrack =4$}
 \caption[~]{The three possible configurations in the 2nd layer, with
 the number of reconstructed tracks mentioned.
The bullets show the true, unknown to the experimentalist, position of
the leptons when crossing that layer.
\label{fig:schema:next:layer} }
 \end{center}
\end{figure}
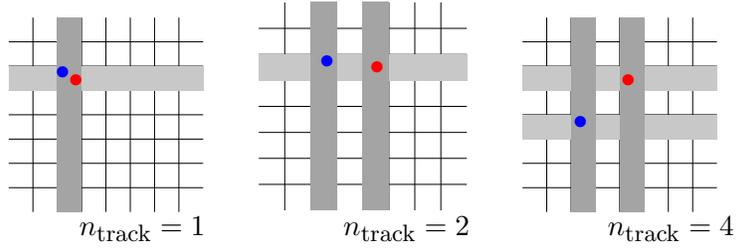

I name $n_i$ the number of clusters in layer $i$ and $n_{xi}$ and
$n_{yi}$ the number of clusters in wafers $x$ and $y$ of layer $i$,
respectively, $i=1, 2$.
\begin{itemize}
\item Possible track crossings are built from the couple of two
 clusters, one $x$ the other $y$ in the same layer.
\item Track candidates are built from pairs of track crossings, one in
 the first layer the other in the second.
\item In the case where 2 track candidates were found, 1 photon candidate is
 built from that pair. 
In the case where 4 track candidates were found, 2 photon candidates are
built from those, using track pairs that have no cluster in common.

As alluded in the previous section, the reliability of the matching of
tracks that would be measured, even perfectly, in the rest of the
detector (angles and momenta) with the hits in the second layer, is
considered to be dubious enough, in this study, that I consider that
the momenta of the track candidates are not known, and therefore I
obtain the direction of a reconstructed photon as the bisectrix of
the directions of its two tracks.
\end{itemize}

\begin{figure}[t]
\begin{center}
\iftoggle{couleur}{
\includegraphics[width=0.98\linewidth]{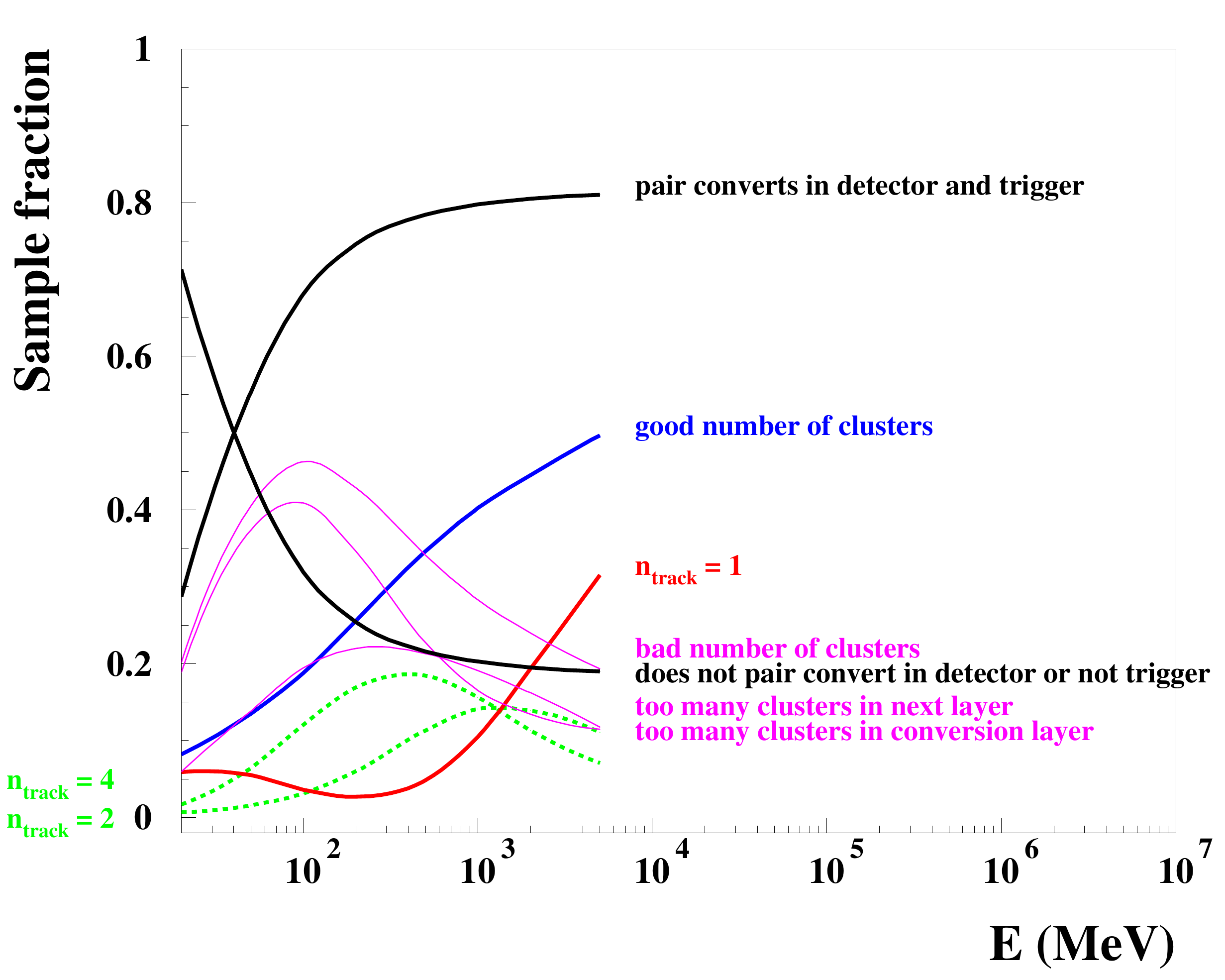}
}{
 \includegraphics[width=0.98\linewidth]{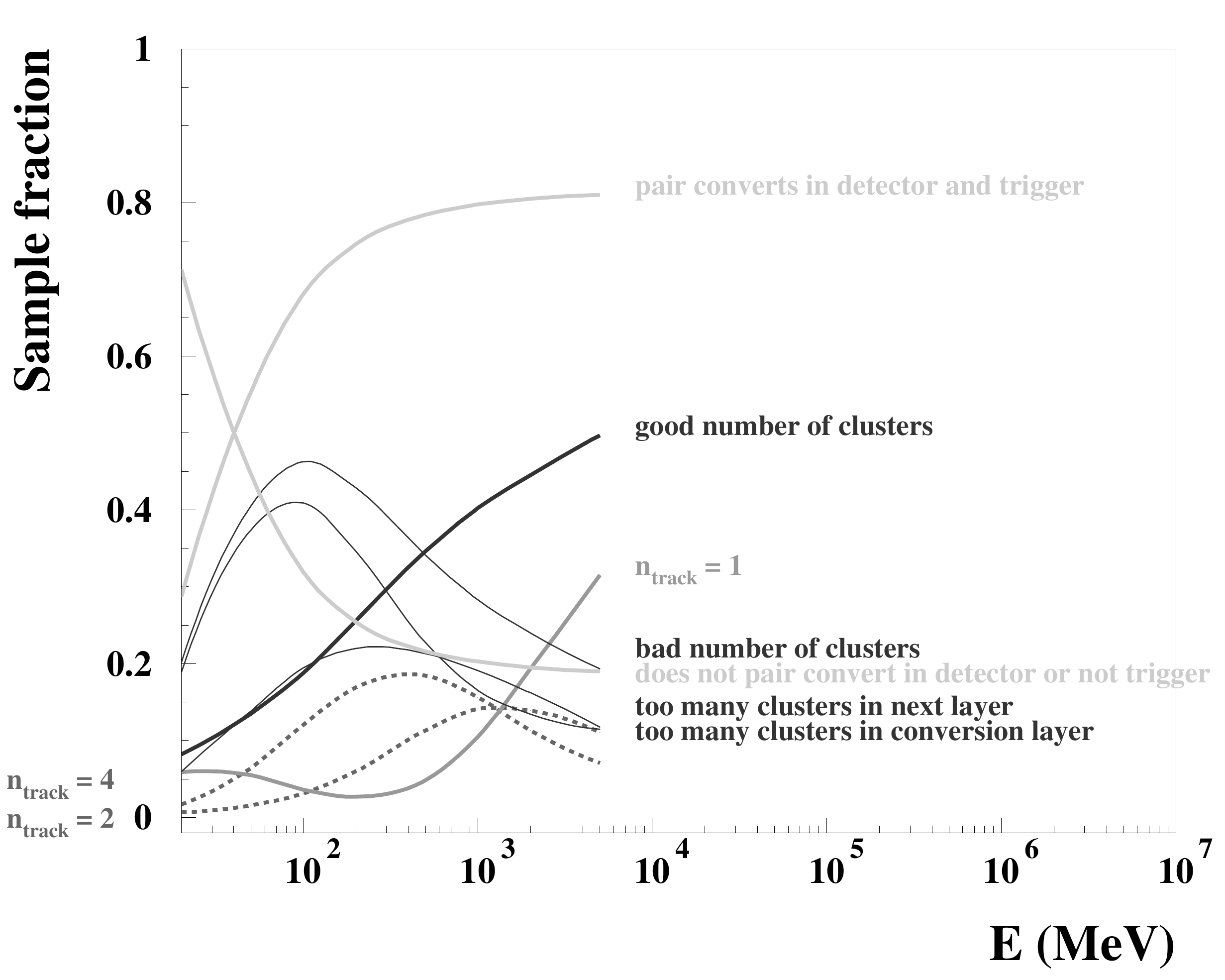}
 }
\caption{Event fractions as a function of incident photon energy.
 Dashed curves show the fractions for the useful event categories
 (good number of clusters, $\ntrack=2$ or $\ntrack=4$).
 Thinner curves show the fractions for the event categories with a bad number of clusters.
 \label{fig:event:fraction}}
\end{center}
\end{figure}

So there are two classes of events that are useful for polarimetry
(Fig. \ref{fig:schema:next:layer}):

~

\begin{minipage}{\textwidth}
 %\noindent
 %\scriptsize
 % \footnotesize
 \small
%\hspace{-1cm}
\begin{tabular}{lll}
 $(n_{x2}=1 \cap n_{y2}=2) \cup (n_{x2}=2 \cap n_{y2}=1)$ & $\ntrack =2$ & $n_{\gamma} =1$ \\
% \\ \hline
$n_{x2}=2 \cap n_{y2}=2$ & $\ntrack =4$ & $n_{\gamma} =2$ \\ 
\end{tabular}
\end{minipage}

~

\begin{itemize}
\item In the case of an $n_{\gamma} =1$ event, the two tracks ``fall''
in the same cluster of one and only one wafer of the 2nd layer
(Fig. \ref{fig:schema:next:layer}, centre).
The azimuthal angle cannot really be ``measured'' then, only a $0\degree$
or a $90\degree$ value can be assigned, in the LAT frame, depending on
the orientation of the wafer that collected two clusters.
Some amount of polarimetry information is still present, though, in
that assignment, as we shall see.

\item In the case of $n_{\gamma} =2$, only one photon candidate is
 genuine, and the other is a combinatorial fake
 (Fig. \ref{fig:schema:next:layer}, right).
The polar angles of these two candidates, with respect to the known
direction of the source, are the same, so that variable cannot be used
for a discrimination.
Some amount of polarimetry information is still present, though, as
already alluded to in the discussion about ``track matching'' in Sect.
\ref{sec:Measurement}.
\end{itemize}

As conversions in a lower wafer are identified from the absence of a
signal in the corresponding upper wafer that is measuring, say, the
$x$ coordinate, the reconstruction of the event is modified: 
\begin{itemize}
\item If $n_{x2}=1$, the projection of the unit vector in the ($x, z$)
plane is taken to be that on the known direction of the incoming photon;

\item If $n_{x2}=2$, the directions of the tracks are taken so that
 the known direction of the incoming photon is their bisectrix;
\end{itemize}

The variation of the fraction of conversion events for various
configurations, and in particular for the $\ntrack =2$ and
$\ntrack =4$ classes that are of interest to this work (dashed
curves), is shown in Fig. \ref{fig:event:fraction} as a function of
the incident photon energy.

\section{Definition of the event azimuthal angle}
\label{sec:Def:Azim}

\begin{figure}[htbp!]
 \begin{center}
 \setlength{\unitlength}{1.375pt}
 \hfill
 \begin{picture}(300,300)(-210,0)
 % \hfill
 \put(-210,0) {
 \includegraphics[width=9.6cm]{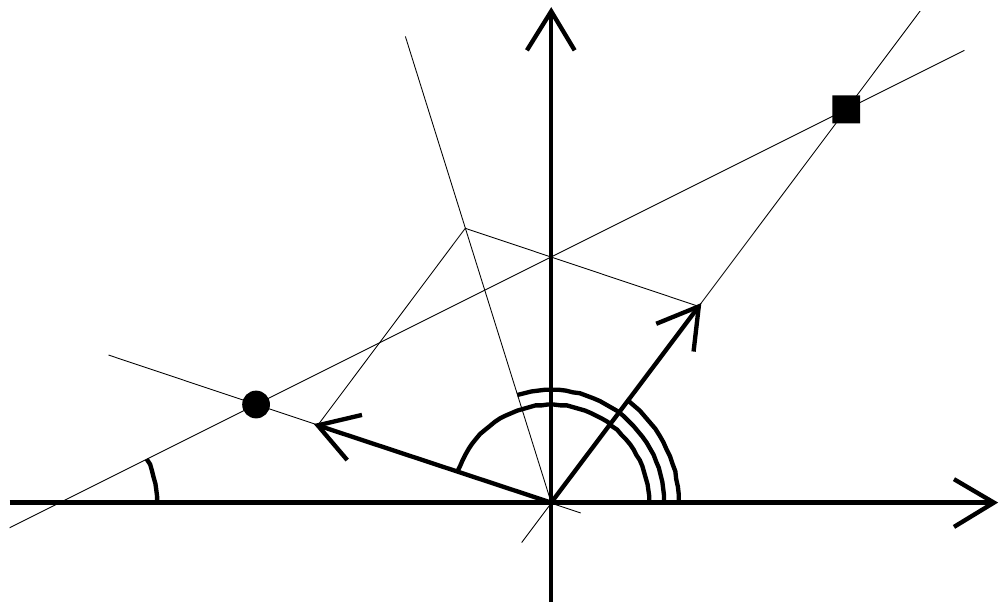}
 }
 \put(0,30){$x$}
 \put(-90,130){$y$}
 \put(-165,25){$\omega$}
 \put(-40,107){$e^+$}
 \put(-157,47){$e^-$}
 \put(-70,30){$\varphi_+$}
 \put(-125,35){$\varphi_-$}
 \put(-86,48){$\phi$}
 \put(-86,8){$\gamma$}
 \end{picture}
 \caption{Azimuthal angles definition in the momentum plane perpendicular to the direction of the incident photon (adapted from \cite{Bogdan:1998}).
 $\varphi_+$ and $\varphi_-$ are the azimuthal angles of the positron and of the electron, respectively, $\phi$ is their bisectrix, and $\omega$ is the
 azimuthal angle defined by \cite{Bogdan:1998}.
 \label{fig:event:les:angles}}
\end{center}
\end{figure}

\begin{figure*}[htbp!]
\iftoggle{couleur}{
\includegraphics[width=0.32\linewidth]{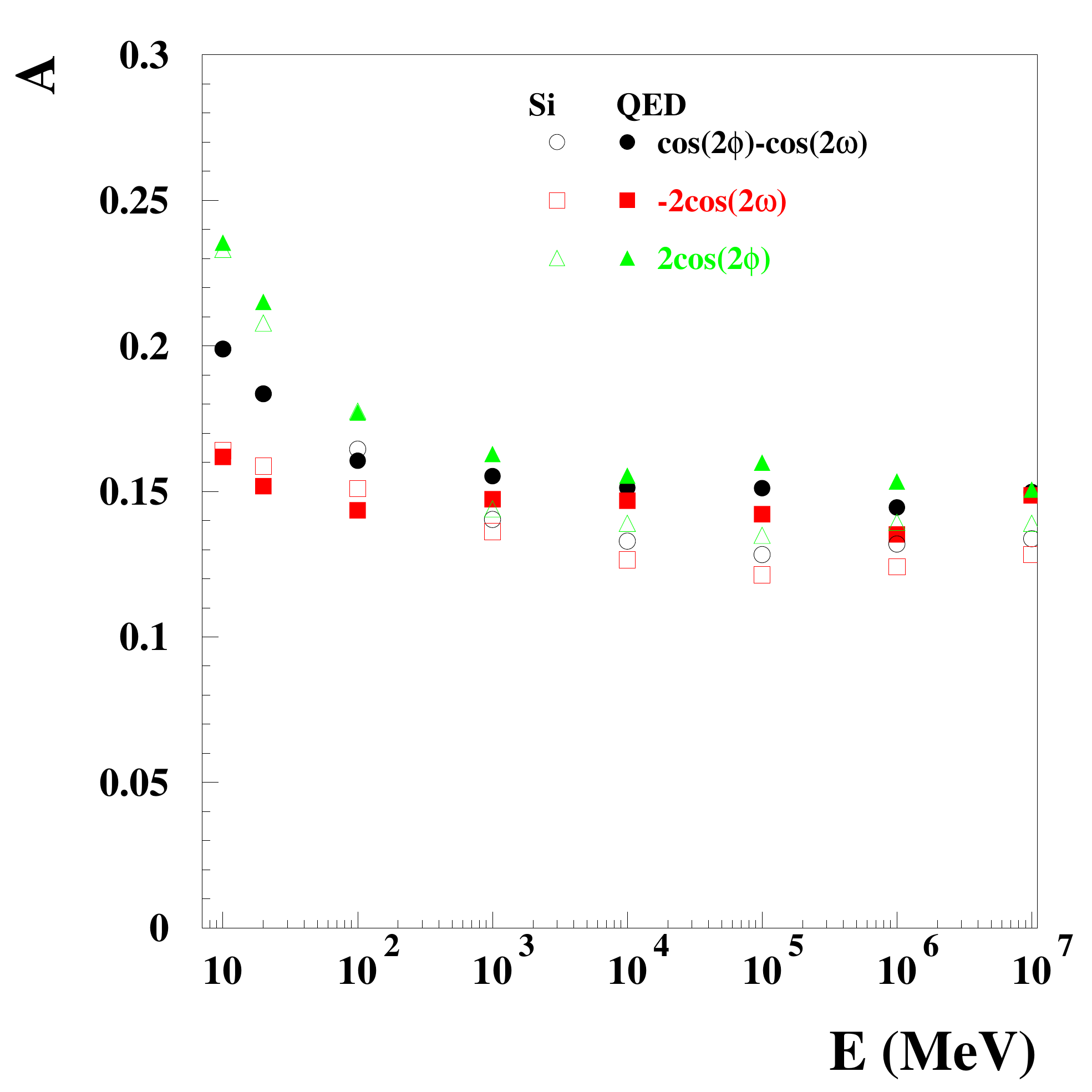}
\includegraphics[width=0.32\linewidth]{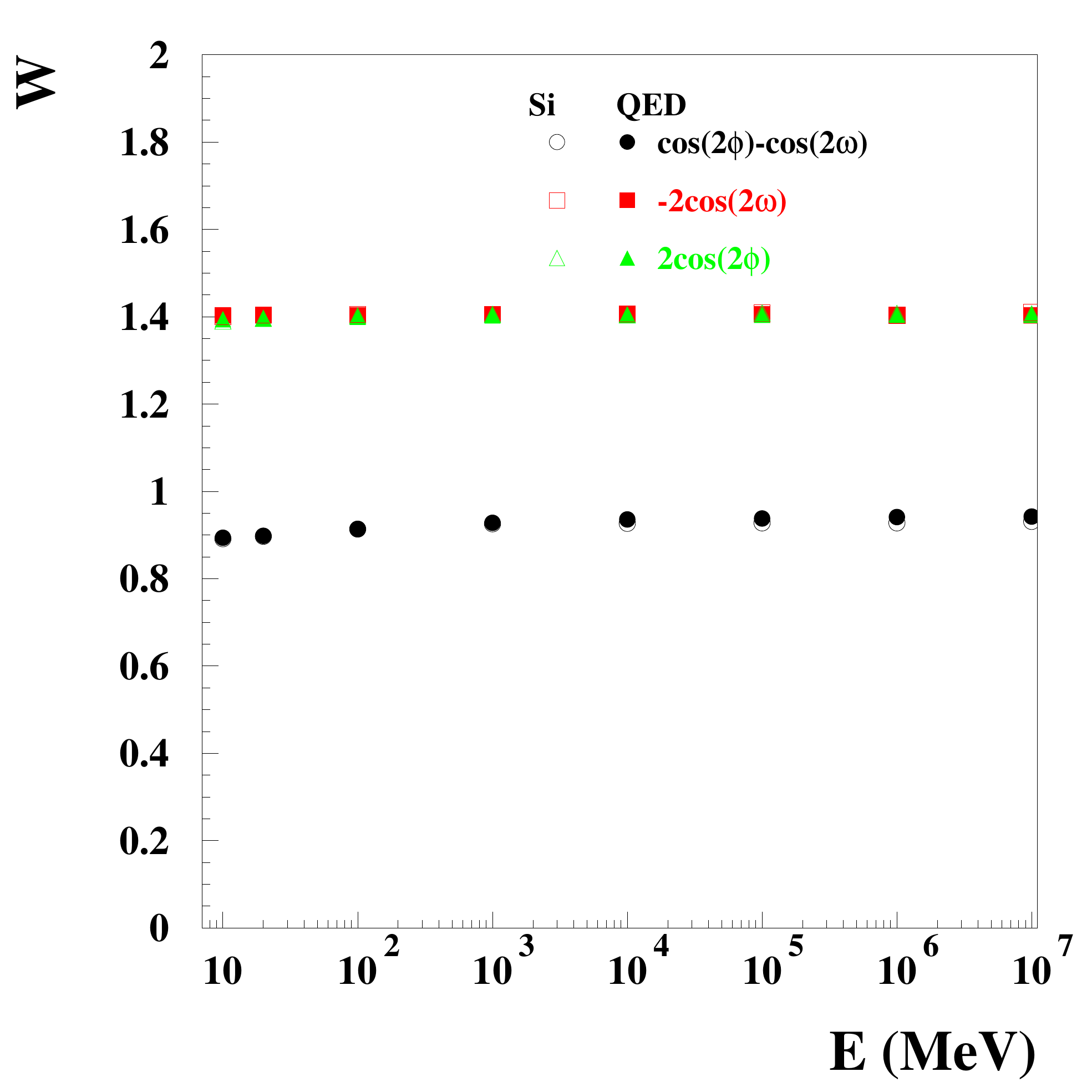}
\includegraphics[width=0.32\linewidth]{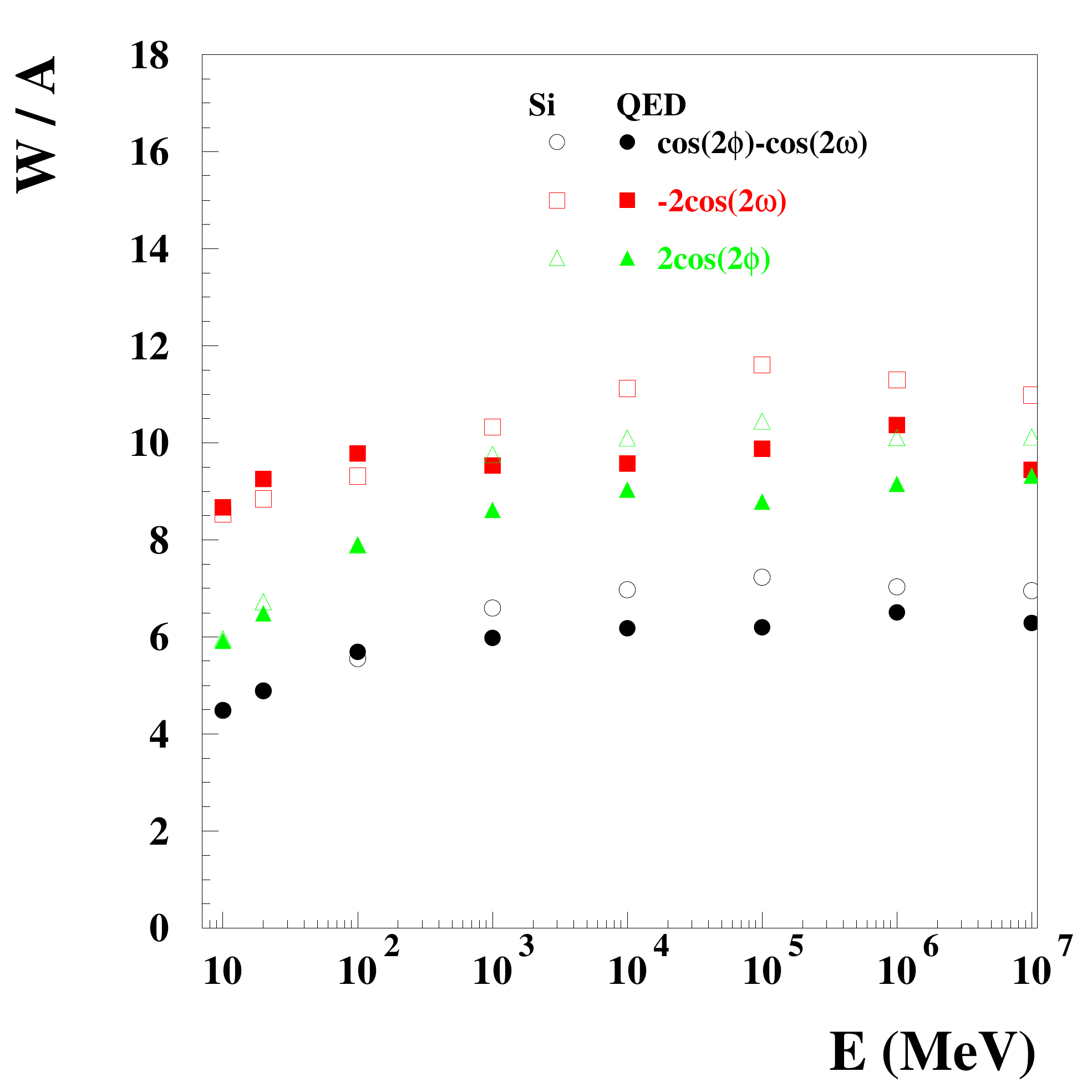}
}{
\includegraphics[width=0.32\linewidth]{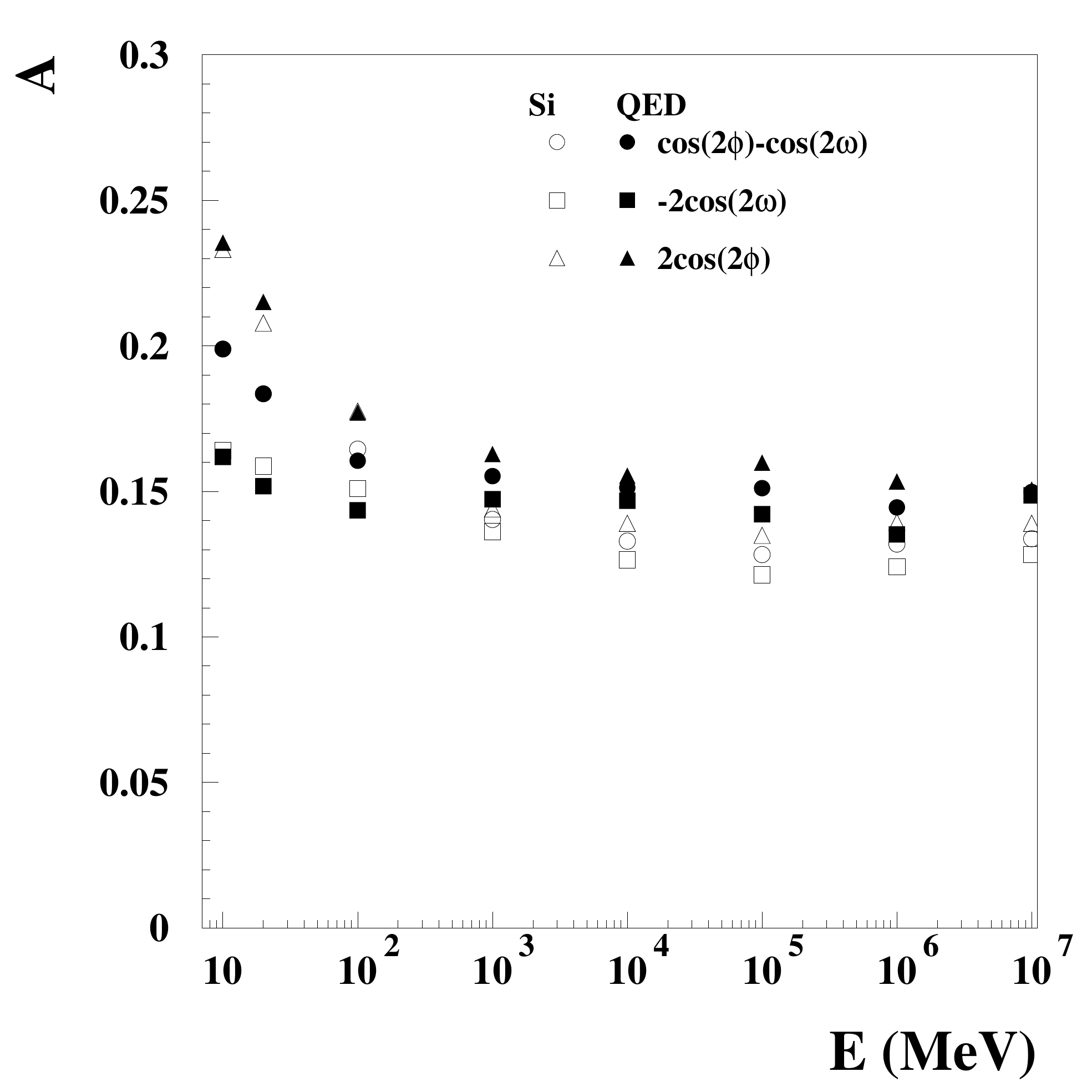}
\includegraphics[width=0.32\linewidth]{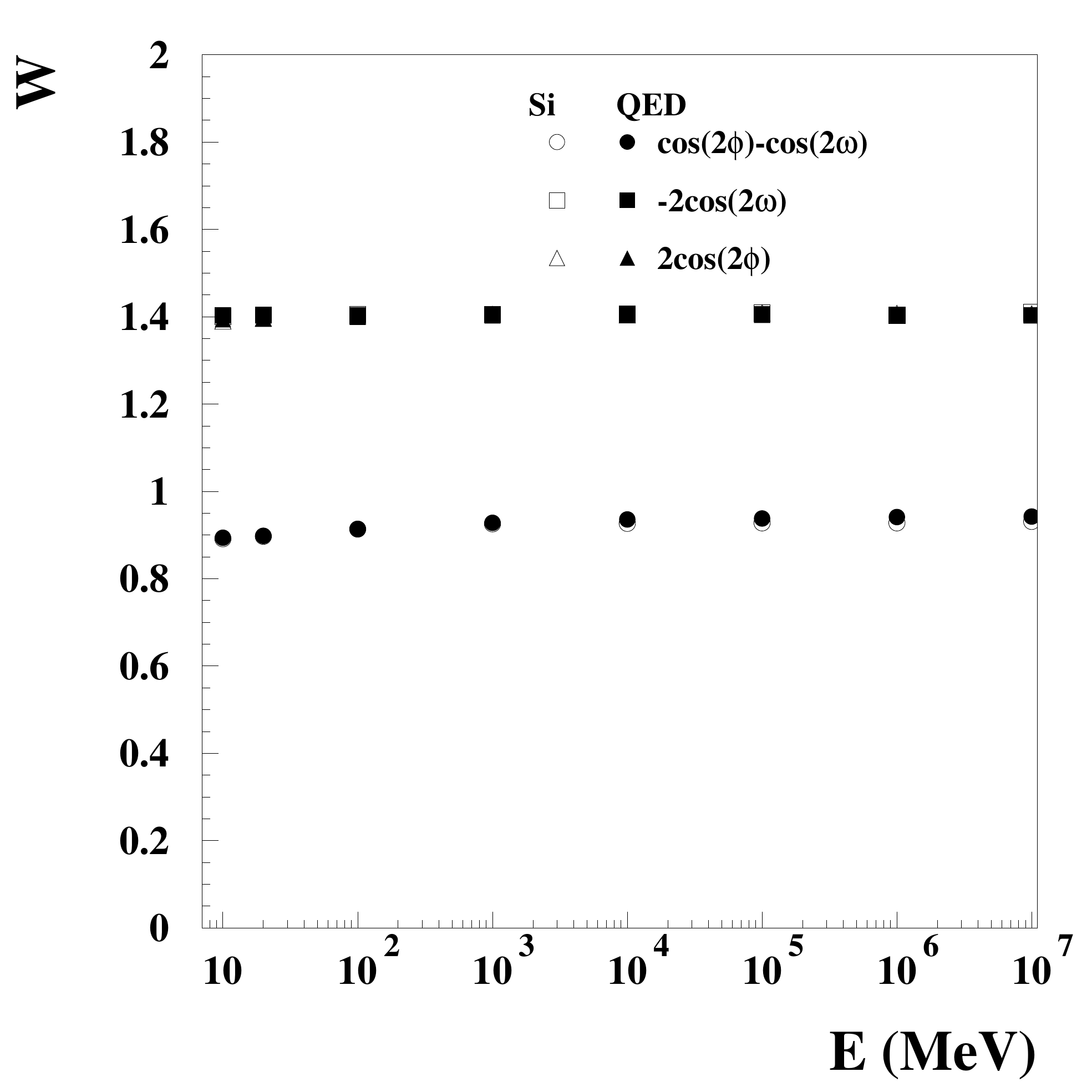}
\includegraphics[width=0.32\linewidth]{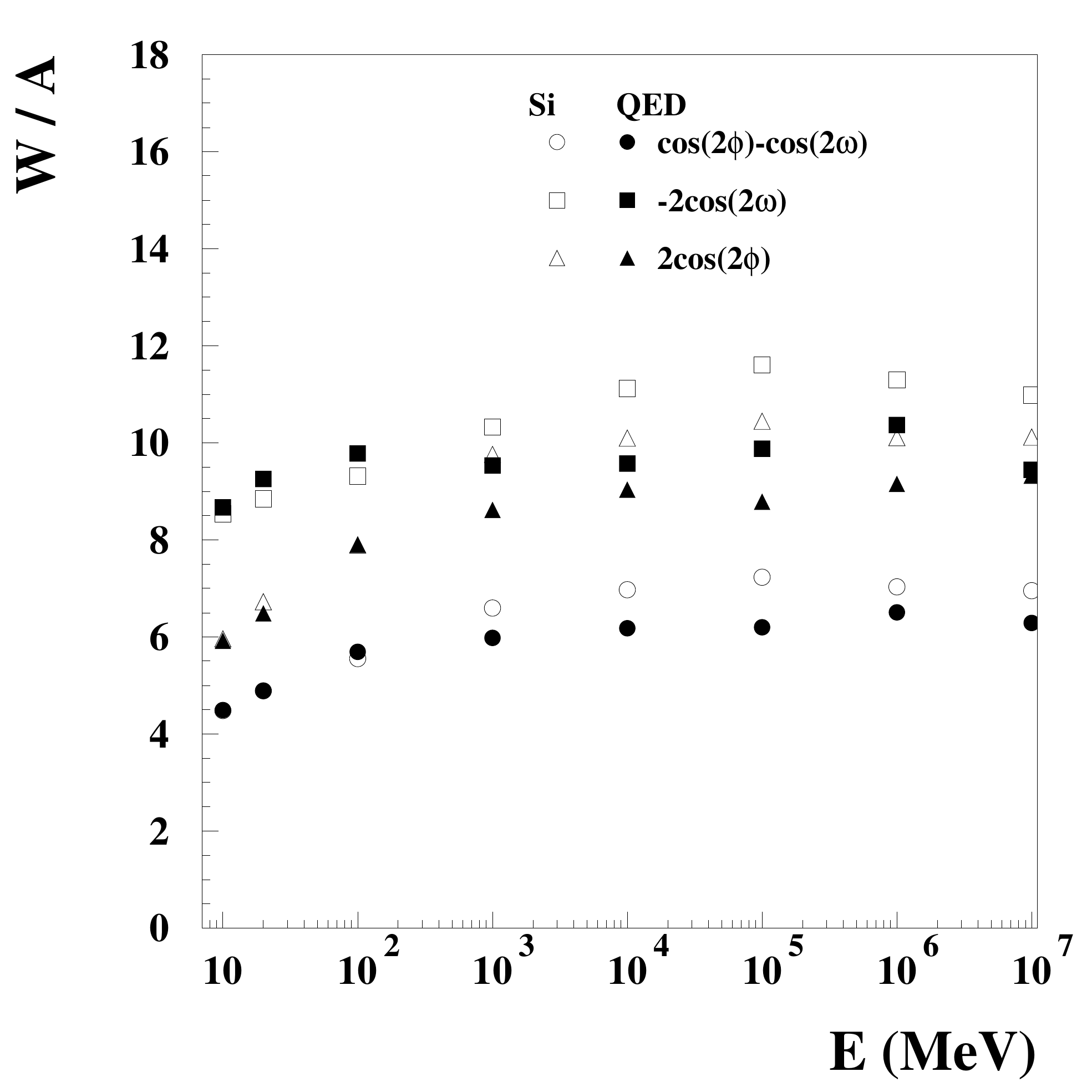}
}
\caption[~]{Study of the variation of the polarisation asymmetry at
generator level, for nuclear conversion of gamma rays
on a naked silicon nucleus (``QED'', full symbols) and
on a silicon atom (``Si'', open symbols).
Asymmetry (left),
R.M.S. width (centre),
and ratio of width to asymmetry (right). 
 \label{fig:combinaison:angles}}
\end{figure*}

For a three-particle final state as is the case for a photon
conversion to a pair, even if the recoiling nucleus cannot be
detected, there was no natural, unique way to define ``the'' azimuthal
angle of an event, that was named generically $\varphi$ in
eq. (\ref{eq:1D}).
Since the pioneering work by Wojtsekhowski \cite{Bogdan:1998}, a
common practice has been to use the angle, $\omega$, defined as ``the
angle between the polarisation plane of the incident photon and the
vector that connects the crossing points of the positron and the
electron in the detector plane, which is perpendicular to the photon
momentum'' \cite{Bogdan:1998}, see Fig.~\ref{fig:event:les:angles}.

It was later discovered that another definition, the bisectrix
of the azimuthal angles of the electron and of the positron,
$\phi \equiv (\varphi_+ + \varphi_-)/2$,
yielded a higher value of the polarisation asymmetry and, more
importantly, a value that was found to be compatible with the
published low-energy and high-energy asymptotic expressions for $A$
\cite{Gros:2016dmp} (with a phase sign change with respect to $\omega$).

It was also found that there is still room for improvement, as an
``optimal'' measurement using the whole (five dimensional)
differential cross section instead of the mere one-dimensional
azimuthal-angle differential cross section yielded a precision of the
measurement improved by a factor of two to three
(Fig. 21 of \cite{Bernard:2013jea}, Fig. 3 of \cite{Gros:2016dmp}):
the value of the polarisation asymmetry is found to be similar to that
of the 1D analysis, but the uncertainty is smaller.

I present results obtained with various definitions of the event
azimuthal angle in the next sections.

It should be noted that following the usual notations, I refer to the
azimuthal angle of ``the positron'' and of ``the electron'', while in
active targets without a magnetic field, the electric charge of the
tracks cannot be measured.
This misuse of language is actually innocuous, as an exchange would be
equivalent to adding $\pi$ to $\varphi$, that is, $2\pi$ to $2\varphi$, and
the cosine (and sine) would therefore be left unchanged.

\section{The measurement on out-of-the-generator 4-vectors}
\label{sec:MC:4vectors}

In this work, I compute asymmetries with the moments method
\cite{Bernard:2013jea,Gros:2016dmp}, that is, as the expectation
value, $\E(w)$, of an optimal weight, $w$, chosen so that $\E(w) = A
\times P$, and computed as the average value of the distribution of
$w$ for the generated sample.
When an optimal weight is used, the moments method is equivalent to a
likelihood fit \cite{Tkachov:2000xq}.
The moments method has been routinely used in partial wave and
amplitude analyses in particle physics and nuclear physics, until
the development of
maximisation tools such as MINUIT favoured the use of likelihood
analyses (\cite{Tkachov:2000xq} and references therein).

Some differences can be noticed though, between the two methods, for
example background noise contribution(s) is(are) taken into account in
likelihood analyses by including their probability density functions
(pdf) in the global maximised pdf, while for moments analyses only the
moments are needed.
Background subtraction can then be performed by injecting events from
the sidebands, or from the external regions on the region of interest
(RoI), with a negative weight.

Let us first perform the measurement on simulated samples generated
from the bare event generator, i.e. without any detector effects.
Figure \ref{fig:combinaison:angles} shows the variation of the
polarisation asymmetry $A$, as a function of incident photon energy
from event samples generated with fully linearly polarised beams
($P=1$) on silicon.
We have
\begin{equation}
 \sum_{i}^{N} w_i = N \times A \times P .
 \label{eq:moments:method:0}
\end{equation}

For the 1D differential cross section (eq. (\ref{eq:1D})), the
optimal weight is $w \equiv 2\cos{2\varphi}$
\cite{Bernard:2013jea,Gros:2016dmp}.
The centre plot shows the RMS width, $W$, of the distribution of $w$.
For $A$ small (which will be the case here, as we shall see), the
width is $W \approx \sqrt{2}$ (eq. (19) of \cite{Gros:2016dmp}).
The precision of the measurement of the modulation factor,
$\sigma_{A \times P}$ is related to the width
\begin{equation}
W = \sqrt{N} ~ \sigma_{A \times P} .
\end{equation}

For a measurement on a source of unknown polarisation fraction, $P$,
the effective value of $A$ must be known either from Monte Carlo
studies or from prototype calibration on a beam with known
polarisation fraction.
The uncertainty of the measurement of $P$ is 
\begin{equation}
 \sigma_{P} = \gfrac{\sigma_{A \times P}}{A} , 
\label{eq:uncertainty:P}
\end{equation}
 that is, 
\begin{equation}
 \sigma_{P} \approx \gfrac{W}{A \sqrt{N} }. 
\label{eq:uncertainty:P2}
\end{equation}

\begin{figure*}[htbp!]
\begin{center}
\iftoggle{couleur}{
\includegraphics[width=\linewidth]{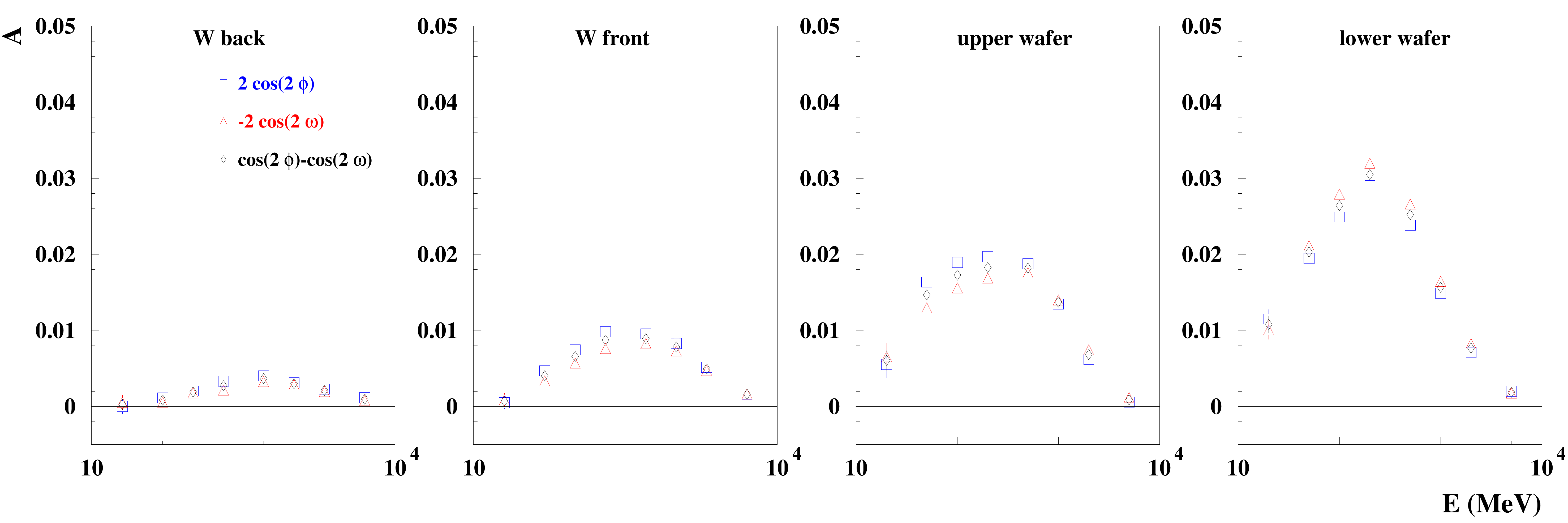}
\includegraphics[width=\linewidth]{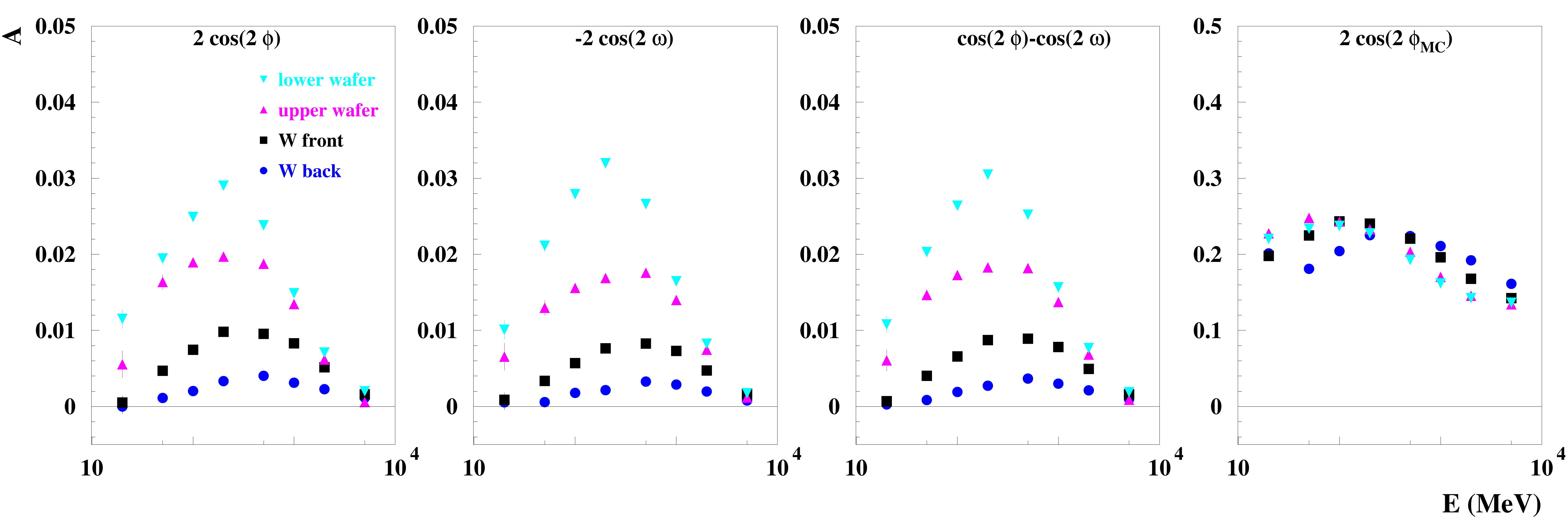}
\includegraphics[width=\linewidth]{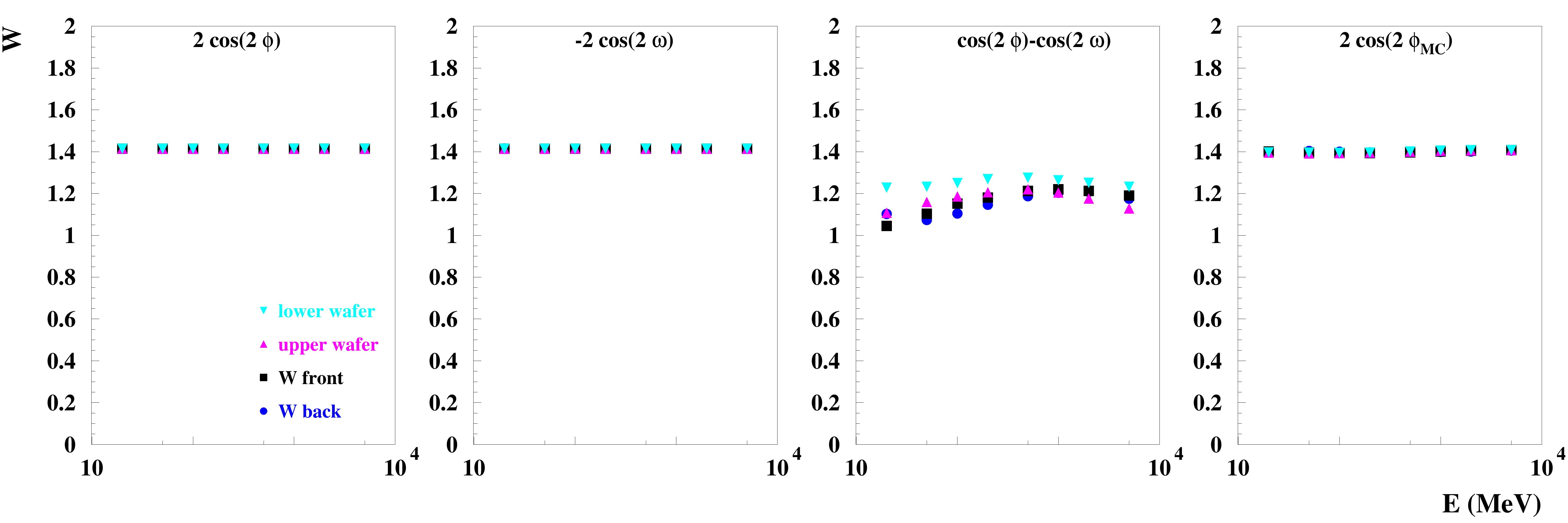}
}{
\includegraphics[width=\linewidth]{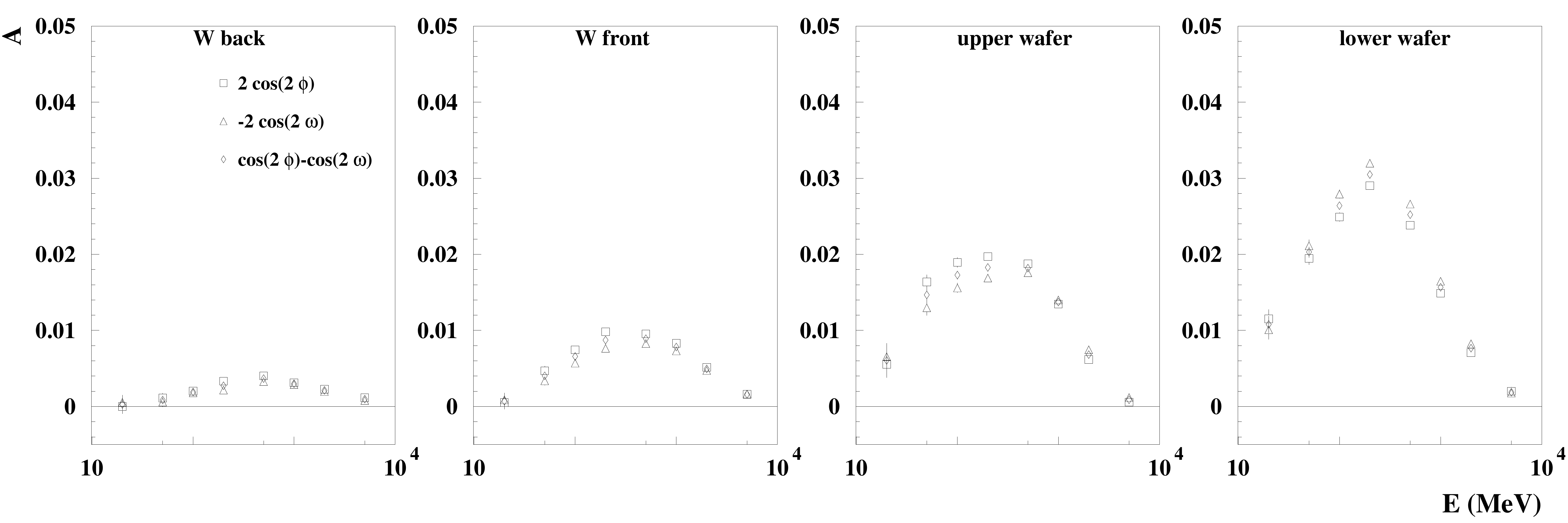}
\includegraphics[width=\linewidth]{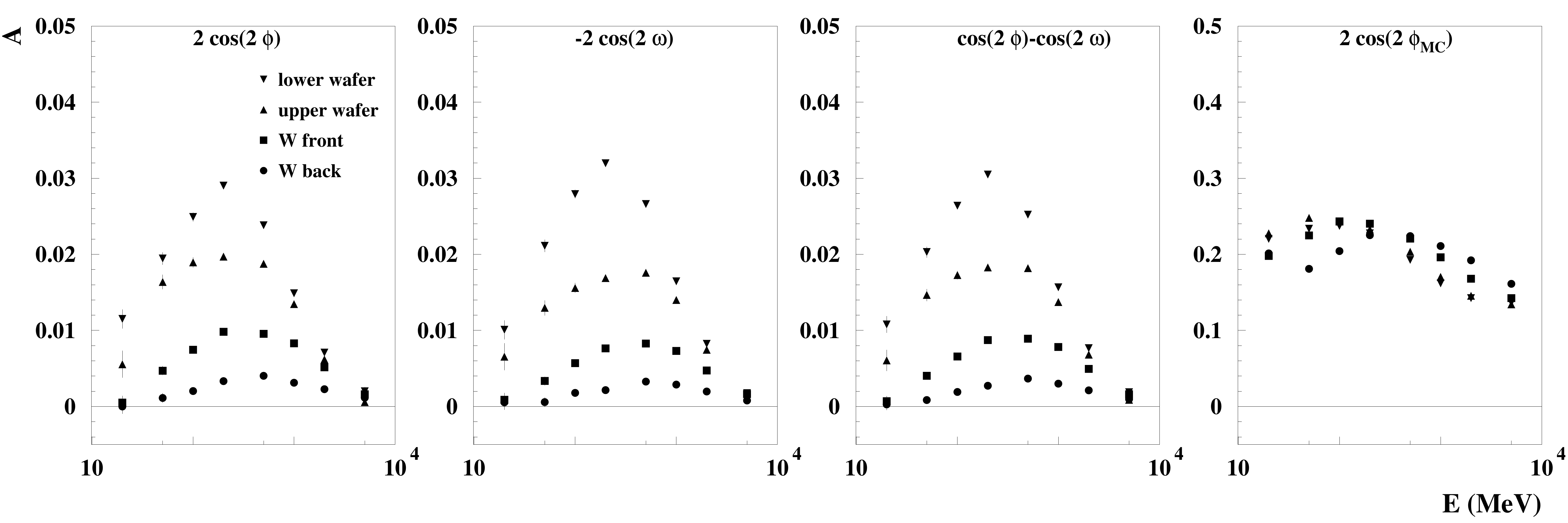}
\includegraphics[width=\linewidth]{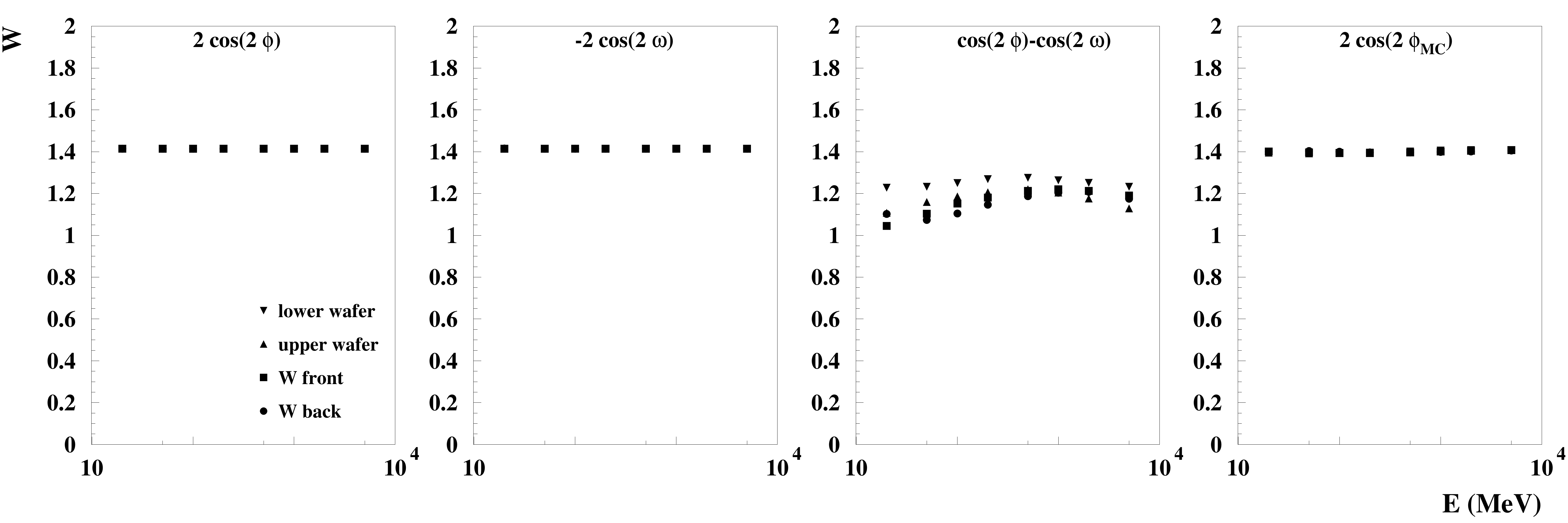}
}
\captionsetup{singlelinecheck=off}
\caption[~]{$\ntrack =2$ events:
Polarisation asymmetry as a function of incident photon energy. \label{fig:aa}
 \begin{itemize}
 \item
 {\bf Upper row}: For various
 conversion locations, from left plot to right plot: W back, W front, upper wafer, lower wafer, and various measurement methods (weights
 $2\cos(2\phi)$ (open square),
 $-2\cos(2\omega)$ (open triangle),
 $\cos(2\phi) - \cos(2\omega)$ (open diamond).
 \item
 {\bf Centre row}: For various methods, from left plot to right plot: 
 $2\cos(2\phi)$,
 $-2\cos(2\omega)$,
 $\cos(2\phi) - \cos(2\omega)$ and for the Monte Carlo generator 
 $2\cos(2\phi_{\MC})$,
 and various conversion locations
 W back (bullet),
 W front (full square),
 upper wafer (upward triangle),
 lower wafer (downward triangle), 
 \item
 {\bf Bottom row}: The width of the distribution.
\end{itemize}
Notice the different vertical scales for the measured
($\phi$ or $\omega$) and the MC values of the azimuthal angle.
}
\end{center}
\end{figure*}

$\sigma_{A \times P}$ is found to be independent, at first order 
\footnote{See eq. (19) of \cite{Gros:2016dmp}.
 A full treatment of the next-to-leading order terms in the
 computation of the uncertainty leads to an additional factor 3/2 in
 the correction term, see eq. (30) of \cite{Besset:1979sh}.
 J. Pretz, private communication, 2018).} ,
of the value of $P$.

Let us examine the performance of the measurement at Monte Carlo
level, that is, without taking any detector effect into account.
From event samples of simulated pair conversions of fully polarised photons:
\begin{itemize}
\item
The variation of $A$ with $E$ shows the known (eq. (16) and Fig. 3 of
\cite{Gros:2016dmp})
increase at low energy (Fig. \ref{fig:combinaison:angles} left).

\item The polarisation asymmetry is found to be smaller for conversion
 on atoms (``Si'') than on isolated nuclei (``QED'') at high energy
 where the screening of the field of the nucleus by the electron
 cloud is effective.
This was expected, as screening suppresses the amount of very low $q$
events who are carrying a higher polarisation content
(Fig. 11 bottom left of \cite{Bernard:2013jea}).

\item The polarisation asymmetry is found to be larger with a
 $\phi$-based weight than with a $\omega$-based weight as was already
 known \cite{Gros:2016dmp}.
Using the average of the cosines as the weight obviously just yields a
result which is the average of the individual measurements.
 
\item Widths are energy-independent to first order
(Fig. \ref{fig:combinaison:angles} centre) and are found to have the
 same value
($W \approx \sqrt{2}$) for both $\phi$- and $\omega$-based weights.
For the combination, with weight $\cos(2\phi) - \cos(2\omega)$, the width is
$\approx 1$, which is a sign of uncorrelated measurements.

\item Finally the relative width plot shows that the combination is
 more precise than any of the individual measurements, by a factor of
 $\approx \sqrt{2}$ (Fig. \ref{fig:combinaison:angles} right).
\end{itemize}

\section{$\ntrack =2$: Polarisation asymmetry from reconstructed events}
\label{sec:Polarisation:Asymmetry}

We now examine the results for events reconstructed from the test
detector, for events with $\ntrack =2$ reconstructed tracks.
The asymmetry measured using either $\phi$- or $\omega$-based weights
or their combination, or the MC $\phi$ value, for various conversion
locations (lower wafer, upper wafer, front W and back W) is shown, as
a function of incident photon energy in Fig. \ref{fig:aa}.
\begin{itemize}
\item The polarisation asymmetry is found to be small, peaking
 at $A \approx 0.03$ for a conversion in the lower wafer, which
 corresponds to a dilution of $D \approx 0.13$.
The peaking energy of about 200\,MeV is found to be commensurate with
the critical energy $E_c$.
The usable energy range ranges (sic) from 20\,MeV to 2\,GeV at maximum, that
is, for a conversion in the lower wafer.

\item The polarisation asymmetry obviously decreases when the amount
 of multiple scattering increases, but it is interesting to note that
 even for conversion in a tungsten foil, some sensitivity is still present.
We can surmise that the issue of deciphering conversions in an upper
wafer from conversions in a ``front'' tungsten foil might be
less critical than anticipated.
 
\item 
It is also interesting to note that in addition to the known energy
dependence of the MC-generator asymmetry, a small additional variation
can be seen, that depends also on the conversion location
 (Fig. \ref{fig:aa}, centre row, right plot);
this is most likely due to indirect selection effects, as the value of
$A$ varies with the kinematic variables in the conversion
(Fig. 11 of \cite{Bernard:2013jea}).

\item 
The widths of the weight distributions show the nominal value of
$\sqrt{2}$ for a simple cosine-based weight.
For the combination though, $w =\cos(2\phi) - \cos(2\omega)$, the
width is found to be larger than 1 in contrast with the results of the
previous section (Fig. \ref{fig:combinaison:angles}), based on the use
of generator-level angles.
This is the indication of a correlation between the two angles, induced
by detector effects;
\end{itemize}

\section{Variation of the polarisation asymmetry with kinematic observables}
\label{sec:A:variation}

In this section I study the variation of the polarisation asymmetry as
a function of several observables that are available for each event:
\begin{itemize}
\item
 $\cos{\theta_{\LAT}}$;

\item
 $\theta$, the polar angle between the known position of the source
 and the reconstructed direction of the photon candidate.
\end{itemize}

As the distributions of $\theta$ vary violently with photon
energy and with conversion location, and as they have extended tails,
I use a function $g(\theta)$ in place of $\theta$ itself, devised so
that the statistics be of the same order of magnitude for all bins.
Function $g(\theta)$ is documented in the Appendix.
Figure \ref{fig:variation} shows the polarisation asymmetry,
measured using $w= 2\cos(2\phi)$, of reconstructed events as a
function of $g(\theta)$ and of $\cos{\theta_{\LAT}}$.
\begin{itemize}
\item
The polarisation asymmetry decreases with $g(\theta)$, either due to
the known decrease at high $q$
(Fig. 11 left bottom of \cite{Bernard:2013jea})
or because these events have undergone a lot of multiple scattering;
or both.
 
\item
 The polarisation asymmetry increases with $\cos{\theta_{\LAT}}$,
 i.e. decreases with $\theta_{\LAT}$,
 again due to multiple scattering;
\end{itemize}

\begin{figure*}[htbp!]
 \begin{center}
\iftoggle{couleur}{
 \includegraphics[width=\linewidth]{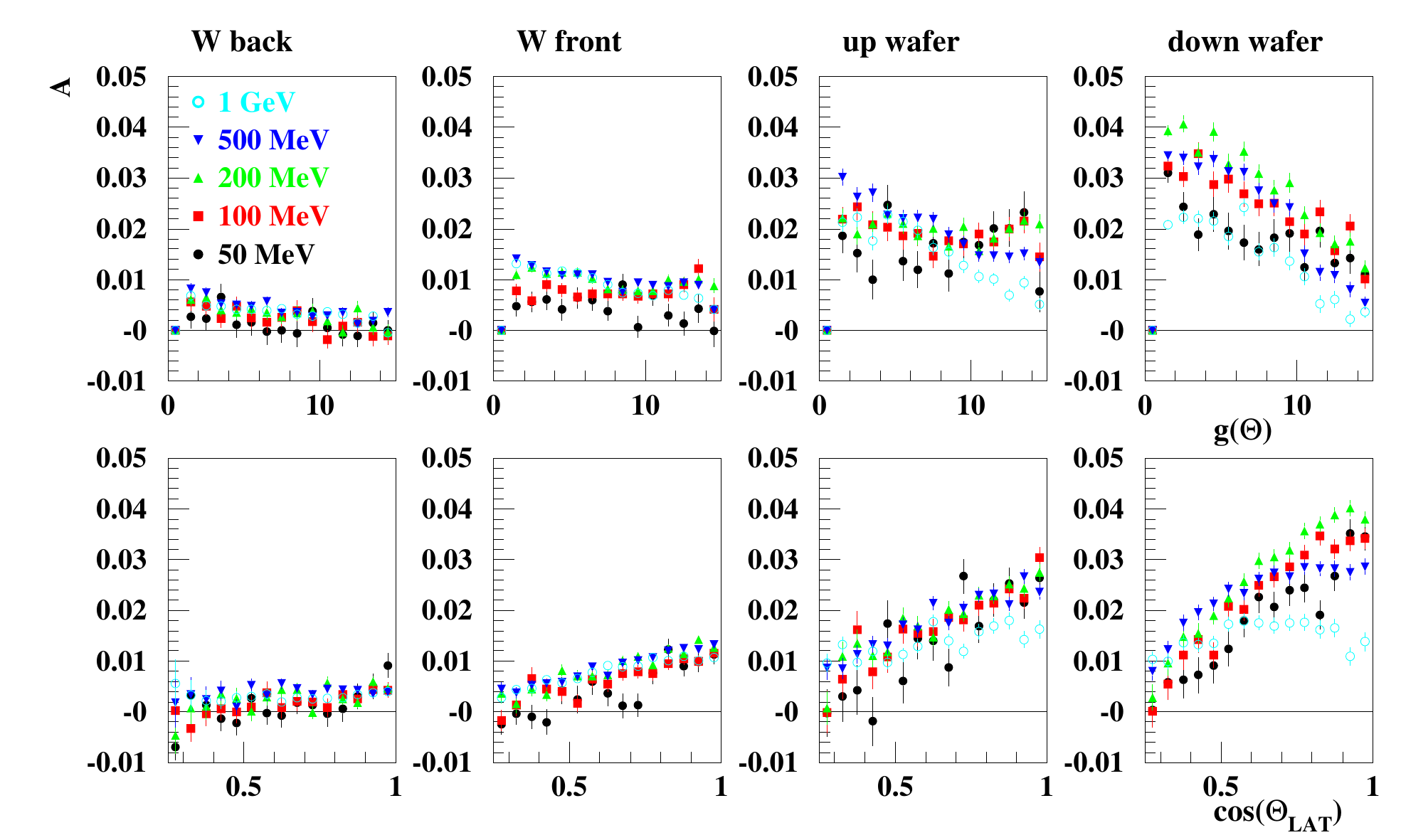}
}{
 \includegraphics[width=\linewidth]{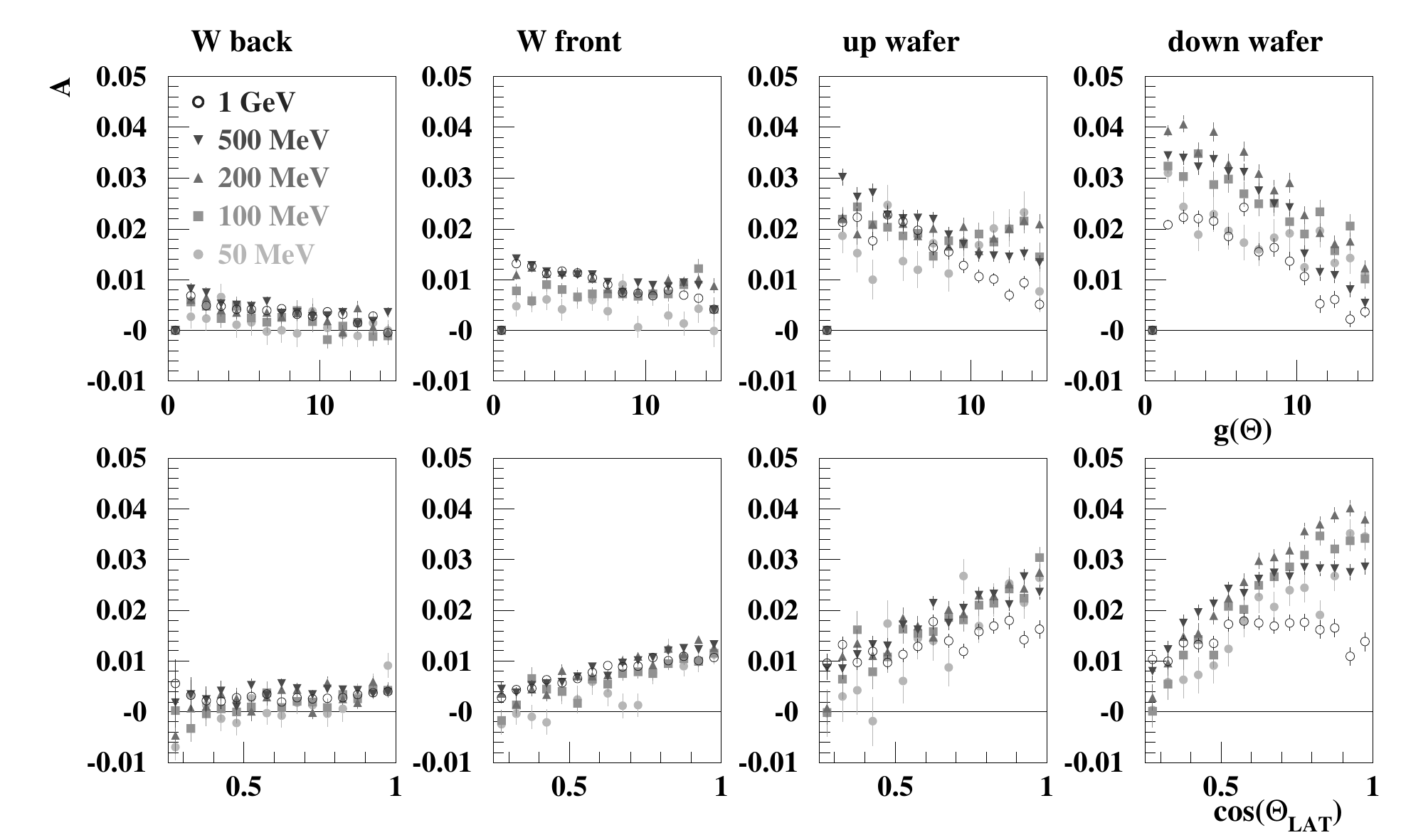}
 }
 \caption{$\ntrack =2$ events: Variation of the polarisation
 asymmetry as a function of event observables
 $g(\theta)$ (top row) and
 $\cos{\theta_{\LAT}}$ (bottom row),
 for various conversion locations
 (from left to right: W back, W front, upper wafer and bottom wafer)
 and for various photon energies
($w=2\cos(2\phi)$).
 \label{fig:variation}}
 \end{center}
\end{figure*}

\begin{figure*}[htbp!]
\begin{center}
 \iftoggle{couleur}{
 \includegraphics[width=\linewidth]{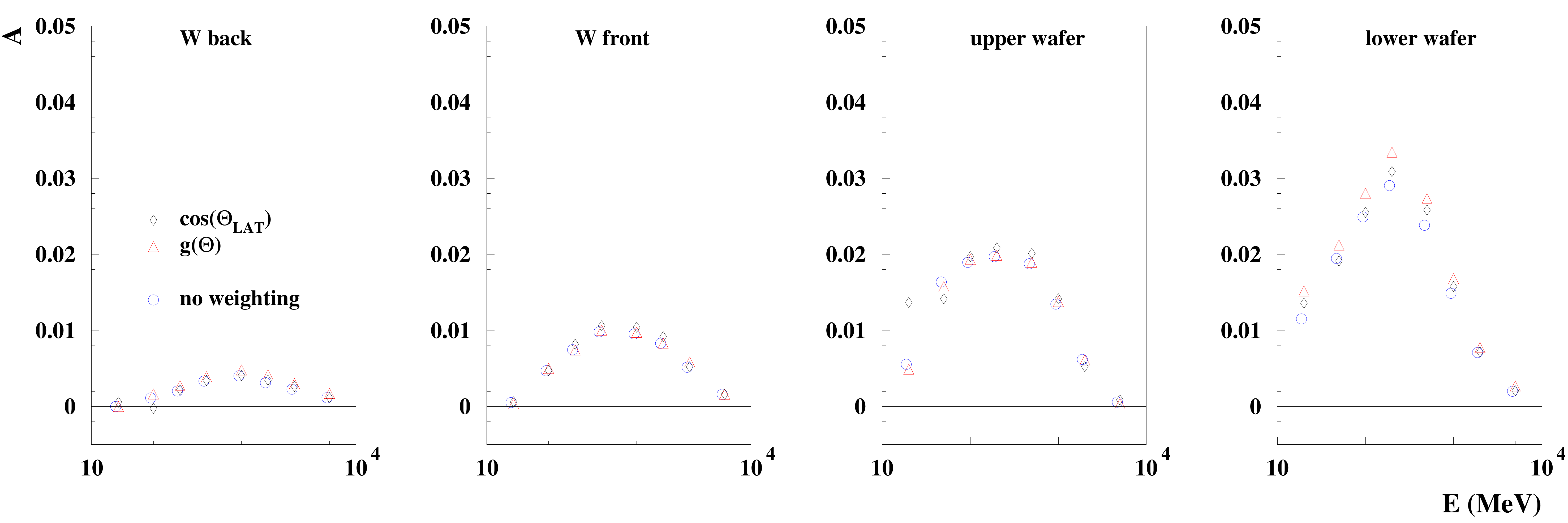}
}{
\includegraphics[width=\linewidth]{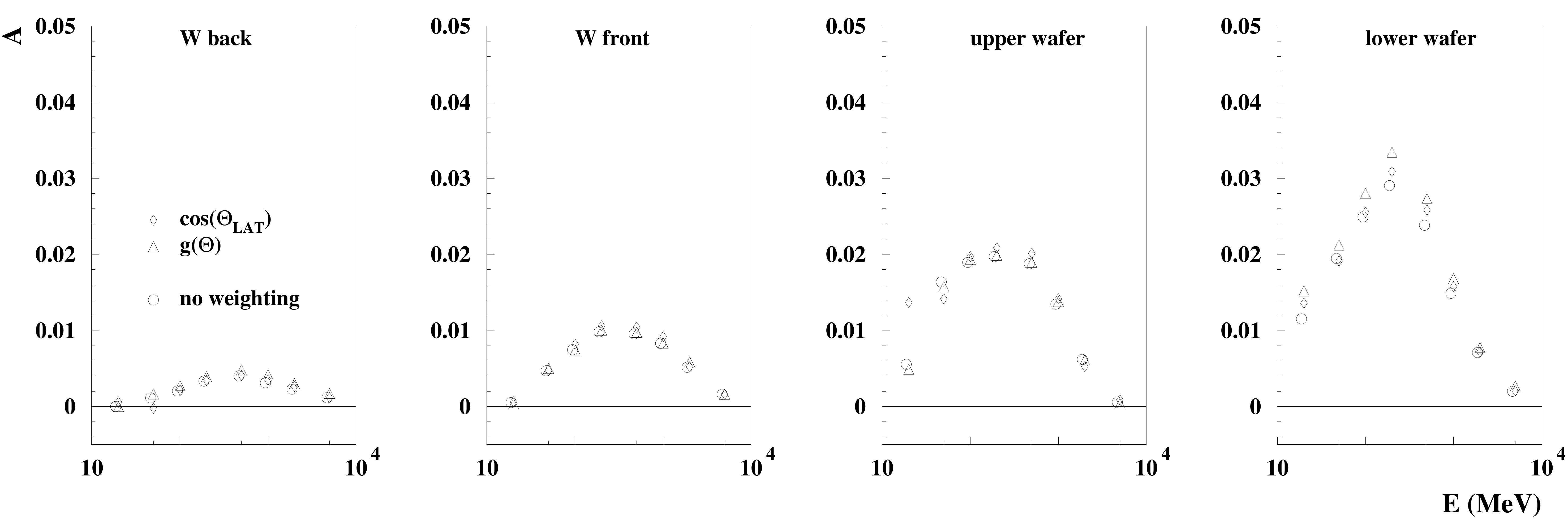}
 }
\caption[~]{Polarisation asymmetry as a function of incident photon
 energy, and various conversion locations, without and with
 weighting 
($\ntrack =2$ events, $w=2\cos(2\phi)$).
Small horizontal shifts have been applied to improve on the readability.
 \label{fig:pond}}
\end{center}
\end{figure*}

\begin{figure*}[htbp!]
\begin{center}
\iftoggle{couleur}{
\includegraphics[width=\linewidth]{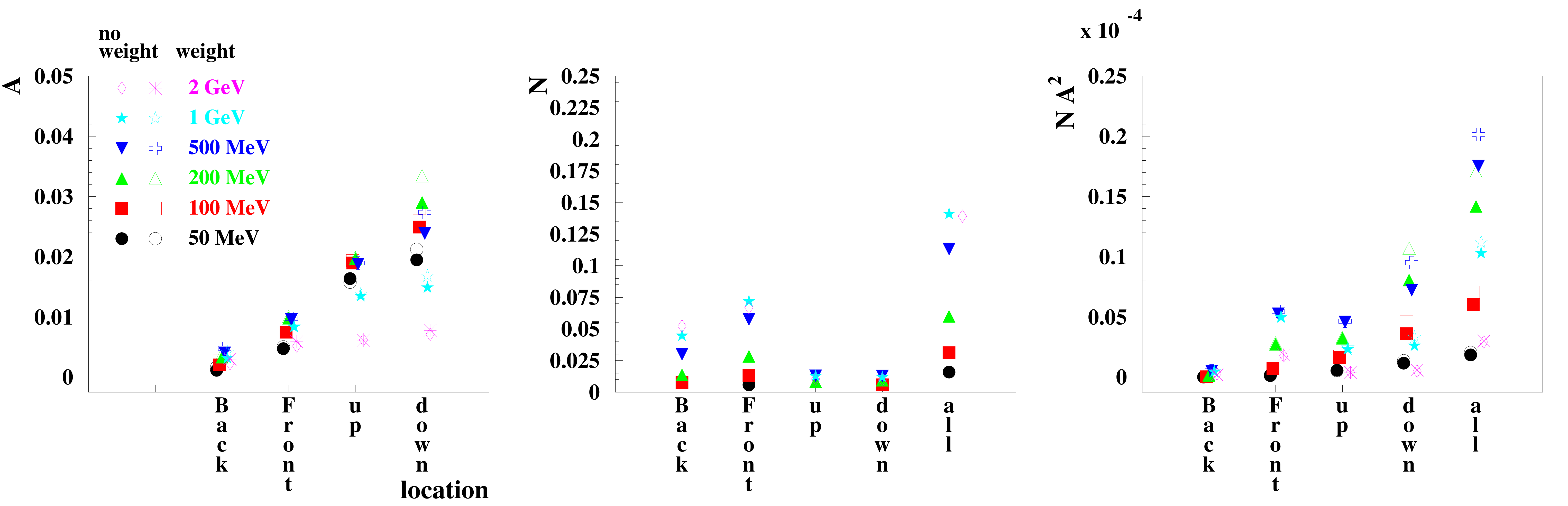}
}{
 \includegraphics[width=\linewidth]{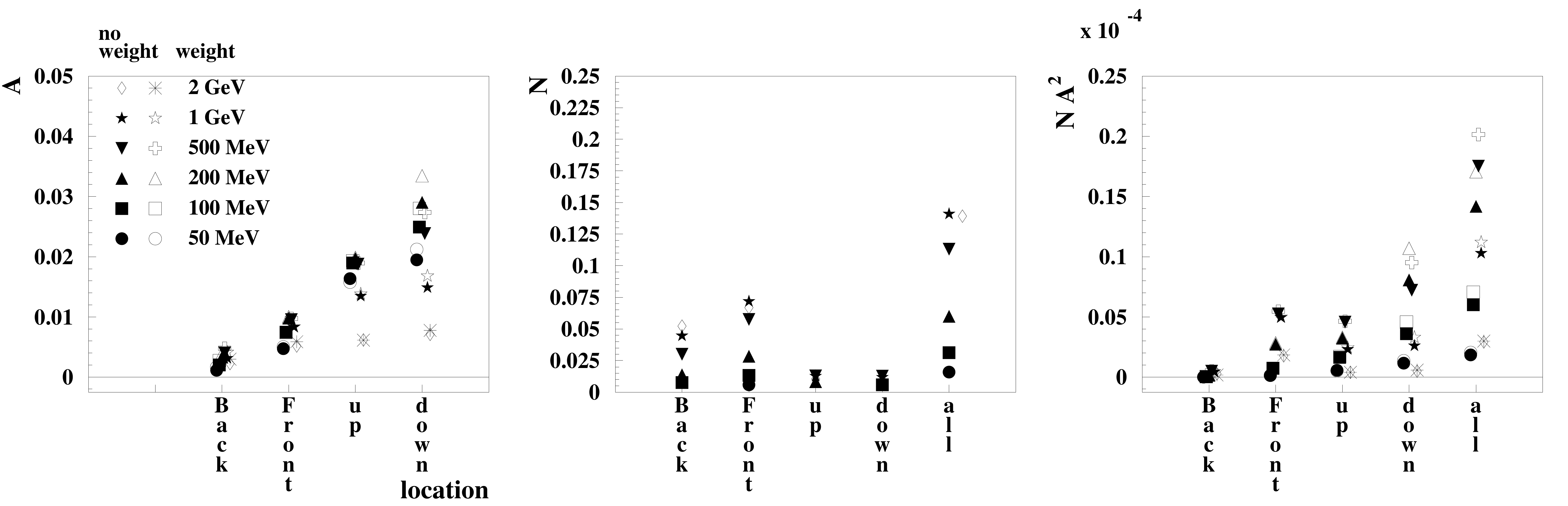}
 }
\caption[~]{$\ntrack =2$ events:
Polarisation asymmetry $A$ (left), number of events $N$ (centre) and value of 
$N_k A^2_k$ as a function of the conversion location
(from ``W back'' ($k=1$) to
``down wafer'' ($k=4$)) and their sum ``all''.
Without and with $g(\theta)$ weighting, for each location,
with global and $g(\theta)$ weighting, for all locations combined.
The numbers of events are normalised to 1 incident photon on the detector.
\label{fig:trd} }
\end{center}
\end{figure*}

\section{Better than cutting: weighting}
\label{sec:Weighting}

Given the variation of the average value of the polarisation
asymmetry as a function of event observables seen in the
previous section, one might be tempted to select events in some range
of these variables.
An improved precision can be reached, though, by weighting events
according to the polarisation asymmetry as averaged in Fig.
\ref{fig:variation}.
Later I combine events from different samples too, for example, with
 different conversion locations or of different energies.

Let us consider $K$ event samples, $k = 1 \cdots K$, each of which
with number of events $N_k$ and providing a measurement of $P$ equal
to $P_k$ with variance $V_k$.
For separate event samples, the measurements
are uncorrelated and an optimal combination is achieved by weighting
each measurement by the inverse of its variance:
\begin{equation}
P = \left( \sum_k\gfrac{P_k}{V_k} \right) / \left( \sum_k\gfrac{1}{V_k} \right)
 .
\end{equation}
From eq. (\ref{eq:uncertainty:P2}) and the fact that the RMS widths
of the $w$ weight distributions for various samples are found to be
identical, $W_k = W$, we have:
\begin{equation}
 \gfrac{1}{V_k} = \gfrac{N_k A_k^2}{W^2}
 ,
\end{equation}
where $A_k$ is the average effective polarisation asymmetry of sample $k$, so
\begin{equation}
P = \gfrac{ \sum_k P_k N_k A_k^2 }{ \sum_k N_k A_k^2 } 
 ,
\end{equation}
and given eq. (\ref{eq:moments:method:0}),
\begin{equation}
 \sum_{i}^{N_k} w_{i,k} = N_k A_k P_k ,
 \label{eq:moments:method:k}
\end{equation}
we have 
\begin{equation}
 P = \gfrac{ \sum_k A_k \sum_{i}^{N_k} w_{i,k} }{ \sum_k N_k A_k^2 }
 ,
 \label{eq:P:weighted:weight}
\end{equation}
that is, 
\begin{equation}
 P = \sum_k \sum_{i}^{N_k} w'_{i,k} 
 ,
\end{equation}
the weighted weights (sic) for event $i$ of sample $k$, $w'_{i,k}$, being 
\begin{equation}
 w'_{i,k} = \gfrac{ A_k w_{i,k} }{ \sum_\ell N_\ell A_\ell^2 }
 .
\end{equation}

In practice, given the fact that if a weight $w$ is an optimal weight,
then $a \times w$, $a>0$, is also an optimal weight, we can forget the
denominator: an optimal combination of samples can be achieved by
weighting each event by the known average value of the polarisation
asymmetry of its sample, obtained from the analysis of MC samples
(see, e.g., \cite{Kislat:2014sdf}).
\begin{equation}
 w''_{i,k} = { A_k w_{i,k} } .
\end{equation}

The variance of $P$ is obtained from eq. (\ref{eq:P:weighted:weight})
\begin{eqnarray}
 \Var(P) = \gfrac
 { \sum_k A_k^2 \sum_{i}^{N_k} \Var(w_{i,k}) }
 { \left(\sum_k N_k A_k^2 \right)^2}
% = \gfrac { \sum_k A_k^2 N_k W^2} { \left(\sum_k N_k A_k^2 \right)^2}
 = \gfrac
 { W^2}
 { \sum_k N_k A_k^2 }
 ,
 \label{eq:uncertainty:var}
\end{eqnarray}

and the precision of the measurement is then
\begin{equation}
 {\sigma_P} = 
 \gfrac{W}{\sqrt{ \sum_k A_k^2 N_k }}
 .
 \label{eq:uncertainty:comb}
\end{equation}

For each conversion location,
the variation
of the polarisation asymmetry of $A$ with energy is presented in
Fig. \ref{fig:pond} for each weighting scheme
(no weighting; weighting based on $g(\theta)$ or on
$\cos{\theta_{\LAT}}$).
A sanity cut of $A_k > 0.001$ is applied in these combinations,
something that induces a loss in statistics, especially at low
energies.

An optimal weighting scheme would obviously be based on a two-fold
segmentation of the $g(\theta)$, $\cos{\theta_{\LAT}}$ space.
This was not possible due to the limited statistics of the Monte Carlo
samples.
Attempts to weighting events with products of weights based on the two
variables brought some minor further improvement on
single-configuration event samples (plots not shown), but induced
deleterious consequences when combining samples (e.g. with various
conversion locations) that is presented in the next section, was
attempted.

\begin{figure}[htbp!]
\begin{center}
\iftoggle{couleur}{
\includegraphics[width=0.8\linewidth]{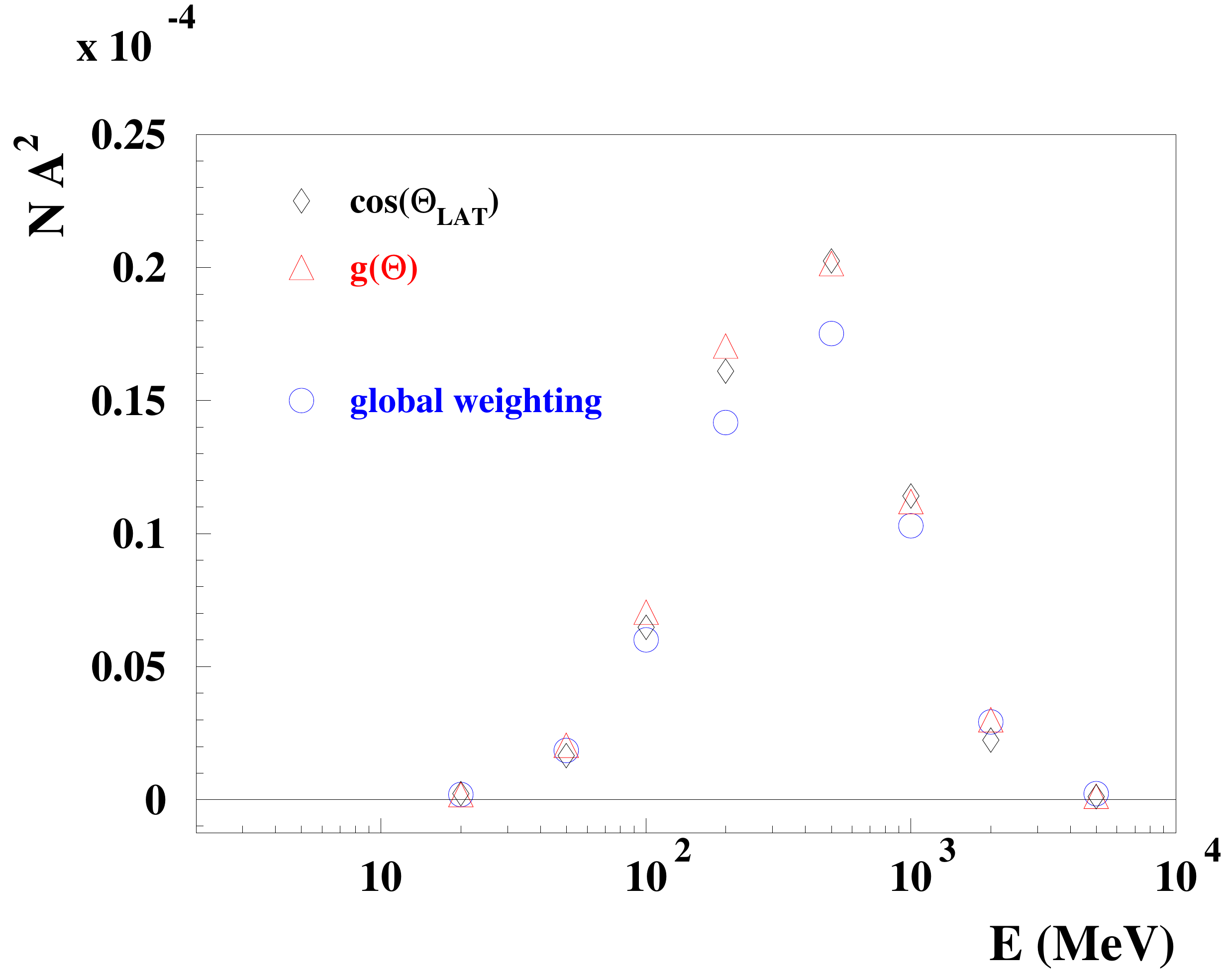}
}{
\includegraphics[width=0.8\linewidth]{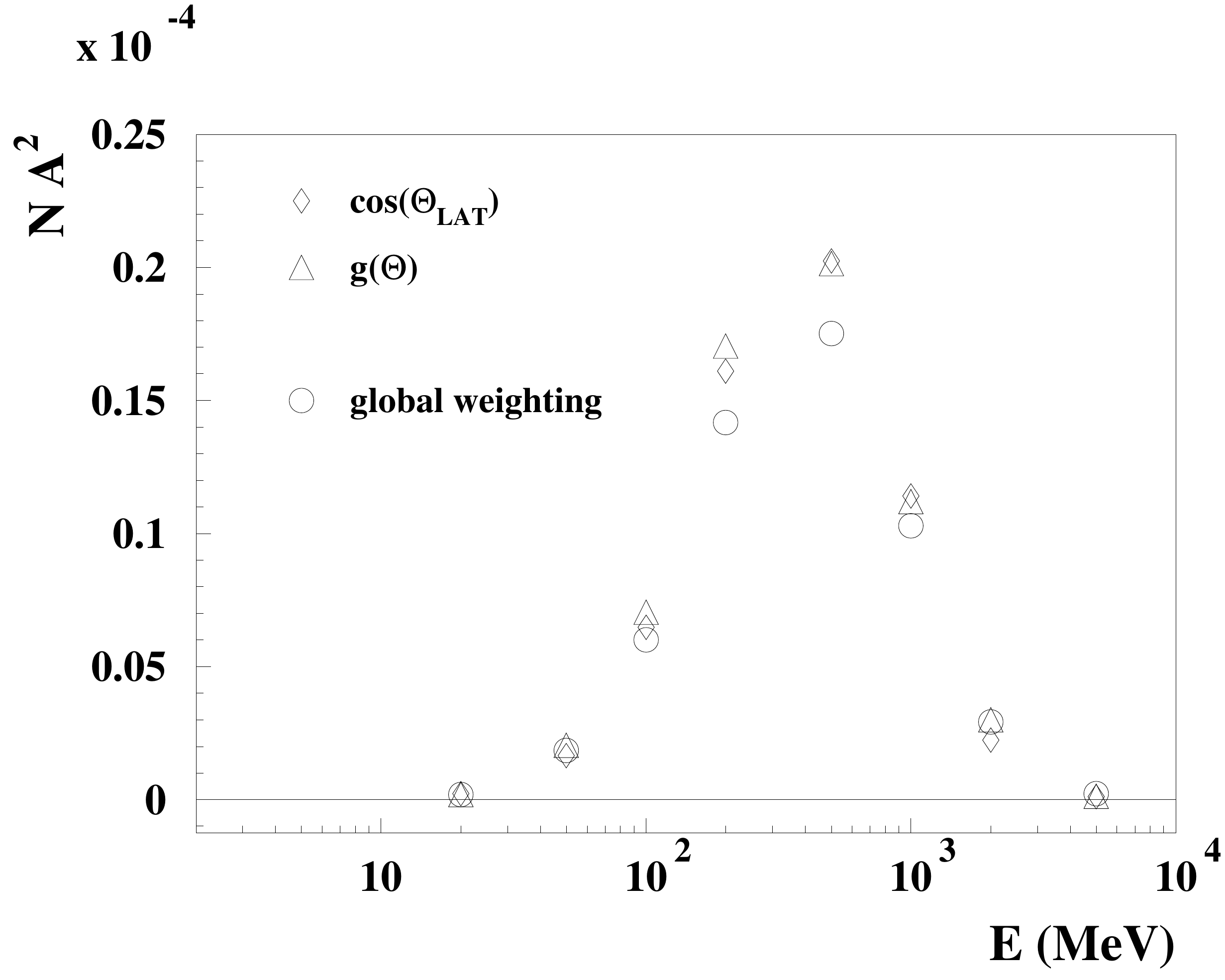}
 }
\caption[~]{$\ntrack =2$ events: Value of $\sum_k N_k A^2_k$
(i.e., all conversion locations combined) as a function of the
 conversion location for various weighting schemes, normalised to one
 incident photon on the detector.
\label{fig:tre} }
\end{center}
\end{figure}

\section{Combination of samples}
\label{sec:Combination}

The same weighting method that was used to take benefit of the
variation of the effective polarisation asymmetry with event variables
in the last section can also be used to combine events from different
samples, such as that with different conversion locations.
The values of $A_k$, of $N_k$ and of $N_k A^2_k$ are shown in
Fig. \ref{fig:trd}, for $k=1$ (W back) to $k=4$ (down wafer), and
their optimal combination (all).
Results with $g(\theta)$ weighting (open symbols) are compared to that
with a global weight for each conversion location.
\begin{itemize}
\item
For $10^9$ $500\,\mega\electronvolt$ incident photons, for example,
$\approx 119.\,10^{6}$
$\ntrack =2$ photons are useable for pair polarimetry with
$\sum_{k=1}^{4} N_k A^2_k \approx 23000$ and $\sigma_P \approx 0.0093$.
\item
For $10^7$ $500\,\mega\electronvolt$ incident photons, 
$\approx 1.19\,10^{6}$
$\ntrack =2$ photons are useable for pair polarimetry with
$\sum_{k=1}^{4} N_k A^2_k \approx 230$ and $\sigma_P \approx 0.093$.
\item The number of events (larger for W than for Si) compensates
 the better asymmetry for Si, to some extent
 (except for ``back'' events).
\end{itemize}

Figure \ref{fig:tre} shows the overall sensitivity, in terms of
$\sum_k N_k A^2_k$, as a function of energy, for various weighting
schemes.

\section{Events with 4 reconstructed tracks}
\label{sec:4tracks}

As mentioned already, the reconstruction of pair conversions in a
two-wafer, single-sided SSDs scheme, for events with two separate
clusters in each of the two directions, $x$ and $y$, in the second
layer, suffers from a two-fold ambiguity
(See Fig. \ref{fig:schema:next:layer} right).

\subsection{Ambiguity: a simple 1D Toy model}
\label{subsec:ambiguity}

Let us consider first a simple model with photons impinging
perpendicular to the detector plane, with an azimuthal distribution in the
sky frame given by (eq. (\ref{eq:1D})), and without any detector
effects.
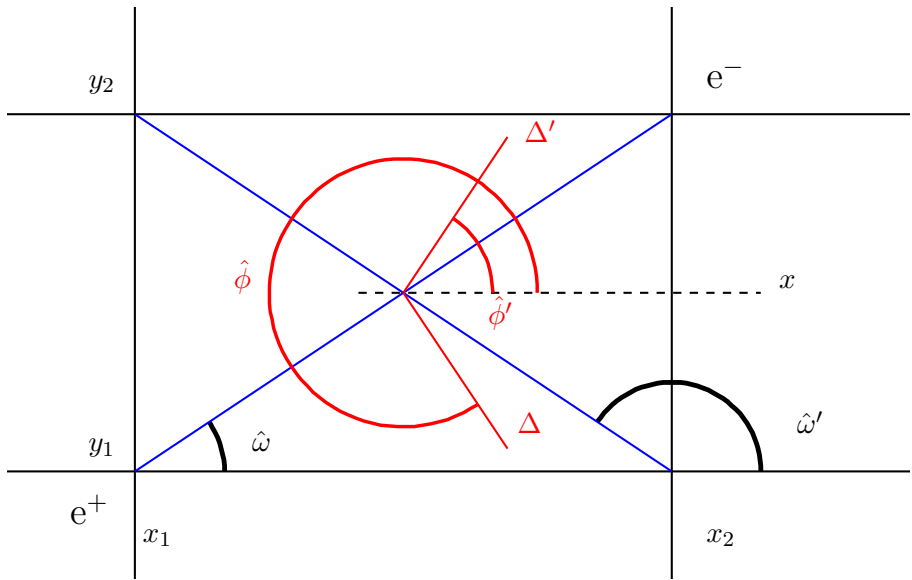
\begin{figure}[htbp!]
 \begin{center}
\setlength{\unitlength}{1.375pt}
 \begin{picture}(300,300)(-255,0)
 % \hfill
 \put(-255,0) {
 \includegraphics[width=12cm, trim=140 180 100 180, clip]{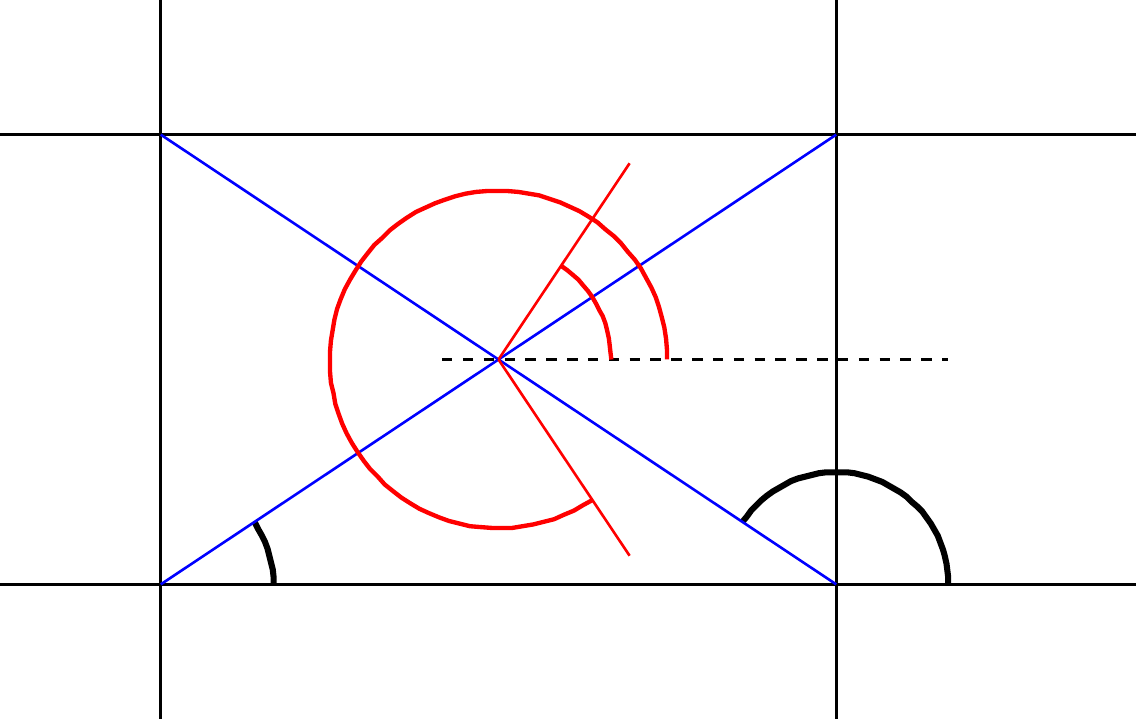}
 }
 \put(-40,80){$x$}
 \put(-60,10){$x_2$}
 \put(-215,10){$x_1$}
 \put(-230,135){$y_2$}
 \put(-230,35){$y_1$}
 \put(-60,135){\Large e$^-$}
 \put(-235,15){\Large e$^+$}
 \put(-35,40){$\hat \omega'$}
 \put(-185,35){$\hat \omega$}
 \put(-120,70){\Red{$\hat \phi'$}}
 \put(-190,80){\Red{$\hat \phi$}}
 \put(-112,40){\Red{$\Delta$}}
 \put(-110,120){\Red{$\Delta'$}}
\end{picture}
\caption{One-dimensional toy model:
Schema of the azimuthal angle definitions for a 4-reconstructed-track event.
The position of the electron and of the positron in the azimuthal plane are indicated by e$^-$ and e$^+$.
The thin horizontal and vertical straight lines denote the hit strips in the second detector.
The correct and wrong azimuthal angles of the event are indicated, for
the two definitions of the azimuthal angle used in this study,
Wojtsekhowski's $\varphi \equiv \omega$ \cite{Bogdan:1998}
and
Gros's $\varphi \equiv \phi$ \cite{Gros:2016dmp}.
Lines $\Delta$ and $\Delta'$ show the azimuthal directions of the bisectrix for the correctly and incorrectly reconstructed pairs, respectively.
\label{fig:4tracks:schema}
 }
 \end{center}
\end{figure}
The azimuthal plane is parallel to the detector wafers.
The azimuthal angle of the correctly reconstructed event in the LAT
frame, $\hat \varphi$, is related to the azimuthal angle in the sky
frame, $\varphi$, by
\begin{eqnarray}
\hat \varphi & = & \varphi + \delta ,
\end{eqnarray}
where $\delta$ documents the azimuthal angle of the LAT wrt the sky
frame.
The azimuthal angle of the wrongly reconstructed candidate in the LAT
frame, $\hat \varphi'$ is obtained from the angle of the correctly
reconstructed candidate, modulo $\pi$
(Fig. \ref{fig:4tracks:schema}), by:
\begin{eqnarray}
 \hat \varphi' & = & \pi - \hat \varphi ;
\end{eqnarray}
hence, back to the sky frame,
\begin{eqnarray}
 \varphi' & = & \hat \varphi' - \delta \nonumber \\
 & = & \pi - \varphi - 2 \delta. 
\end{eqnarray}

\begin{figure}[htbp!]
 \begin{center}
\iftoggle{couleur}{
 \includegraphics[width=0.49\linewidth]{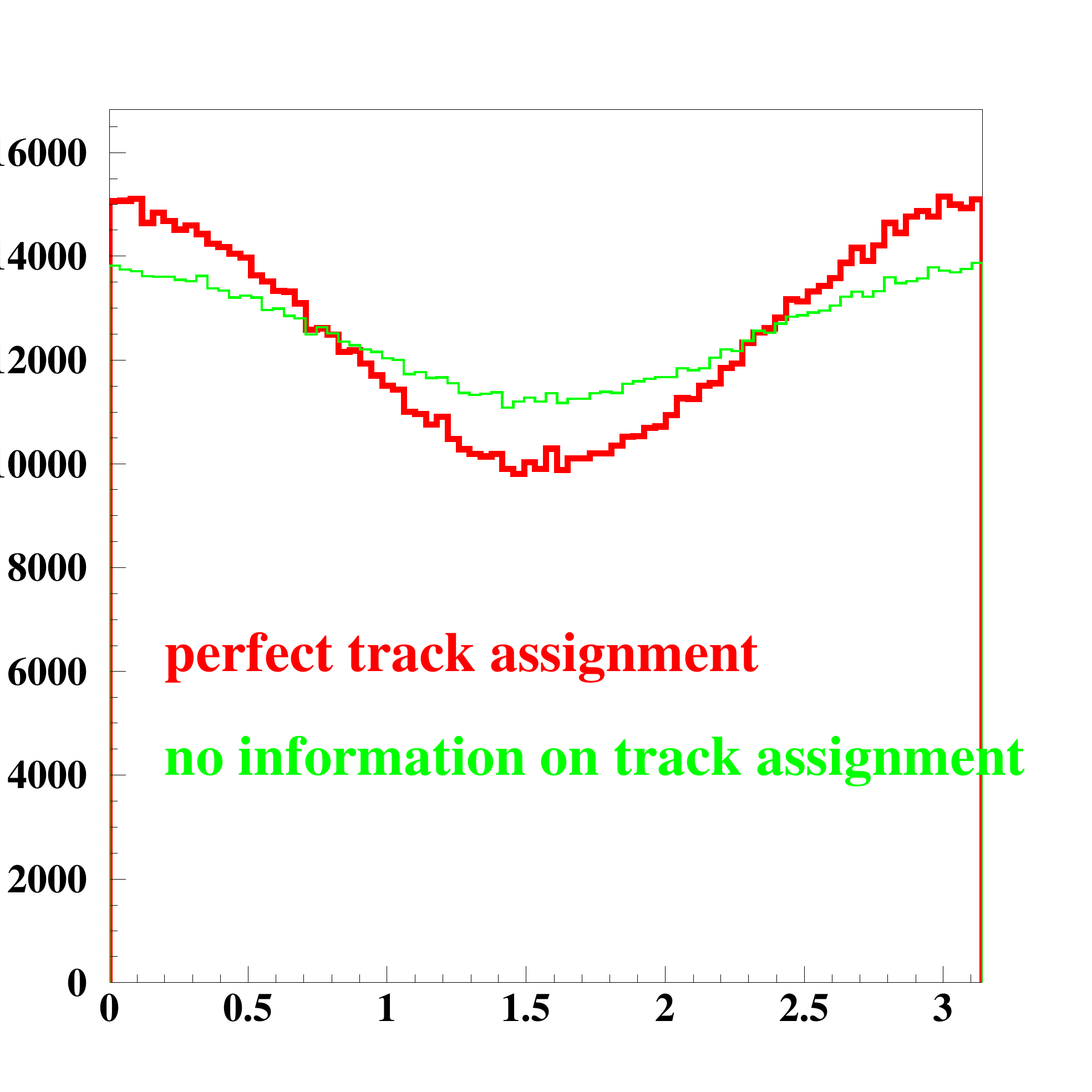}
 \put(-45,0){$\varphi$}
 \hfill
 \includegraphics[width=0.49\linewidth]{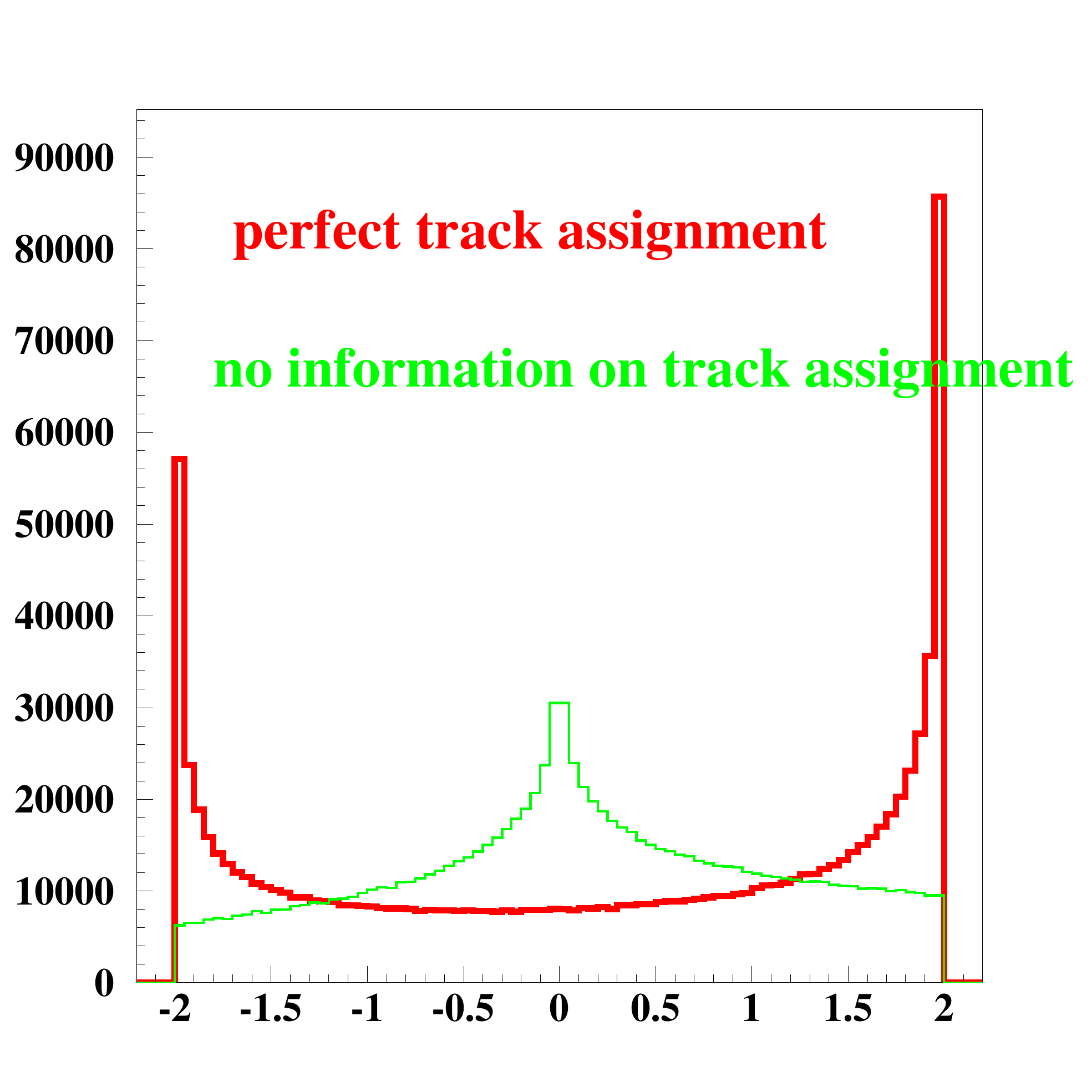}
 \put(-45,0){$w$}
}{
 \includegraphics[width=0.49\linewidth]{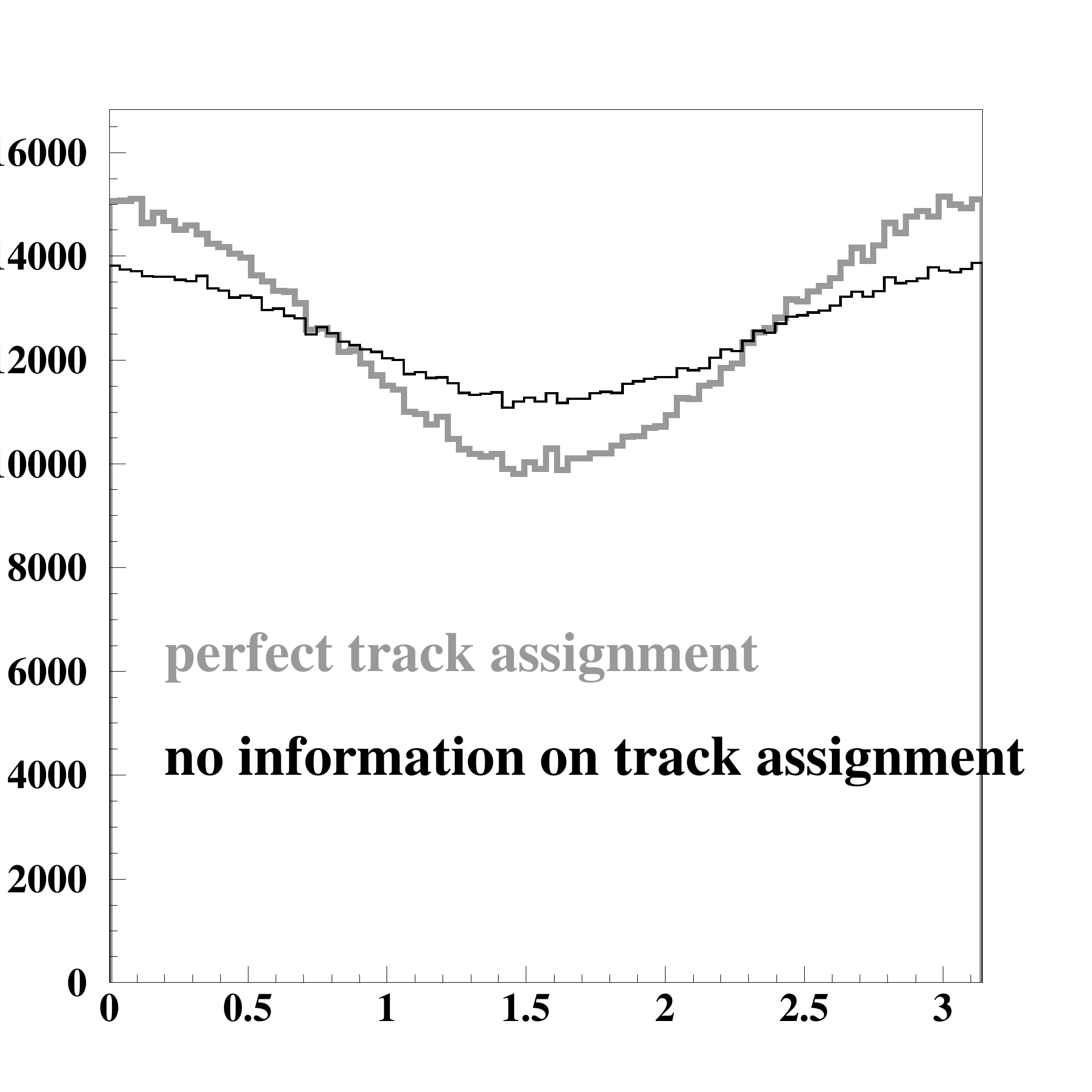}
 \put(-45,0){$\varphi$}
 \hfill
 \includegraphics[width=0.49\linewidth]{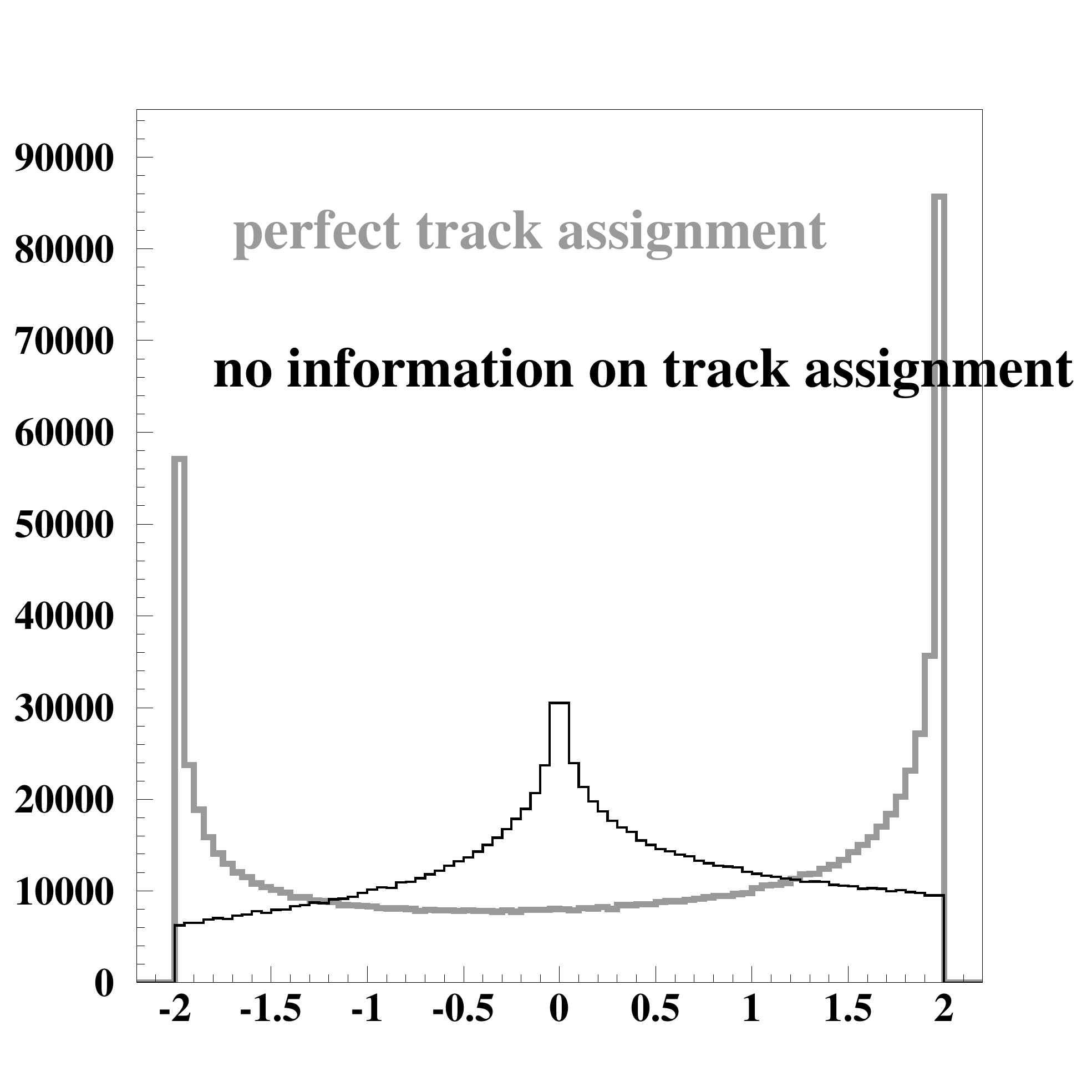}
 \put(-45,0){$w$}
 }
 \end{center} 
\caption{ $\varphi$ (left plot) and $w$ (right plot) distributions
 from a toy MC sample generated from the 1D differential cross
 section of (eq. (\ref{eq:1D})) with $A=0.2$ and $P=1$.
Thick histogram: original distribution, i.e. with perfect track
assignment available.
Thin histogram: using both candidates with equal weights, as when no
information on track assignment is available.
Adapted from \cite{Philippe}, see pages 39-40 of
\cite{Bernard:SciNeGHE2016}.
\label{fig:4tracks:ToyMC}
 }
\end{figure}

When no information is available to decipher which track pair is the
correct one, both possible values are used with $1/2$ weighting, and
the original distribution (eq. (\ref{eq:1D})) becomes
\begin{eqnarray}
 \gfrac{\dd^2 N }{ \dd \varphi \dd \delta } & \propto &
\gfrac{1}{2} 
 \left[
 \left(1 + A \times P \cos[2(\varphi-\varphi_0)] \right) + \right. \nonumber \\
 & & \left.
 ~ ~ ~ ~
 \left(1 + A \times P \cos[2(\varphi + 2 \delta +\varphi_0)] \right)
 \right] .
\end{eqnarray}

Under an isotropic exposure, for each value of $ \varphi $, the
second term of the second line
cancels upon integration on $\delta$, and we are left with
\begin{equation}
 \gfrac{\dd N }{ \dd \varphi} \propto
 \left(1 + \gfrac{A \times P}{2} \left[ \cos(2(\varphi-\varphi_0)) \right] \right)
 ,
\end{equation}

which means that the additional dilution factor induced by the
two-fold ambiguity in the 4-reconstructed-track events is of $1/2$.
This is illustrated with a toy-MC simulation in Fig. \ref{fig:4tracks:ToyMC}.

In case some non-zero efficiency to assign the correctly reconstructed
track pair would be available, the dilution is found to vary linearly
with that efficiency, from $D=0.5$ for zero efficiency to $D=1$ for
perfect assignment \cite{Philippe}.

\begin{figure}[htbp!]
 \begin{center}
 \includegraphics[width=0.9\linewidth]{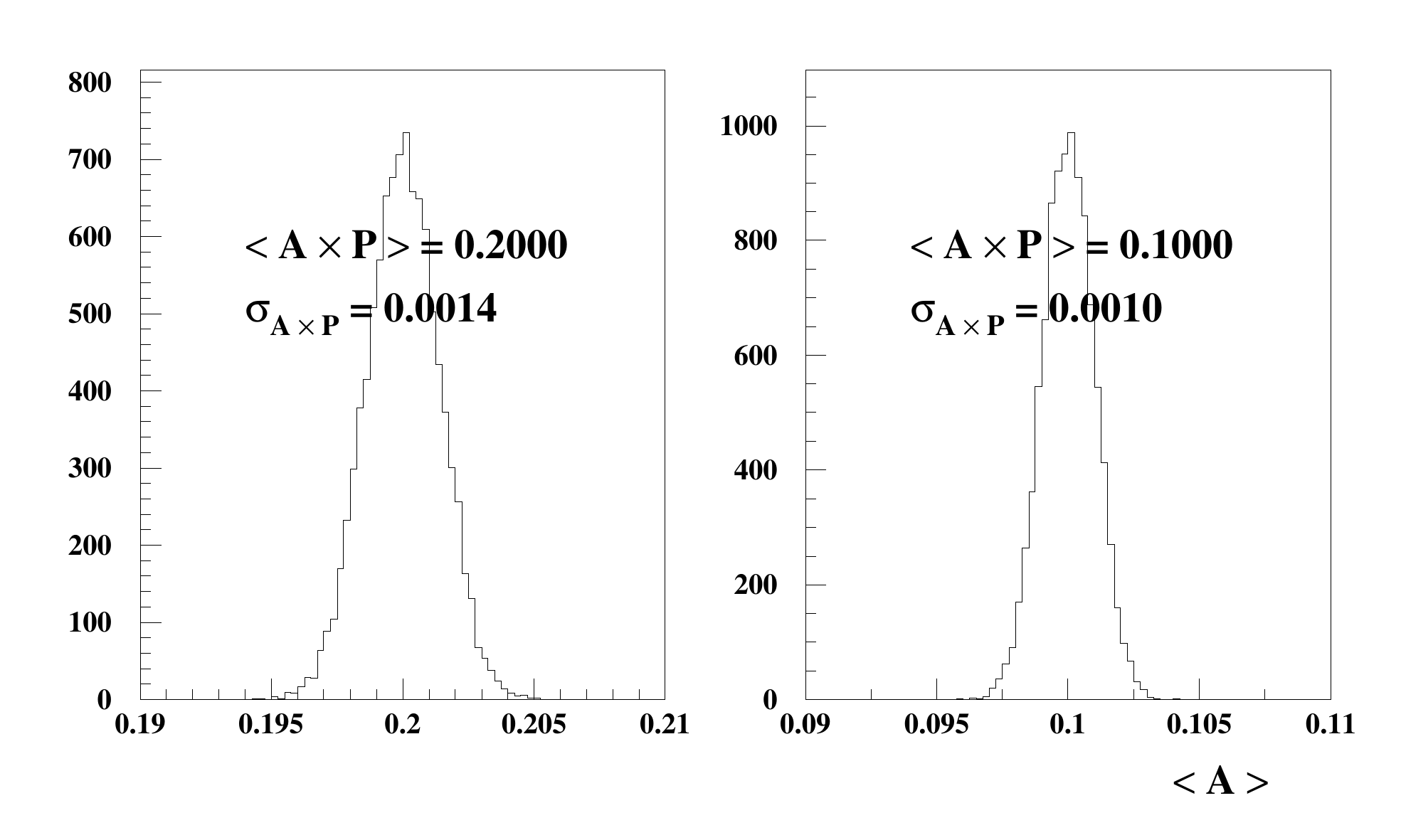}
 \end{center} 
\caption{Results of $10^4$ Toy MC experiments of $10^6$ events each,
 sampled from eq. (\ref{eq:1D}) with $A=0.2$ and $P=1$.
 Left: perfect assignment; right:
 random LAT orientation, and both possible track combinations used.
\label{fig:4tracks:ToyMC:expts}
 }
\end{figure}

\begin{figure}[htbp!]
 \begin{center}
\iftoggle{couleur}{
 \includegraphics[width=0.6\linewidth]{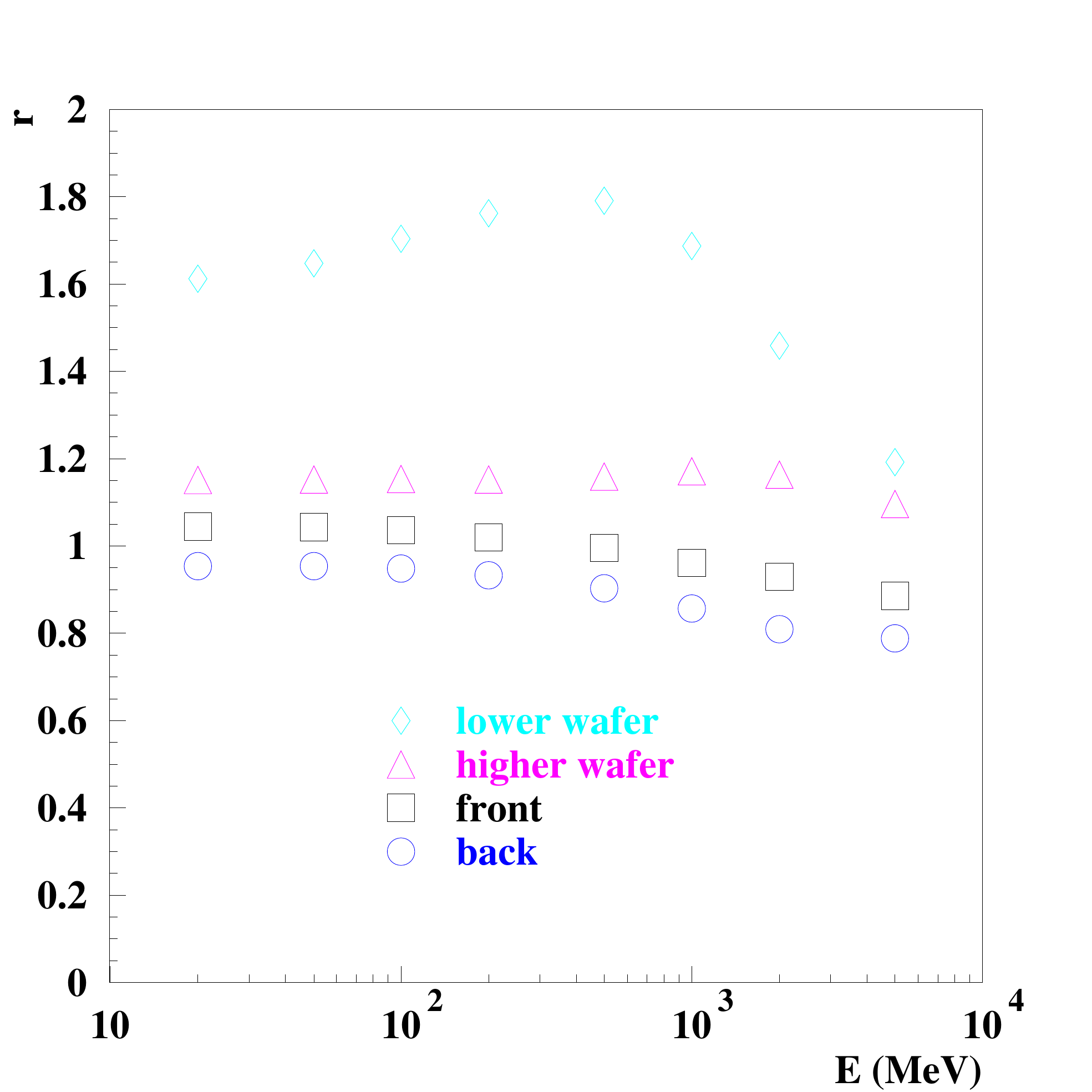}
}{
 \includegraphics[width=0.6\linewidth]{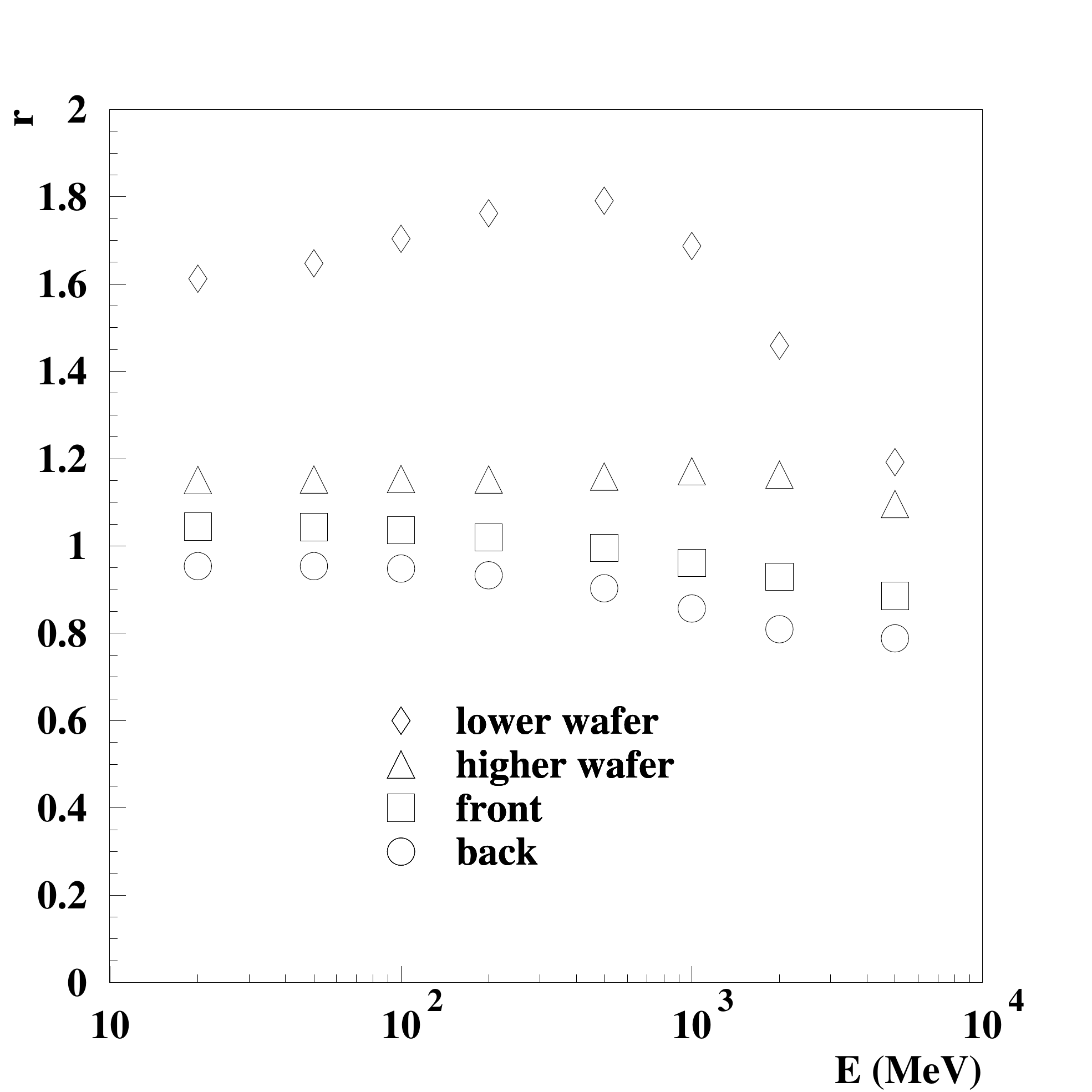}
 }
 \end{center} 
 \caption{Test of a best-candidate assignment based on acoplanarity:
 Ratio, $r$, of the number of
correctly to the number of falsely assigned candidates, as a function
of energy, for various conversion locations.
\label{fig:4tracks:assign}
 }
\end{figure}

Figure \ref{fig:4tracks:ToyMC:expts} shows
the results of a toy MC exercise in which
$10^4$ experiments are simulated with 
$10^6$ events each, sampled from 
eq. (\ref{eq:1D}) with $A=0.2$ and $P=1$.
We obtain 
\begin{itemize}
 \item
 with perfect track assignment (left plot,
 $\langle A \times P \rangle = 0.2000$,
 $\sigma_{A \times P} =0.0014$, that is,
 $\sigma_{A \times P} / \langle A \times P \rangle = 0.007$
 ), or
\item
 with random $\delta$ angle and both assignments used (right plot,
 $\langle A \times P \rangle = 0.1000$,
 $\sigma_{A \times P} =0.0010$, that is,
 $\sigma_{A \times P} / \langle A \times P \rangle = 0.010$
 ),
\end{itemize}
so even though the ambiguity inflicts a $D = 0.5$ dilution of the
polarisation asymmetry, using both possible candidates in the
calculation of the weight results in a reduction of $W$ from $W
\approx \sqrt{2}$ to $W \approx 1$, and therefore the loss in the
precision of the measurement of $P$ due to the ambiguity is only of a
factor of $\sqrt{2}$.

The detailed simulations using the exact differential cross section
and taking detector effects, that have been presented in the previous
sections, confirm the results of the simple 1D Toy model, that is,
$D \approx 0.5$ and $W \approx 1$ (plots not shown).

\begin{figure*}[htbp!]
\begin{center}
\iftoggle{couleur}{
\includegraphics[width=\linewidth]{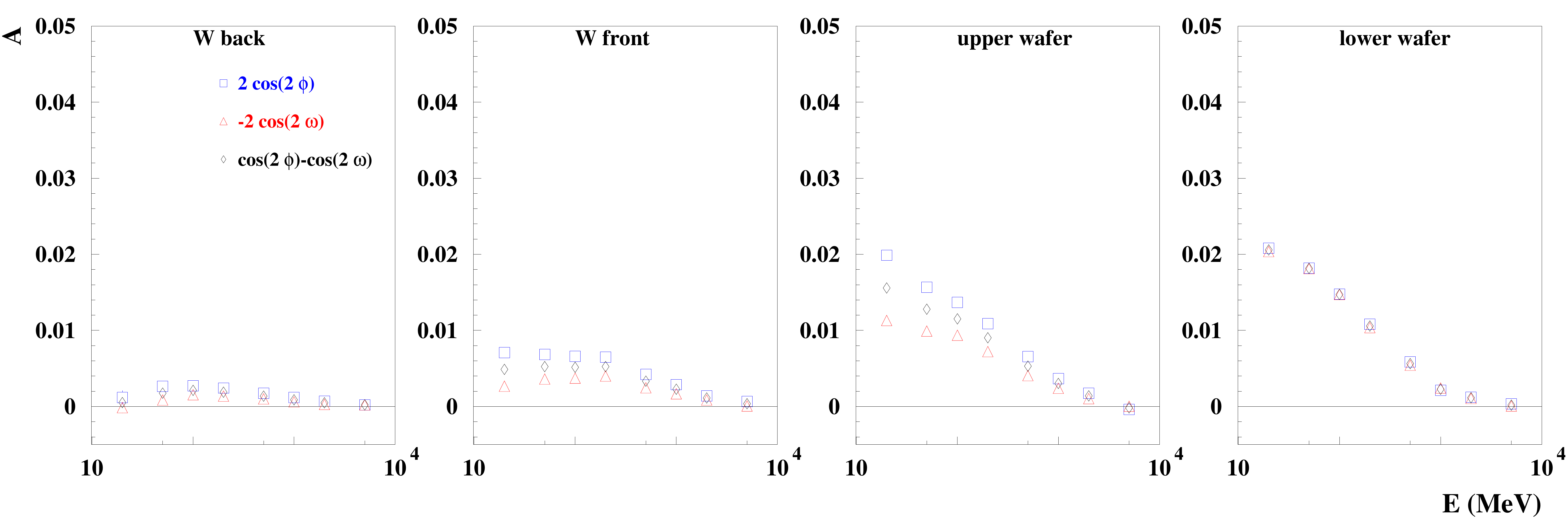}
\includegraphics[width=\linewidth]{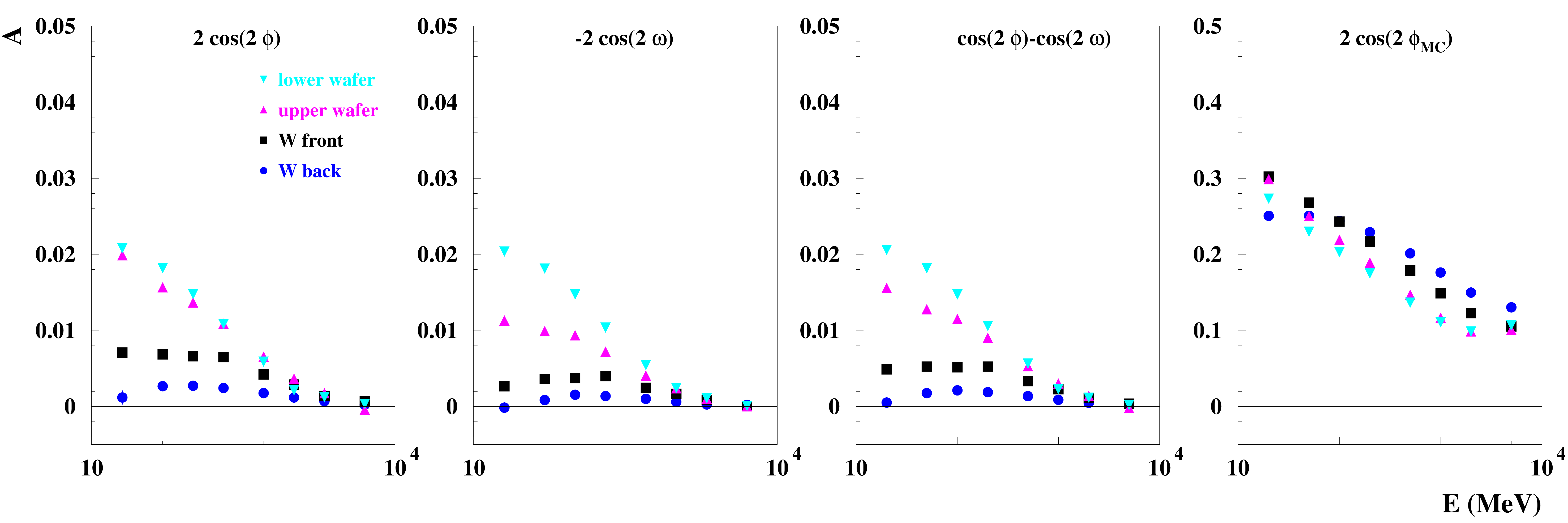}
\includegraphics[width=\linewidth]{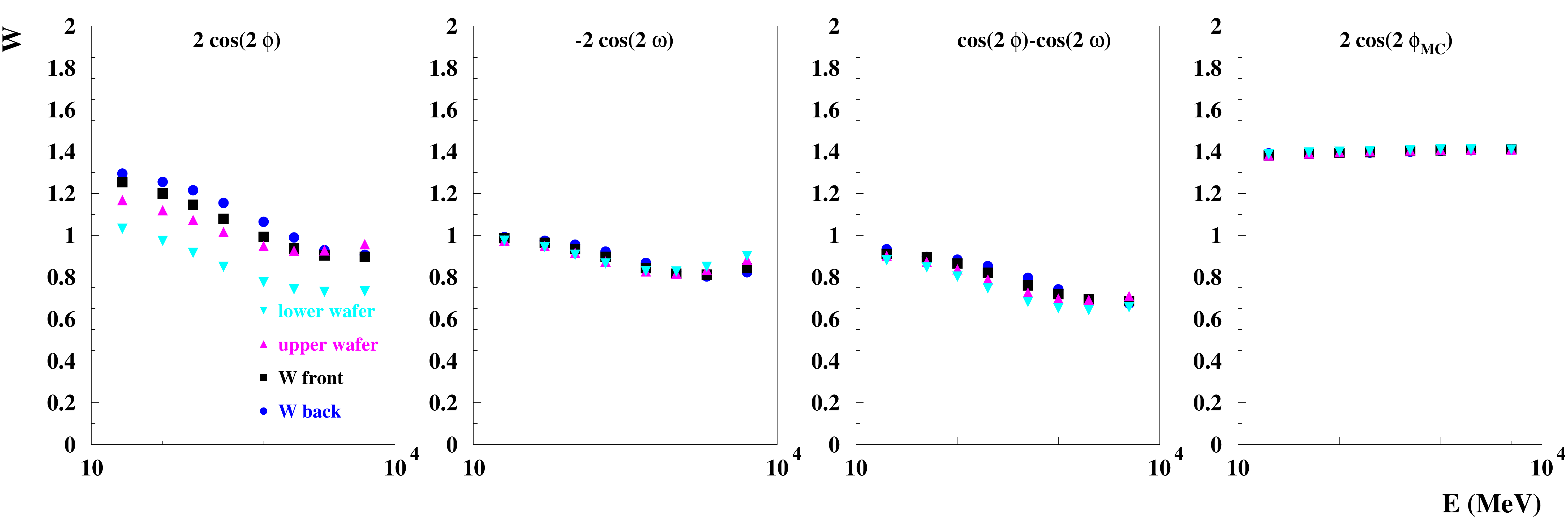}
}{
\includegraphics[width=\linewidth]{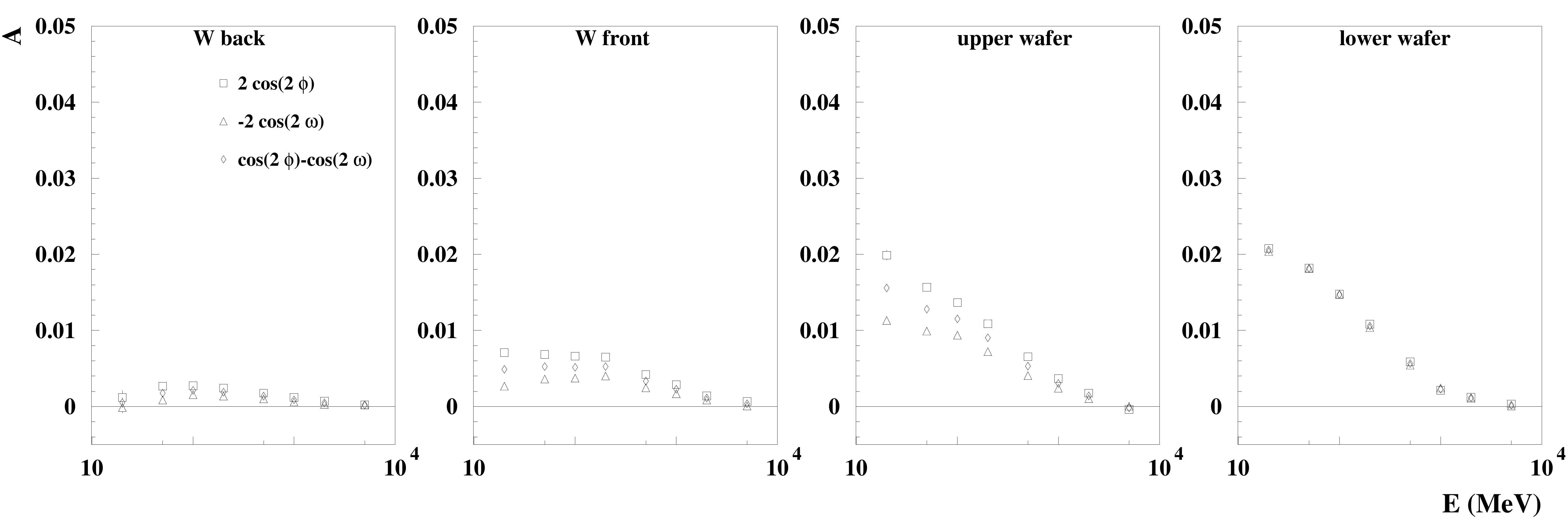}
\includegraphics[width=\linewidth]{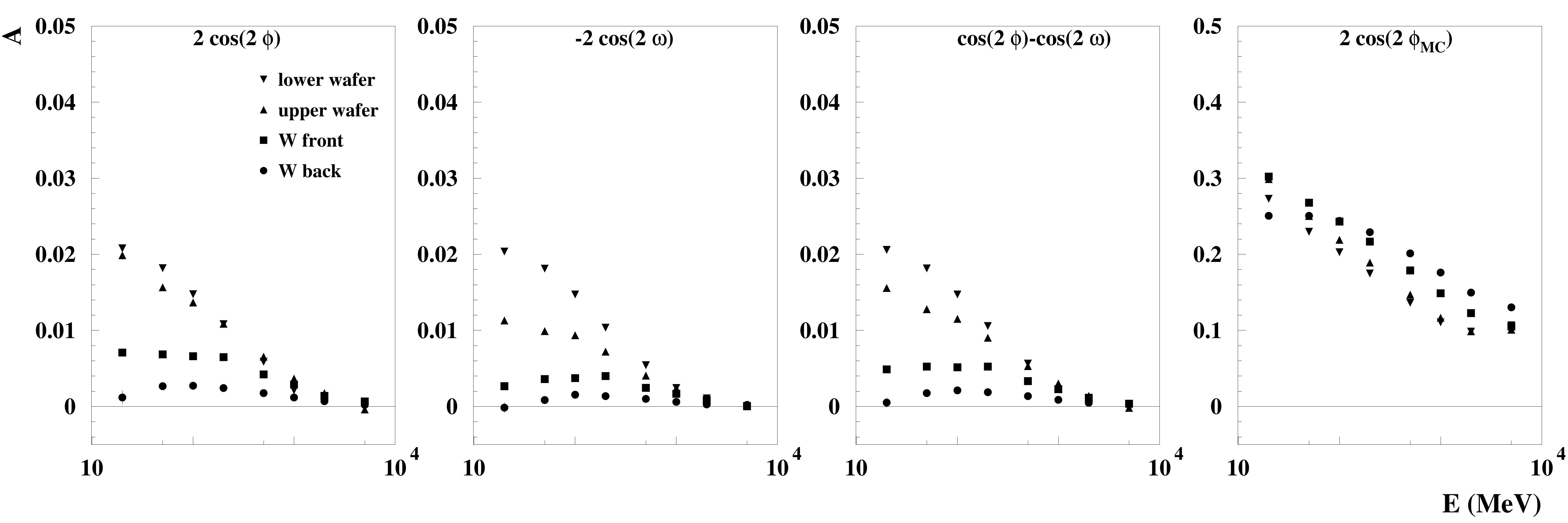}
\includegraphics[width=\linewidth]{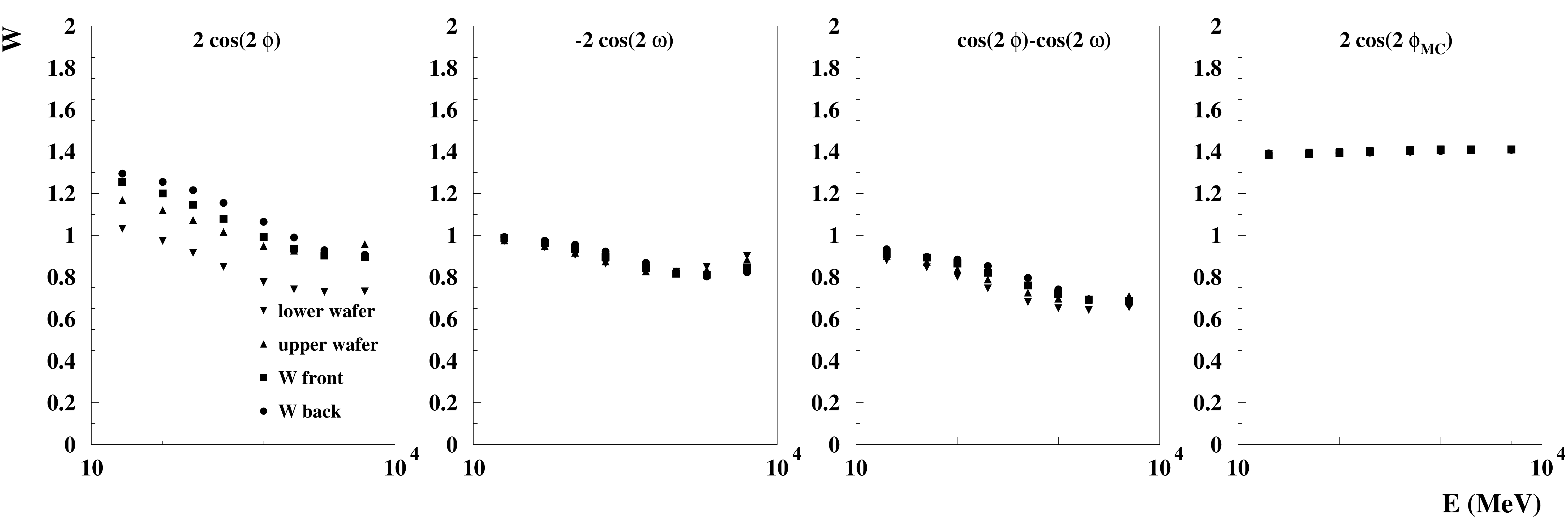}
 }
\captionsetup{singlelinecheck=off}
\caption[~]{$\ntrack =4$ events:
Polarisation asymmetry as a function of incident photon energy. \label{fig:aa:4}
 \begin{itemize}
 \item
 {\bf Upper row}: For various
 conversion locations, from left plot to right plot: W back, W front, upper wafer, lower wafer, and various measurement methods (weights
 $2\cos(2\phi)$ (open square),
 $-2\cos(2\omega)$ (open triangle),
 $\cos(2\phi) - \cos(2\omega)$ (open diamond).
 \item
 {\bf Centre row}: For various methods, from left plot to right plot: 
 $2\cos(2\phi)$,
 $-2\cos(2\omega)$,
 $\cos(2\phi) - \cos(2\omega)$ and for the Monte Carlo generator 
 $2\cos(2\phi_{\MC})$,
 and various conversion locations
 W back (bullet),
 W front (full square),
 upper wafer (upward triangle),
 lower wafer (downward triangle), 
 \item
 {\bf Bottom row}: The width of the distribution.
\end{itemize}
Notice the different vertical scales for the measured
($\phi$ or $\omega$) and the MC values of the azimuthal angle.
}
\end{center}
\end{figure*}

\subsection{Track matching attempt}
\label{subsec:matching:attempt}

The transverse angular kick undergone by the pair due to the recoil
momentum carried away by the nucleus has a 68\,\% containment value of
$\theta_{68} = 1.5 \, \radian [E / \mega\electronvolt]^{-5/4}$
\cite{Bernard:2012uf}, 
while the most probable value of the pair opening angle is
$\hat\theta_{+-} = {1.6\, \radian \, \mega\electronvolt}/{E}$ 
(eq. (\ref{eq:theta:pm}) and \cite{Olsen:1963zz}),
so the kick is smaller than the opening angle on the whole energy range
(we obtain $\theta_{68} =\hat\theta_{+-} $ for
$E = (1.5 / 1.6)^4 \approx 0.8 \, \mega\electronvolt$).
Therefore the two tracks are produced almost back-to-back in the
azimuthal plane, something that could be conserved, to some extent, in
correctly reconstructed photon candidates, but that would be disturbed
for incorrect ones.
Indeed, the spectra of the acoplanarity angle for correct and false
reconstructions show a difference, something that can enable an
assignment: the combination that shows the largest acoplanarity angle
(closest to $\pi$) is chosen.

Figure \ref{fig:4tracks:assign} shows the ratio, $r$, of the number of
correctly to the number of falsely assigned candidates, as a function
of energy, for various conversion locations.
\begin{itemize}
\item
Little improvement is seen with respect to random assignment
($r\approx 1$), except for conversions in the lower wafer and at low energies.

\item
For other locations, multiple scattering is the main source of
acoplanarity and blurs the information initially present in the tracks.

\item
At higher energies, even for a conversion in the lower wafer, $r$
decreases towards unity, most likely because the opening angle at the
vertex is small and multiple scattering, that is then the main contribution
to an effective opening angle large enough to reach $\ntrack =4$,
ruins the information of the acoplanarity angle at vertex for both
candidates.
\end{itemize}

Given these results, I will not use that best-candidate method in the
following.

\begin{figure*}[htbp!]
 \begin{center}
\iftoggle{couleur}{
\includegraphics[width=\linewidth]{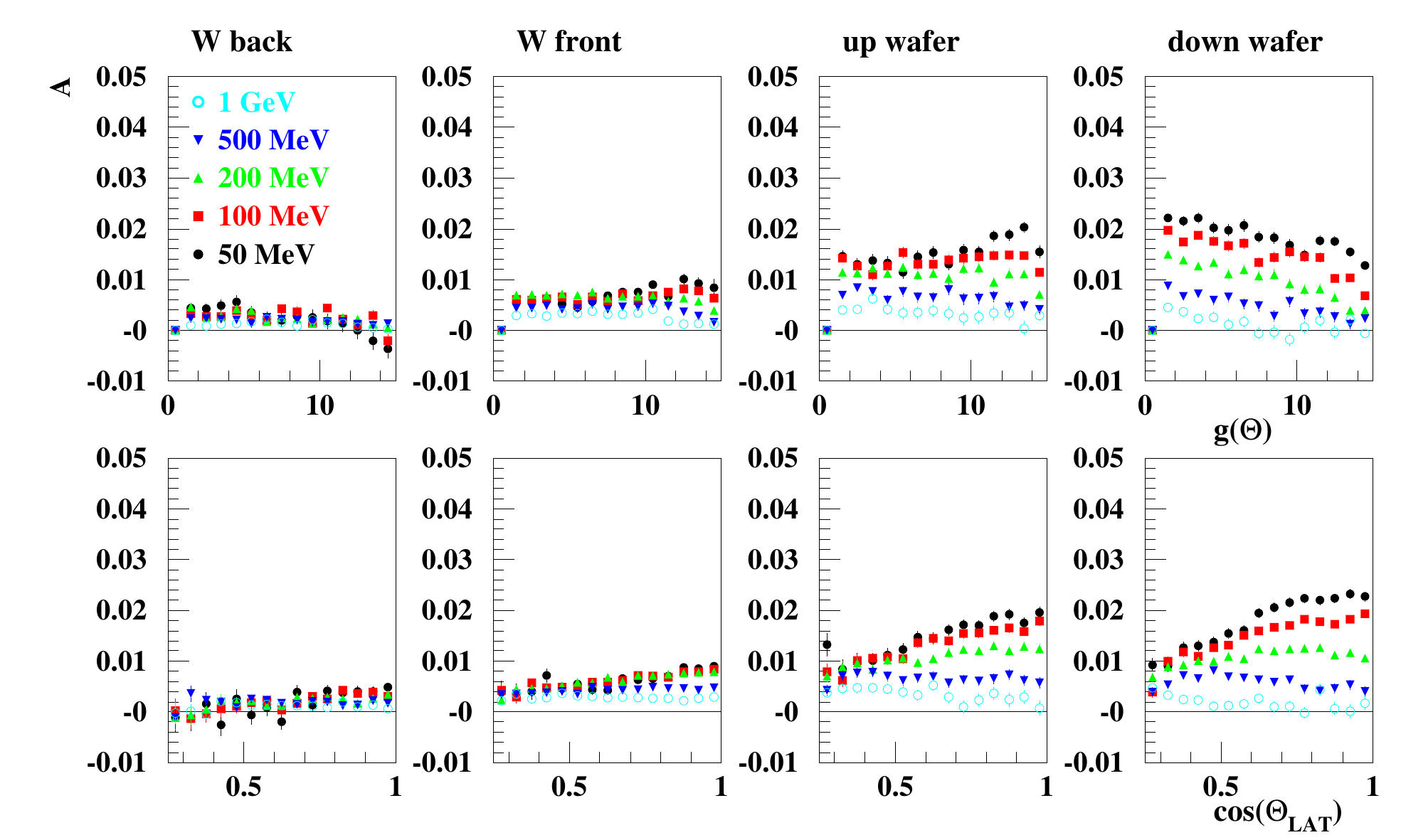}
}{
 \includegraphics[width=\linewidth]{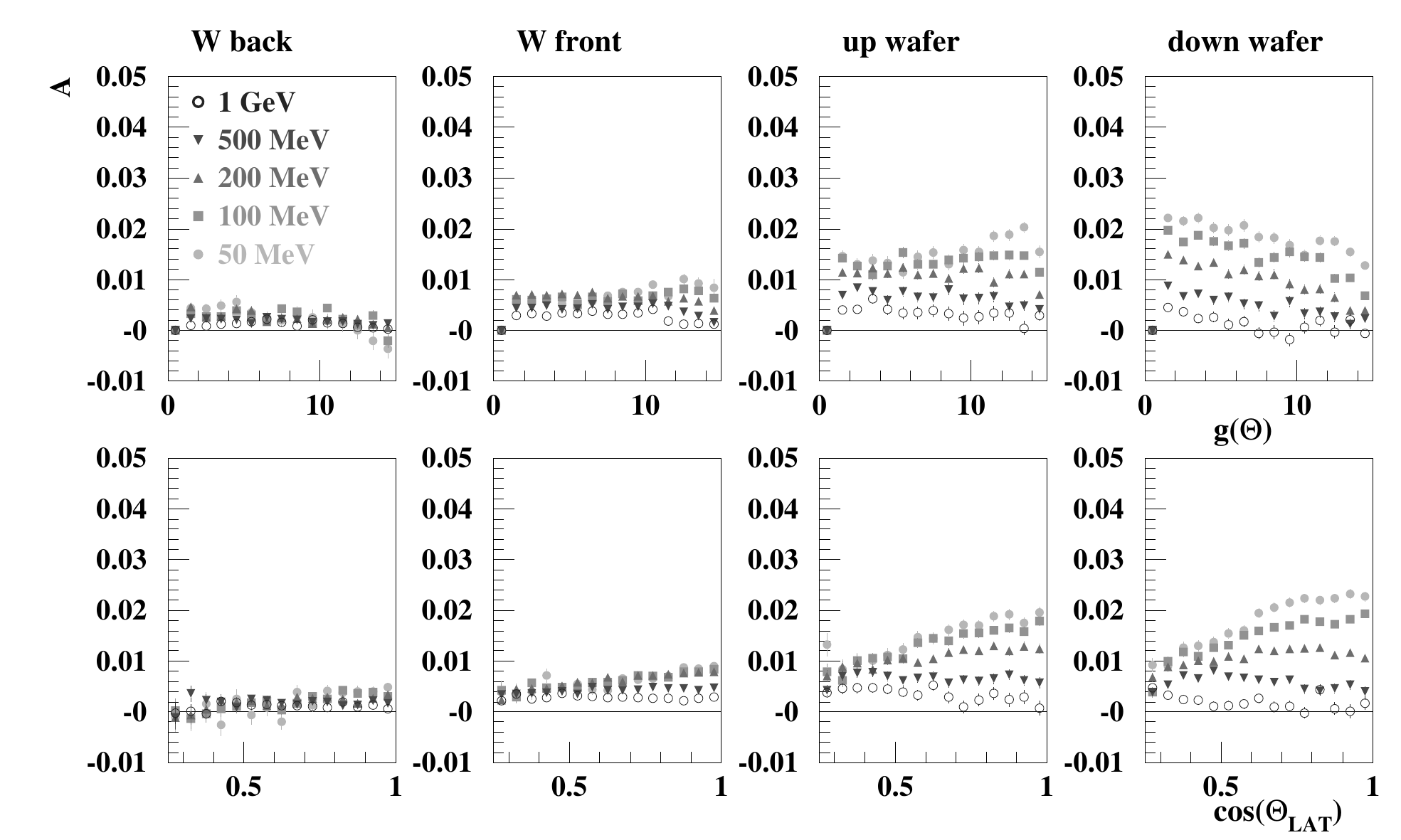}
 }
 \caption{$\ntrack =4$ events: Variation of the polarisation
 asymmetry as a function of event observables
 $g(\theta)$ (top row) and
 $\cos{\theta_{\LAT}}$ (bottom row),
 for various conversion locations
 (from left to right: W back, W front, upper wafer and bottom wafer)
 and for various photon energies.
The measurement is performed using the moments method with weight $2\cos(2\phi)$.
 \label{fig:variation:4}}
 \end{center}
\end{figure*}

\subsection{$\ntrack =4$ events: measurement}

The asymmetry measured using either $\phi$- or $\omega$-based weights
or their combination, or the MC $\phi$ value, for various conversion
locations (lower wafer, upper wafer, front W and back W) is shown as
a function of incident photon energy in
Fig. \ref{fig:aa:4} for $\ntrack =4$ in a way similar to
Fig. \ref{fig:aa} for $\ntrack =2$.
We see again a number of features that we observed for the
$\ntrack =2$ events:
\begin{itemize}
\item A polarisation asymmetry that vary with conversion location,
 improving from ``back'' to ``lower wafer'';
\item A poor improvement of the values of $W$, for a combination of
 $\omega$- and $\varphi$-using measurements, with respect to
 measurements made using either of these azimuthal angle;
\end{itemize}
but we note also some different features:
\begin{itemize}
\item A polarisation asymmetry that decreases with energy,
\item A value of $W$ that is close to unity, as anticipated in subsection
 \ref{subsec:ambiguity}.
\end{itemize}
 
\begin{figure*}[htbp!]
\begin{center}
\iftoggle{couleur}{
\includegraphics[width=\linewidth]{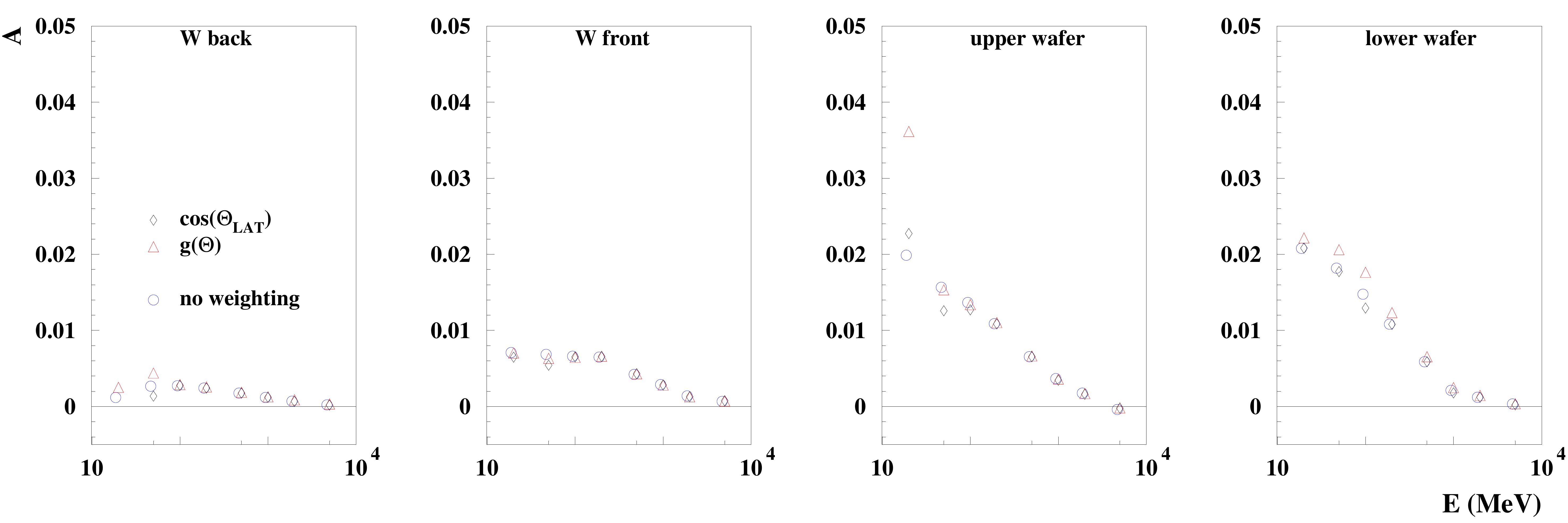}
}{
\includegraphics[width=\linewidth]{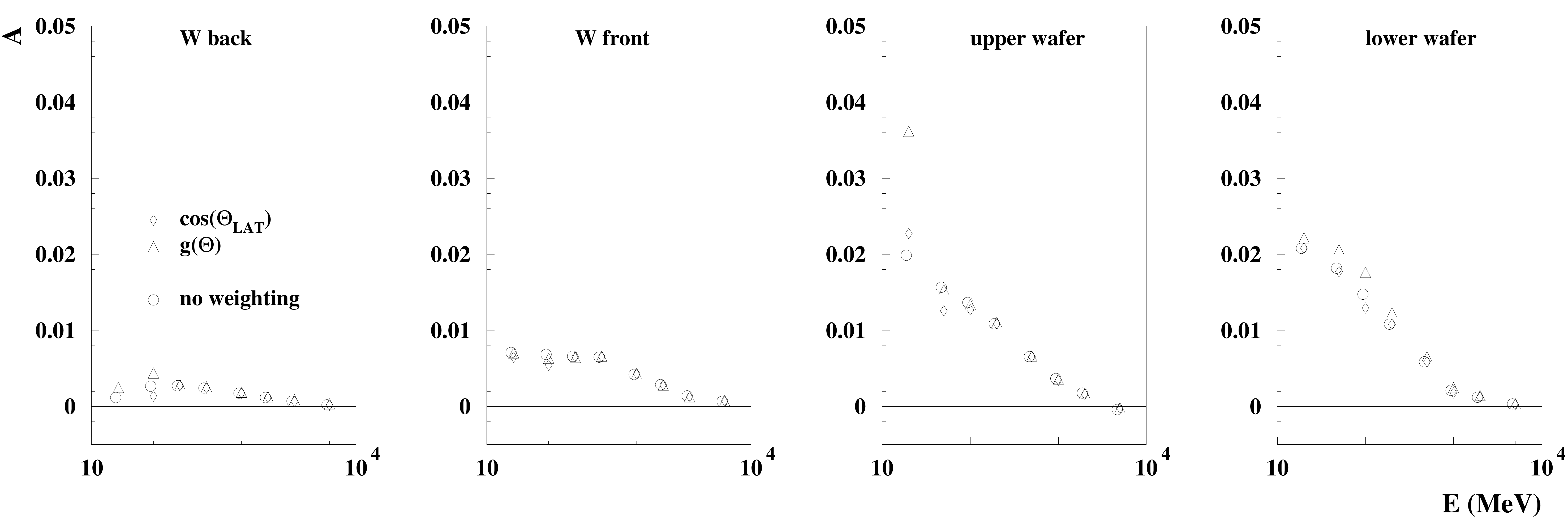}
 }
\caption[~]{Polarisation asymmetry as a function of incident photon
 energy, and various conversion locations, with and without
 weighting.
($\ntrack =4$ events, $2\cos(2\phi)$).
Small horizontal shifts have been applied to improve on the readability.
 \label{fig:pond:4}}
\end{center}
\end{figure*}

The variation with event variables $g(\theta)$ and
$\cos{\theta_{\LAT}}$ is presented in
Fig. \ref{fig:variation:4} for $\ntrack =4$ in a way similar to
Fig. \ref{fig:variation} for $\ntrack =2$.
The variation of the polarisation asymmetry is found to be milder for
$\ntrack =4$ events than for $\ntrack =2$.
Accordingly, weighting events with an average value of the polarisation
asymmetry that depends on the value of one of these variables brings
little improvement, if any (Fig. \ref{fig:pond:4}).

\begin{figure*}[htbp!]
\begin{center}
\iftoggle{couleur}{
\includegraphics[width=\linewidth]{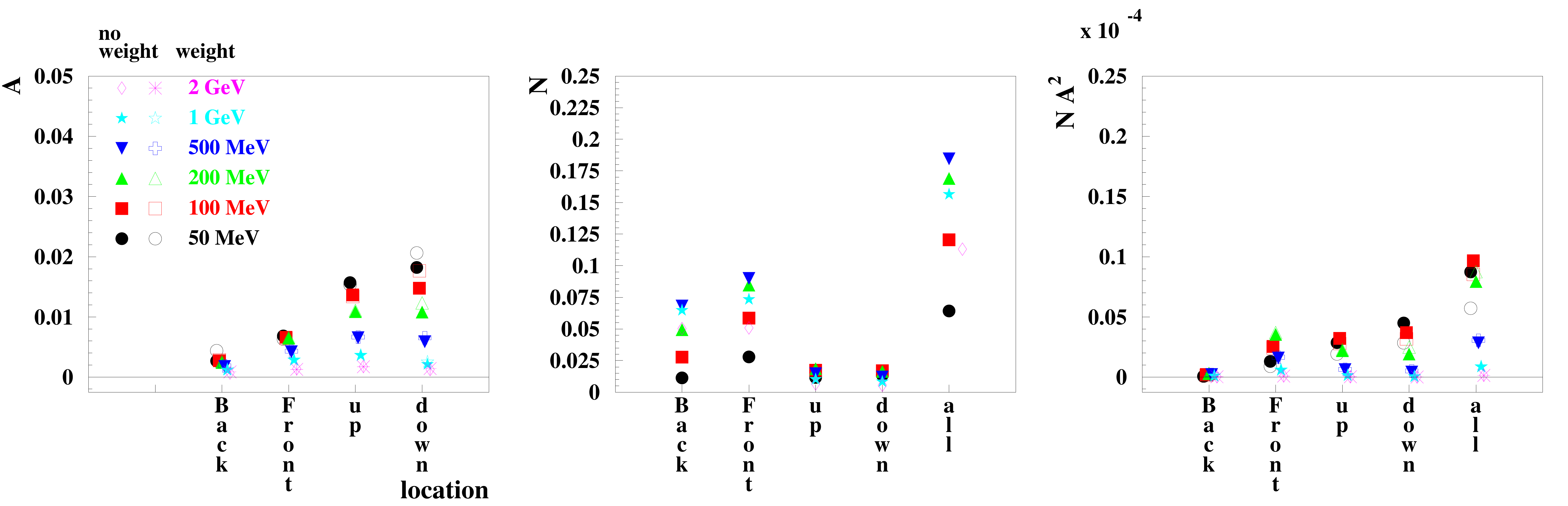}
}{
 \includegraphics[width=\linewidth]{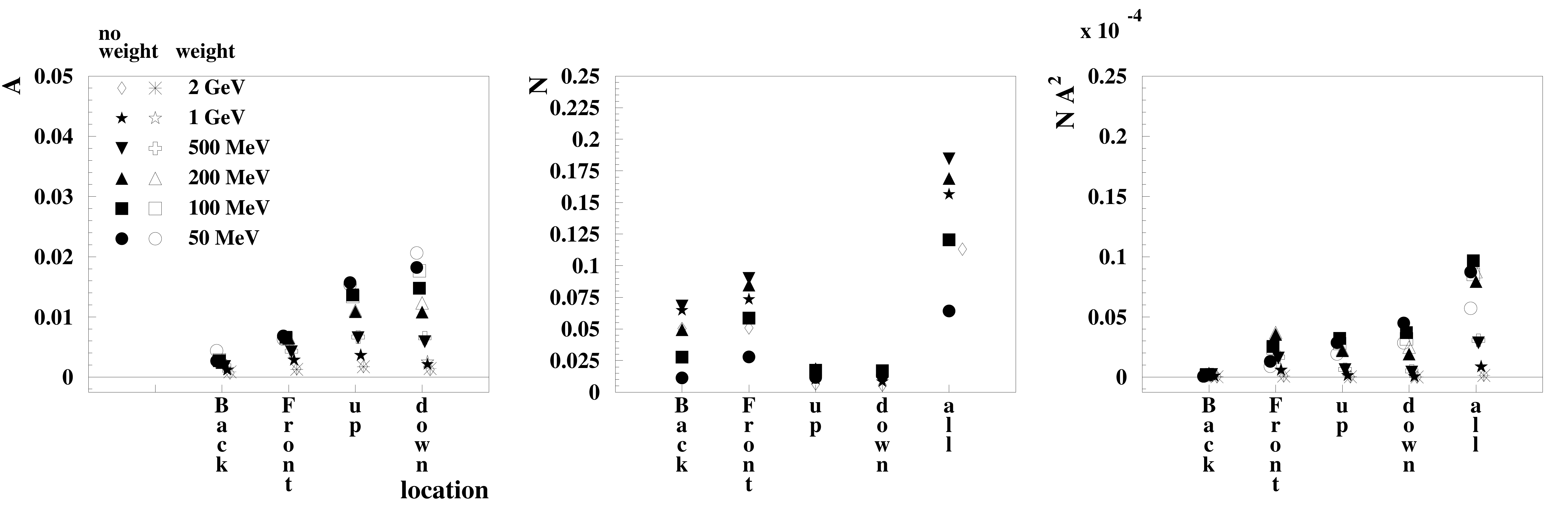}
 }
\caption[~]{$\ntrack =4$ events:
Polarisation asymmetry $A$ (left), number of events $N$ (centre) and value of 
$N_k A^2_k$ as a function of the conversion location
(from ``W back'' ($k=1$) to
``down wafer'' ($k=4$)) and their combination ``all''.
Without and with $g(\theta)$ weighting, for each location,
with global and $g(\theta)$ weighting, for all locations combined.
The number of events is normalised to 1 incident photon on the detector.
\label{fig:trd:4} }
\end{center}
\end{figure*}

The values of $A_k$, of $N_k$ and of $N_k A^2_k$ are shown in
Fig. \ref{fig:trd:4}, for $k=1$ (W back) to $k=4$ (down wafer), and
their optimal combination (all),
in a way similar to Fig. \ref{fig:trd} for $\ntrack =2$.

\begin{figure}[htbp!]
\begin{center}
\iftoggle{couleur}{
\includegraphics[width=0.8\linewidth]{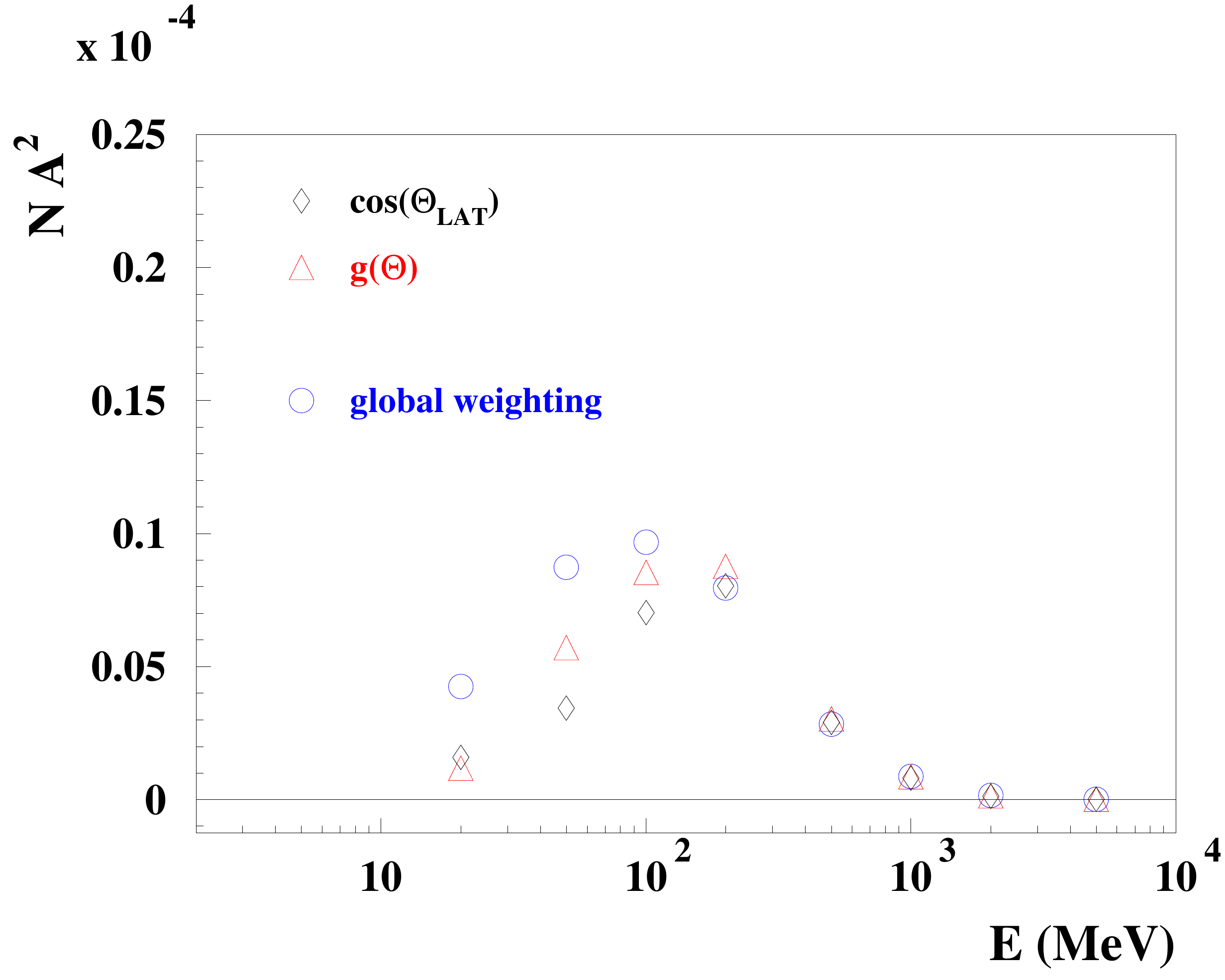}
}{
 \includegraphics[width=0.8\linewidth]{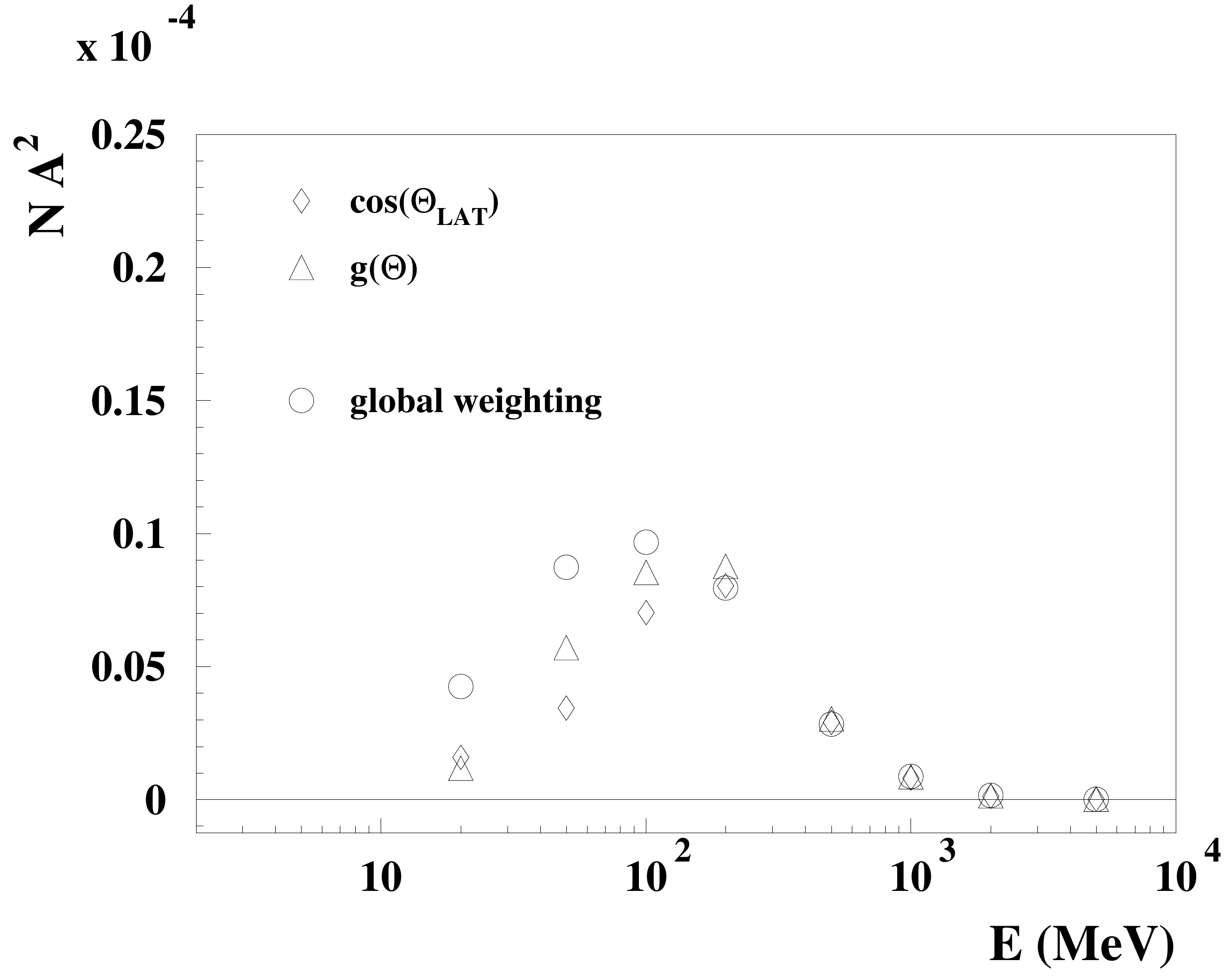}
 }
\caption[~]{$\ntrack =4$ events: Value of $\sum_k N_k A^2_k$
(i.e., all conversion locations combined) as a function of the
 conversion location with various weighting schemes, normalised to one
 incident photon on the detector.
\label{fig:tre:4} }
\end{center}
\end{figure}

Figure \ref{fig:tre:4} shows the overall sensitivity, in terms of
$\sum_k N_k A^2_k$, as a function of energy, for various weighting
schemes, 
in a way similar to Fig. \ref{fig:tre} for $\ntrack =2$.
Most of the sensitivity is clearly at lower energies for $\ntrack =4$ events.
Not only a variable-dependent weighting does not increase the figure
of merit, but at low energies is decreases it, because of the loss in
statistics induced by the sanity cut.

\begin{figure*}[htbp!]
\begin{center}
 \iftoggle{couleur}{
 \hfill
 \includegraphics[width=0.49\linewidth]{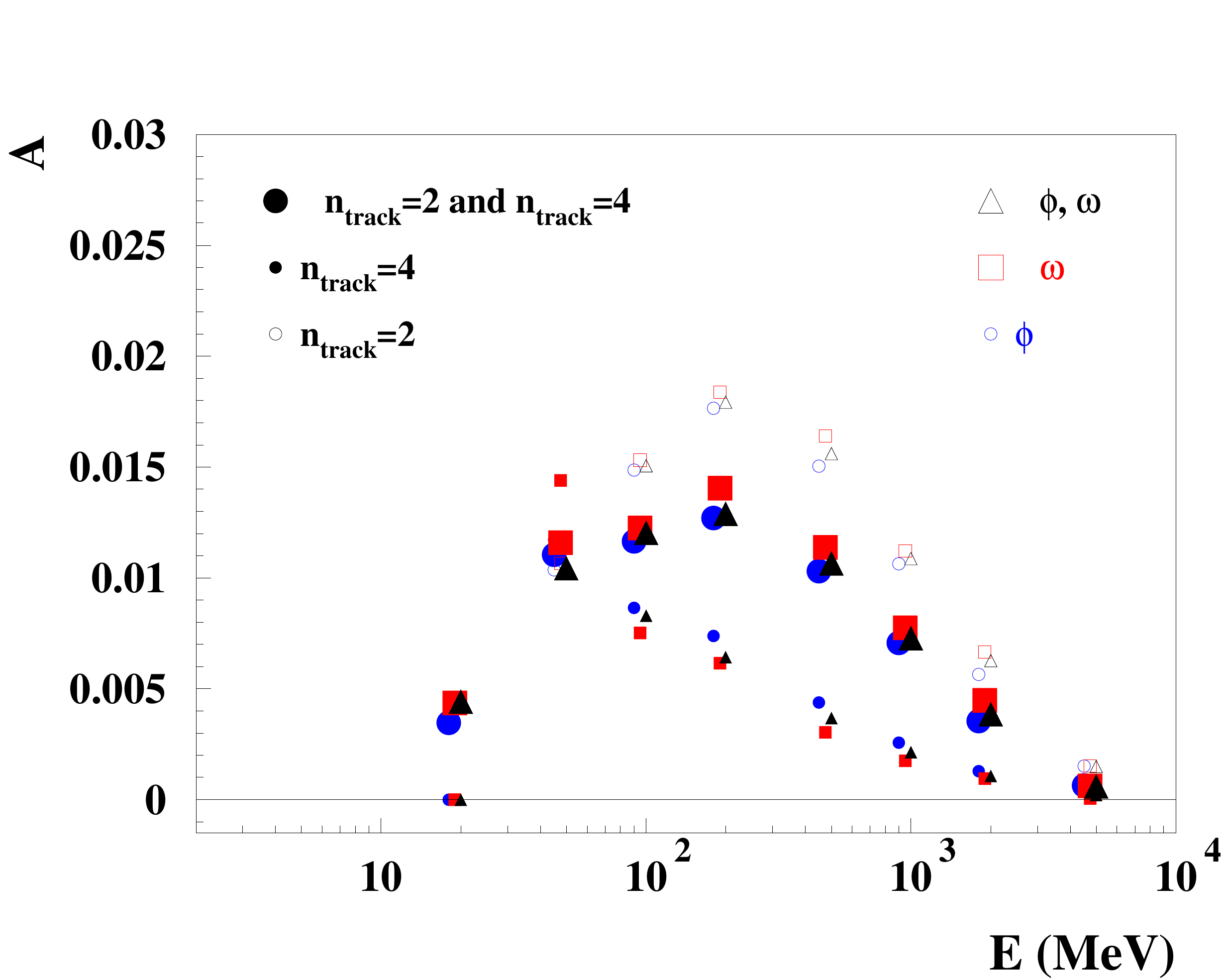}
 \hfill
\includegraphics[width=0.49\linewidth]{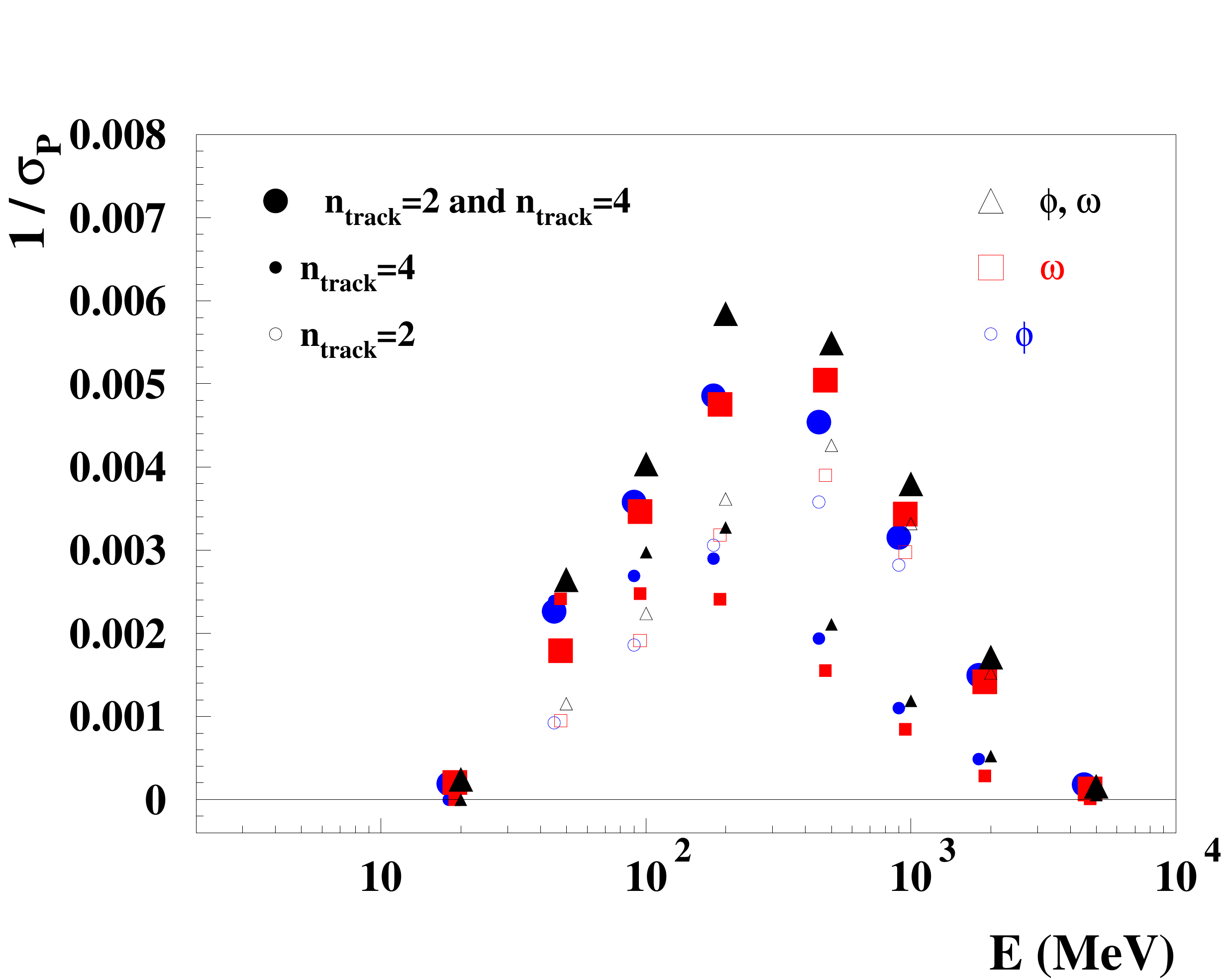}
\hfill
~
}{
 \hfill
 \includegraphics[width=0.49\linewidth]{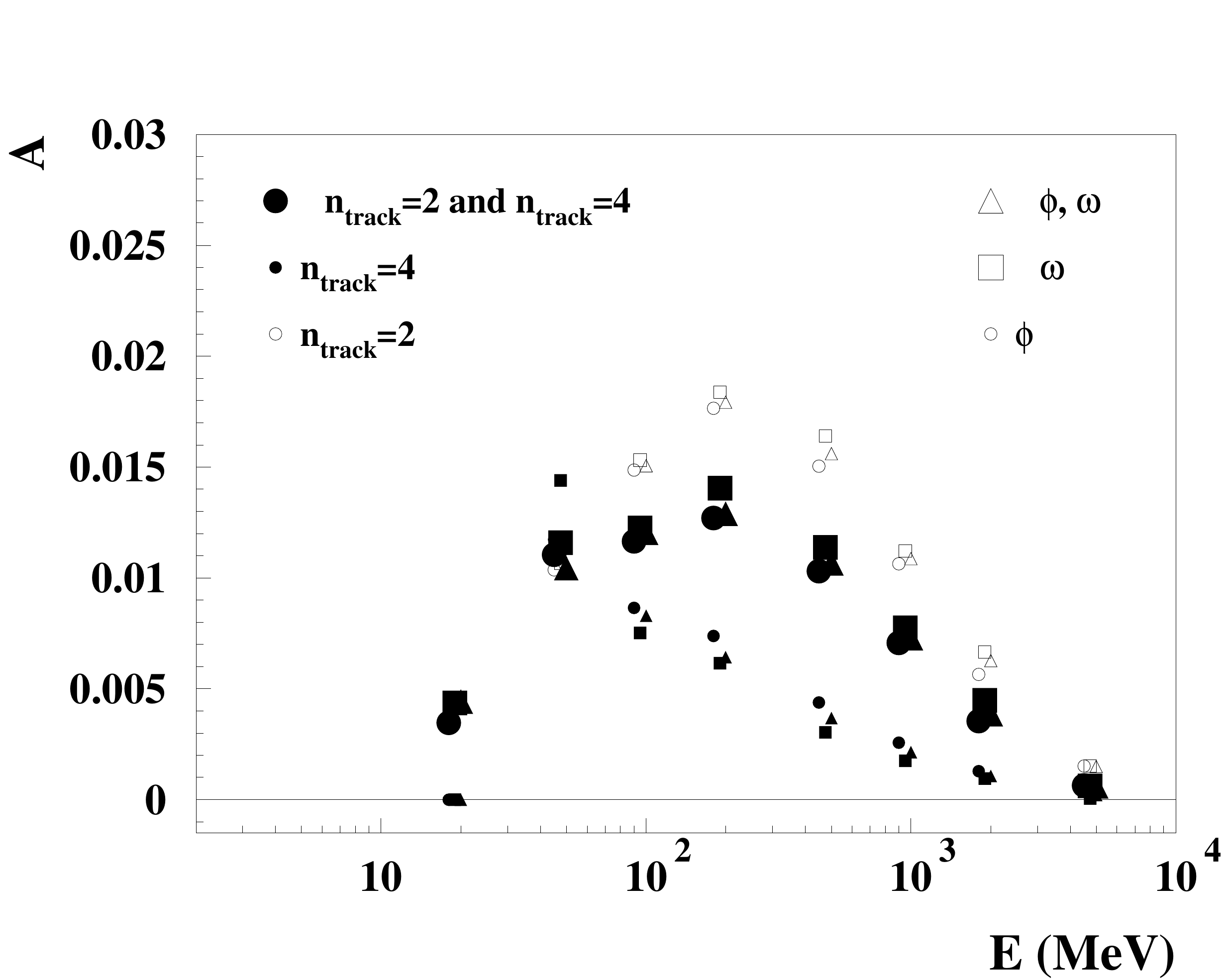}
 \hfill
\includegraphics[width=0.49\linewidth]{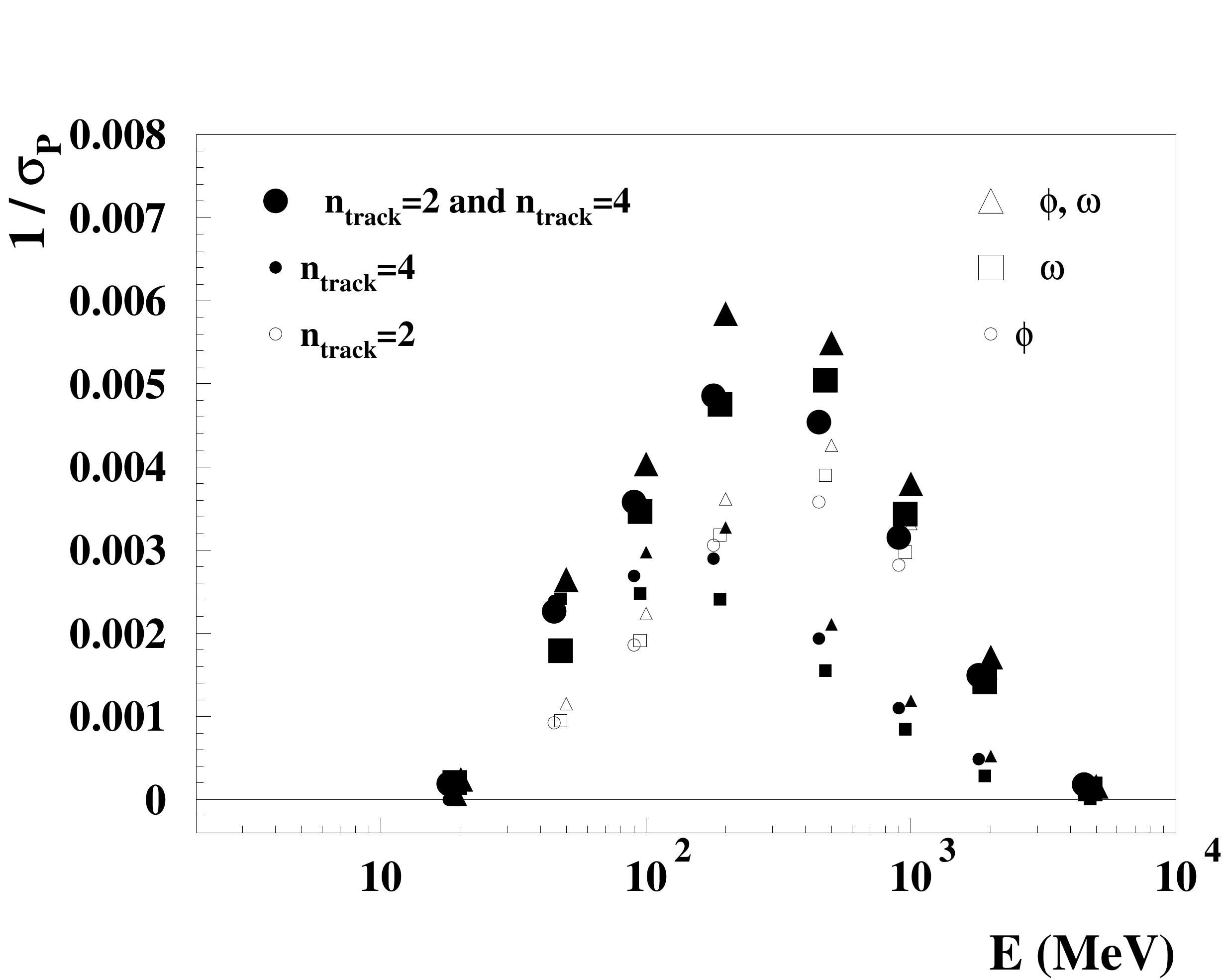}
\hfill
~
 }
\caption[~]{Combination of $\ntrack =2$ and of the $\ntrack =4$
 samples:
polarisation asymmetry (left) and inverse of the precision $\sigma_P$
of the measurement of $P$, normalised to one incident photon (right),
as a function of photon energy.
\label{fig:trf:4:2} }
\end{center}
\end{figure*}

\section{Wrap-up}
\label{sec:wrap}

The next step is the combination of the $\ntrack =2$ and of the
$\ntrack =4$ samples.
For each energy, all events are combined to a single measurement, each
event affected with a weight that depends on the conversion location
and on the number of reconstructed tracks.
Figure \ref{fig:trf:4:2} presents the polarisation asymmetry (left)
and the inverse precision of the 
measurement of $P$, $1 / \sigma_P$,
as a function of $E$.
\begin{itemize}
\item The sensitivity for the 
$\ntrack =2$ sample peaks at a higher energy
($\approx 500 \, \mega\electronvolt $), than the
 $\ntrack =4$ sample ($\approx 100 \, \mega\electronvolt $).

\item The precision of the combined sample peaks at 
($\approx 200 \, \mega\electronvolt $), with 
$1/ \sigma_P \approx 0.006 \times \sqrt{N}$, where $N$ is here the
 total number of photons incident on the detector, that is, 
$\sigma_P \approx 0.005$ for the $N = 10^9$ simulated samples used in
this study.
The effective polarisation asymmetry of the combined sample is
$A \approx 1.4 \,\%$ at the peak.
\end{itemize}

\begin{figure}[htbp!]
\begin{center}
 \includegraphics[width=0.6\linewidth]{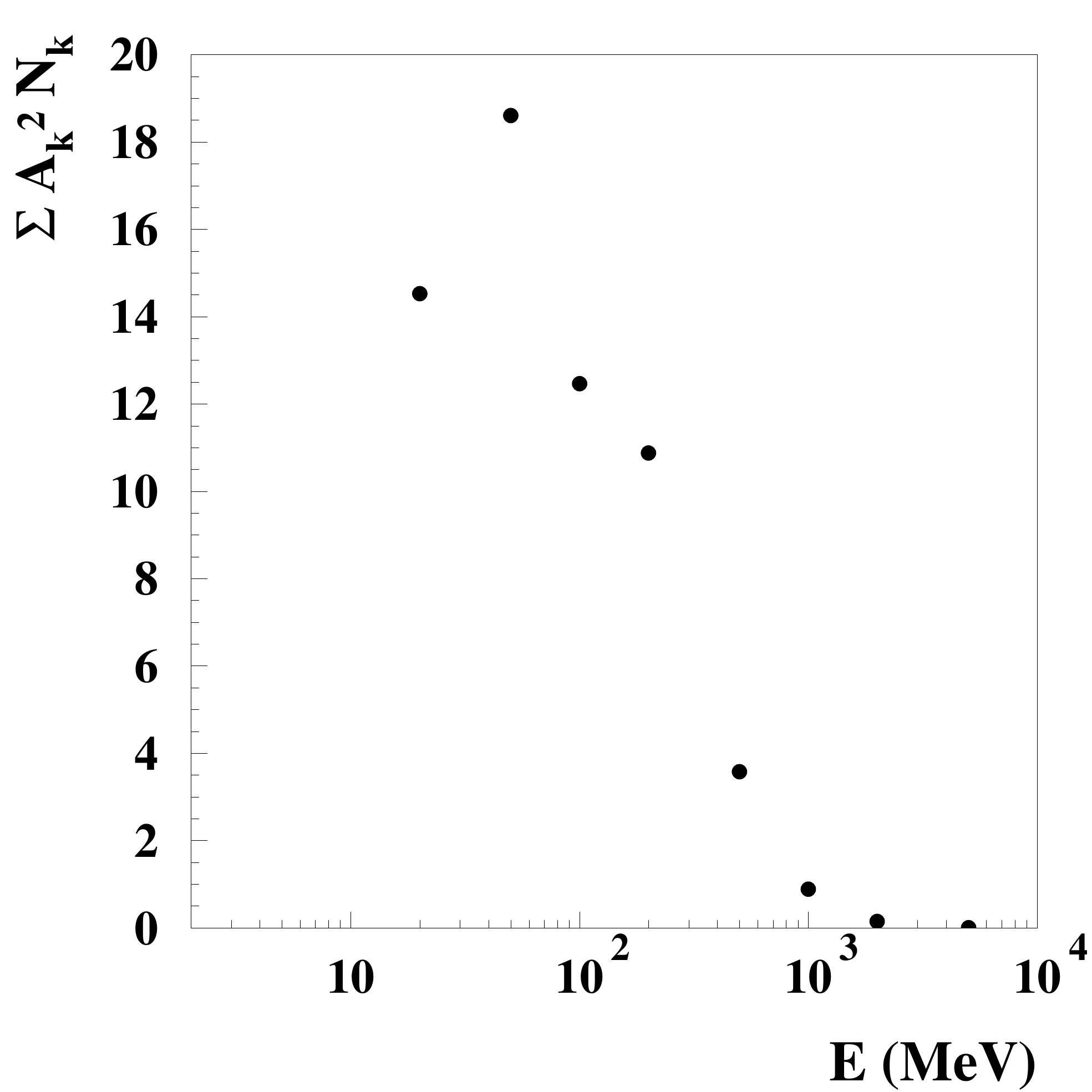}
\caption[~]{Energy distribution of the figure of merit as a function of energy for the 
 $\epsilon \approx 2. \times 10^{11} \centi\meter^2 \second$ exposure measurement of a
$\Gamma = 2$,
 $F_0 = 1.5 \times 10^{-3} \mega\electronvolt \centi\meter^{-2} \second^{-1} $
 bright source.
\label{fig:spectrum} }
\end{center}
\end{figure}

\section{Analysis of a full spectrum}
\label{sec:full:spec}

The energy spectrum is binned, and events from different energy bins
are analysed with the same combination method as presented previously.

Let us first present a simple estimate of the precision 
to be expected for an exposure of about
$\epsilon \approx 2 \times 10^{11} \centi\meter^2 \second$
 observing a bright source with spectral index
$\Gamma = 2$ and
${\dd N} / {\dd E} = F_0 / E^2$, 
$F_0 = 1.5 \times 10^{-3} \mega\electronvolt \centi\meter^{-2} \second^{-1} $,
with $E > 20 \,\mega\electronvolt$.
Supposing an isotropic exposure map, and taking into account the
fraction of the sphere within the $\cos{\theta_{\LAT}} > 0.25$ cut, we
obtain the number of photon incident on the detector in each energy
bin.

Taking into account the fraction of these events that convert to a
pair in the active target (see Fig. \ref{fig:event:fraction}), we
obtain the number of pair-converted-and-triggered photons, of
$N_d = 3 \times 10^6$,
a number
which is commensurate with the number of photons actually detected by
the \FermiLAT\ on the brightest sources of the $\gamma$-ray sky
\cite{Fermi-LAT:2019yla}.

Energy bin limits are defined to be the logarithmic average of the
nominal energies of successive bins, that is 
 20., 31.6, 70.7, 141., 316., 707., 1414., 3162., 7071. MeV.

\subsection{Back-of-the-envelope estimation of the precision}

I first estimate the precision of the measurement of $P$ using the
approximate expression of eq. (\ref{eq:uncertainty:comb}), where the
sum, $\sum_k A_k^2 N_k $ runs on both the conversion location and the
number of reconstructed tracks.
\begin{equation}
 {\sigma_P} = 
 \gfrac{W}{\sqrt{ \sum_k A_k^2 N_k }}
 .
 \label{eq:uncertainty:comb:re}
\end{equation}

I obtain $\sum_k A_k^2 N_k \approx 61$, that is, with $W \approx 1.2$, 
$\sigma_P = 0.15$.

\subsection{Full scale exercise}

A sample of $10^9$ events is generated between 20\,MeV and 5\,GeV
with a $\cos{\theta_{\LAT}} > 0.25$ cut,
with a $\Gamma = 2$ spectral index and $P=1$.

A measurement performed on that huge sample
($\varphi = \phi$,
all conversion locations combined, 
$\ntrack =2$ and $\ntrack =4$ combined,
global weighting, with the weight applied for a given event being
taken from the closest energy (in log space) of the samples described
in Sections \ref{sec:MC:4vectors}-\ref{sec:wrap})
 yields $ A \times P = 0.0129$ and $ \sigma_{A \times P } = 0.00012$, that is $ \sigma_{P} = 0.0096$, and $W = 1.18$.

\begin{figure}[htbp!]
\begin{center}
 \includegraphics[width=0.6\linewidth]{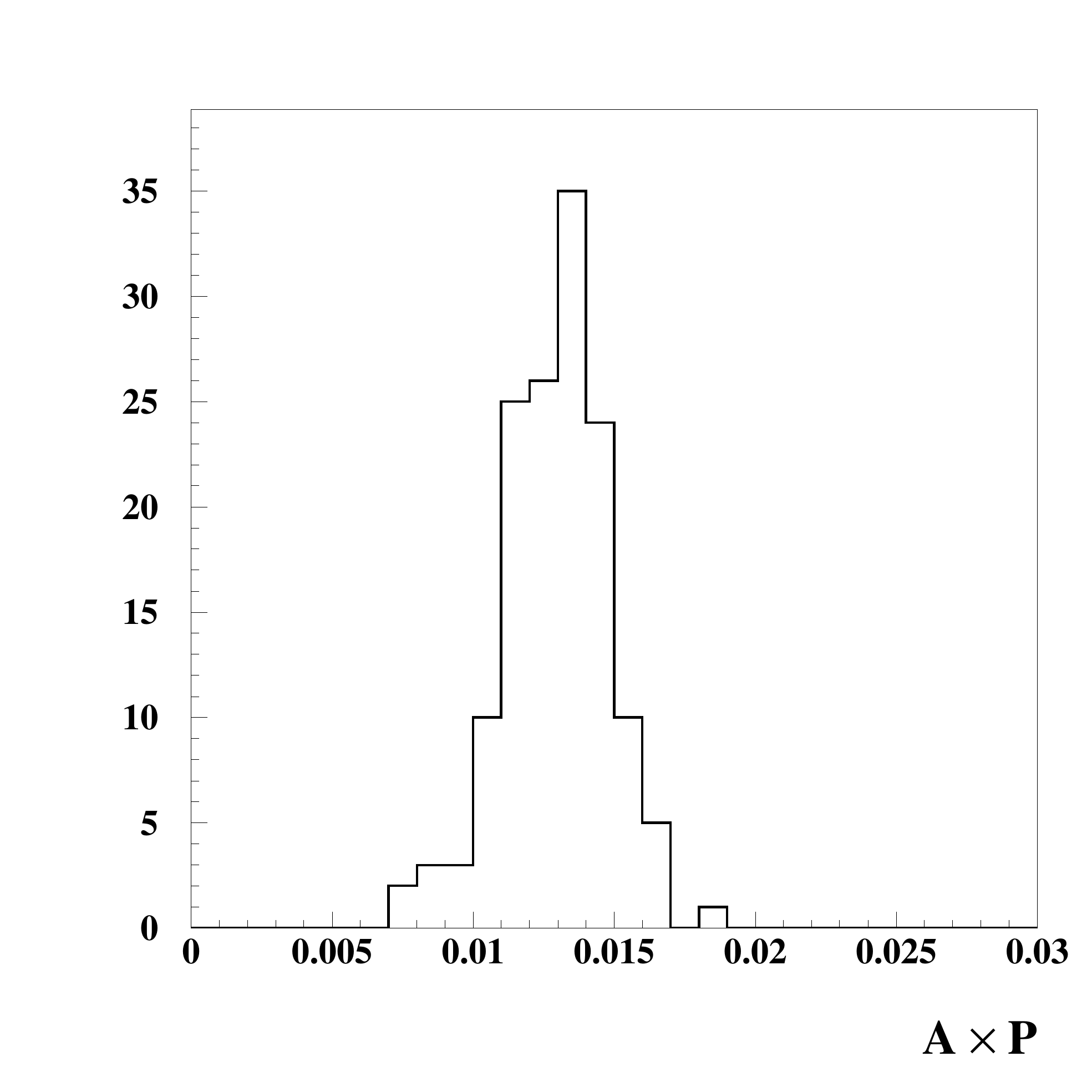}
 \caption[~]{Distribution of the values of $A\times P$ obtained from $n=144$
 samples of photons generated between 20\,MeV and 5\,GeV with a
 $\Gamma = 2$ spectral index and $P=1$.
\label{fig:spec} }
\end{center}
\end{figure}

That sample is then split into $n=144$ subsamples with same sizes, for
which the average number per sample of pair-converted-and-triggered
events is $\approx 3 \times 10^6$, a number that is commensurate to
the number of photons associated to a bright source in the
\FermiLAT\ data.
The R.M.S. width of the $n$ results is found to be
$\sigma_P \approx 0.14$ (Fig. \ref{fig:spec}), that is, compatible
with the approximate value from eq. (\ref{eq:uncertainty:comb}), of
$\sigma_P \approx 0.15$.

\clearpage

\begin{figure*}[htbp!]
 \hfill
\includegraphics[width=0.98\linewidth]{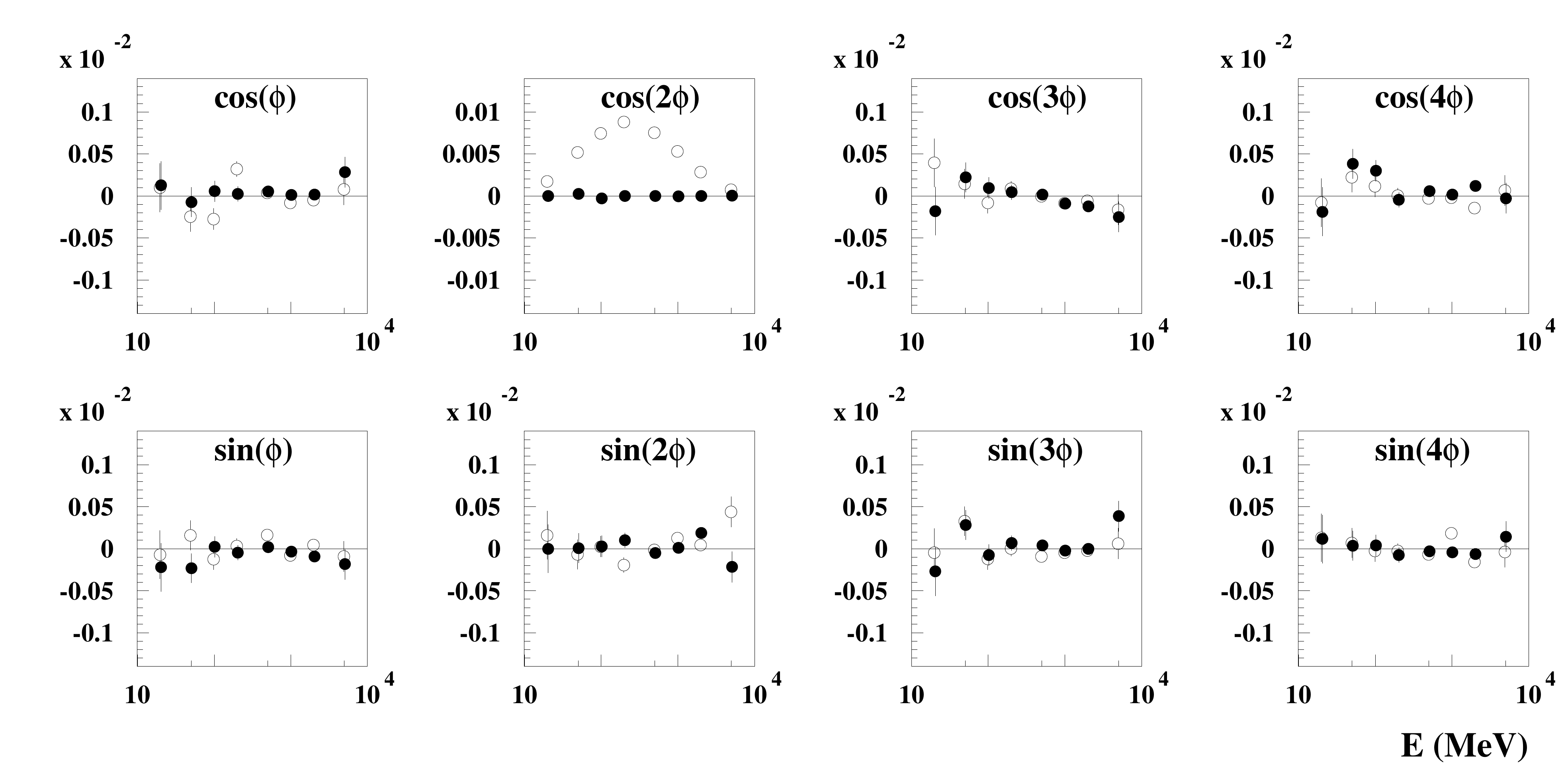}
 \hfill ~
\caption[~]{Study of systematics:
 $\langle \cos(n\phi) \rangle$ (upper row) and 
 $\langle \sin(n\phi) \rangle$ (lower row),
 $n=1 \cdots 4$,
 as a function of photon energy
 ($\varphi_0=0$, $\varphi \equiv \phi $, $\ntrack =2$, global weighting).
Simulations were performed with $P=0$ (bullet) and with $P=1$ (circle). 
Note the different vertical scale for $\cos(2\phi)$.
An isotropic exposure map was used in this simulation and therefore no
distortion from the $\cos(2\phi)$ signal was expected;
actually no significant distortion is found.
 \label{fig:systematics}}
\end{figure*}

\section{Discussion and perspectives}
\label{sec:discussion:perspectives}

I have estimated the potential of a \FermiLAT-like active target to
perform the polarimetry of a bright source of the gamma-ray sky.

From the geometry of the detector (layer spacing, strip pitch) and
from the energy scaling of the distribution of the pair opening angle,
the critical energy around which a significant fraction of the
pair-conversion events are expected to show a large enough opening
angle to yield a measurable value of the azimuthal angle is in the
very lowest part of the LAT sensitivity energy range
(Sect. \ref{sec:Measurement}).

The precision of the measurement of the direction of a track at vertex
is dominated by the precision of the measurements in the two first
layers on most of the energy range relevant to the present study, and,
due to the multiple scattering in the next tungsten foils, it is only
above $1\,\giga\electronvolt/c$ that the angular precision approaches the
homogeneous-detector asymptote for which the use of a Kalman filter is
needed to perform an optimal fit
(Sect. \ref{sec:Tracking}).
Therefore I use only the information from the two first layers in this
analysis (Sect. \ref{sec:Reconstruction}).
No pattern recognition is applied and only events for which no extra
clusters are present and a straightforward reconstruction can be
performed are used.
The main sources of event loss are then
(Fig. \ref{fig:event:fraction}) 
\begin{itemize}
\item At low energies, the requirement that 3 layers in a row are hit
 (an emulation of the basic ingredient of the \FermiLAT\ trigger) and
 the rejection of events with too many clusters in the conversion
 layer ($n_{1x}>1$ or $n_{1y}>1$).

Also, the cross section of pair conversion decreases at low energies
(less than one half that of the high-energy asymptote at 20\,MeV for
tungsten \cite{NIST:gamma}), so a fraction of the incident photons
just sail through the target without interacting.

\item At high energies, the reconstruction of one single track that
 does not carry any usable azimuthal information,
 and the rejection of events with too many clusters in the next layer.
\end{itemize}

The effective polarisation asymmetry is found to be small, and, for conversions in the lower wafer of a layer,
\begin{itemize}
\item 
to peak at $A \approx 0.03$ for $E \approx 200\,\mega\electronvolt$
for $\ntrack =2$ events (Fig. \ref{fig:aa});
\item to decrease as a function of energy for $\ntrack =4$ events
 (Fig. \ref{fig:aa:4}).
\end{itemize}
For conversions in the upper wafer of a layer, or for conversions in a thin
tungsten foil, the effective polarisation asymmetry is only worse.
For conversions in a thick tungsten foil of the so-called ``back''
part of the active target, there is barely any significant asymmetry.

The combination of measurements of $P$ from several samples is
performed in an optimal way by weighting each measurement by its
inverse variance, as is usual, which turns out here to weighing each
event by the polarisation asymmetry of the sample to which it belongs
(Sect. \ref{sec:Weighting}).

I have tested a number of methods in the hope to further improve on
the precision,
\begin{itemize}
\item 
 weight with a weight that depends on one of the observables of the
 event (Sect. \ref{sec:A:variation} and Figs.
 \ref{fig:pond},
\ref{fig:tre},
 \ref{fig:pond:4} and
\ref{fig:tre:4});

\item 
 combine measurements performed with various definitions of the
 azimuthal angle (Sect. \ref{sec:Def:Azim} and Figs.
 \ref{fig:aa},
 \ref{fig:aa:4} and
\ref{fig:trf:4:2});
 
\item in the case of $\ntrack =4$ events, solve the photon
 reconstruction ambiguity by choosing the candidate with the best
 value of the acoplanarity angle
(Subsect. \ref{subsec:matching:attempt}).
\end{itemize}
These attempts have proven to bring a mild improvement only, and
therefore I finally use a simple scheme in which
\begin{itemize}
\item 
Events are weighted with a weight that depends on the conversion
location, on the number of reconstructed tracks and on the photon
energy;
\item
The ``bisectrix'' azimuthal angle is used, $\varphi \equiv \phi$;
\item
For $\ntrack =4$ events, both photon candidates are used with weight $1/2$.
\end{itemize}

Under the approximation of a constant value of the weight width $W$, a
figure of merit is $\sum_k N_k A_k^2 $, which is proportional to
the combined inverse variance (eq. (\ref{eq:uncertainty:var})).
To that measure, conversions in the lower wafer dominate the precision
of a combined measurement, as is expected, but conversion in the upper
wafer and in the thin tungsten foils contributed equally, the higher
statistics for tungsten compensating for the lower effective asymmetry
(Figs. \ref{fig:trd} and \ref{fig:trd:4}).

For event samples simulated at various photon energies with the same
number of incident events, the figure of merit then peaks
at $E \approx 500\,\mega\electronvolt$ for $\ntrack =2$
(Fig. \ref{fig:tre}) and
at $E \approx 100\,\mega\electronvolt$ for $\ntrack =4$
(Fig. \ref{fig:tre:4}).
For realistic spectra with a $\Gamma = 2$ spectral index, peaking
takes place at a much lower energy (Fig. \ref{fig:spectrum}).

Under that simple scheme, a measurement on a bright source of the
gamma-ray sky with an exposure presently available at the \FermiLAT\ 
would yield a precision of $\approx 15 \,\%$.

The main avenue for improvement, besides pursuing the limited-gain
methods explored above, would be the use of the pattern-recognition
software of the experiment to analyse the events with a too large
number of clusters.

A number of issues that are relevant to polarimetry have been left
outside the scope of the present work.
\begin{itemize}
\item Measuring the polarisation angle together with the polarisation
 fraction, using the same moments' method as was used here, can be
 performed based on a combination of moments of cosines and of sines
 (\cite{Kislat:2014sdf} and references therein).
The precision of the measurement of $P$ is left unchanged 
 (eq. (36) of \cite{Kislat:2014sdf}).

\item Establishing the significance of a non-zero polarisation
 fraction in the presence of the nuisance parameter of a free
 polarisation angle can be an issue, as
``Linear polarization is non-negative, i.e. one always measures
something, even for an unpolarized source'' (M. Weisskopf, 2014).
This question has been addressed in \cite{Weisskopf:2010se};
 in an experimental context, see for example \cite{Ozaki:2016gvw}.

 \item
For such small effective polarisation asymmetries, and therefore such
small modulation factors, systematics are obviously expected to be an issue.
\begin{itemize}
 \item
Given the expected value of the uncertainty on the measurement of $P$,
I don't expect multiplicative systematics affecting the asymmetry,
such as those originating from the imprecise knowledge of the mass
model of the active target (e.g., the exact thickness of the wafers),
to be a limiting issue.

 \item
More serious problems could originate from the variation of the
effective asymmetry with $\theta_{\LAT}$
(Figs. \ref{fig:variation} and \ref{fig:variation:4}),
and the possible dependence of the $\theta_{\LAT}$ distribution on
the actual exposure map
(that was taken to be isotropic in the present study).
Any discrepancy between the simulated and the actual exposure map,
any discrepancy between the simulated and the actual 
dependence on $\theta_{\LAT}$ of the trigger/reconstruction/selection
efficiency would end up in data-to-MC differences in the higher-order
spectra.
\end{itemize}

The examination of the higher-order moments,
(Fig. \ref{fig:systematics}) and in particular of their variation with
$\theta_{\LAT}$, will be critical to the control of these biases, as
is usual in high-energy physics.

For polarization measurements performed using a moments method,
background subtraction is usually performed by injecting events from
the sidebands in the sum, with a negative weight.
This should be easy in the present case of a bright source, for which
the signal/noise ratio is large, but care should be taken at low
energies for which the PSF gets worse \cite{Principe:2018lzn}.
\end{itemize}

\section{Conclusion}

The potential for $\gamma$-ray polarimetry in the pair conversion
regime with a SSD-based active target is characterised in quantitative
way.
A non-zero, yet small, effective polarisation asymmetry is
demonstrated for the first time.
The main strenght of this study is the use of validated and simple
tools, and the generation of large MC samples, typically $10^9$ events
per sample, that enabled precision studies of these small asymmetries.
The main weakness is the lack of pattern recognition, so the
analysis was restricted to events with the nominal numbers of clusters
in the two first layers, something that induced a large loss in
statistics, in particular at low energies
(Fig. \ref{fig:event:fraction}).

It is not easy to predict the performance of the AMEGO
\cite{McEnery:2019tcm} and ASTROGAM \cite{DeAngelis:2016slk} projects
from the present study without performing a dedicated analysis.
Due to the absence of high-$Z$ converter foils in the active target,
and thanks to the DSSSD scheme, all pair-conversion events are of the
``lower wafer'' category, in my classification.
The DSSSD scheme makes $(x,y)$ track matching straightforward for
$\ntrack=4$ events.
Taken at face value, the geometry seems to be less favourable than for
the \FermiLAT, with a smaller distance between layers and larger strip
pitches (accordingly their critical energy is lower, see
Tab. \ref{tab:telescopes}), but these detectors
are designed to enable trigger / reconstruction / selection of
pair-conversion events of lower energies than the \FermiLAT.
The silicon sensitive masses are similar, 69\,kg for the \FermiLAT
\cite{Atwood:2007ra}, to be compared to 40\,kg for AMEGO
\cite{DeAngelis:2016slk}.

The SSD part of the AMS-02 tracker should enable a high-performance
$\gamma$-ray polarimetry, with its DSSSD scheme, thin wafers, narrow
readout pitch, and large distance between layers
\cite{Ambrosi:2017gez}.
Unfortunately the smallness of the silicon mass (4.5\,kg)
and the need to select photons that convert in a thin layer just above
an empty space followed by a wafer,
might limit the available statistics for a significant measurement.

Gamma-ray polarimetry with pair conversions with a higher value of
the effective polarisation asymmetry than in the present study, may
have to wait for projects using homogeneous detectors
\cite{Ozaki:2016gvw,Gros:2017wyj} or thinner silicon wafers
\cite{Suljic:2021hfl}.

\clearpage

%\begin{table*}[t] 
% \caption[~]{
 Variables used in the text.
 % \label{tab:vars}}

~
 
\begin{tabular}{lll}
$A$ & (linear) polarisation asymmetry & eq. (\ref{eq:1D})\\
 $D$ & dilution of the polarisation asymmetry & Sect. \ref{sec:Measurement}\\
 $\delta$ & azimuthal angle of the LAT wrt the sky frame. & Sect. \ref{sec:4tracks} \\
 $\Delta$ & W scatterer thickness & Sect. \ref{sec:Tracking} \\
 $\Delta$, $\Delta'$ & azimuthal directions of the bisectrix for 
 & Fig.\ref{fig:4tracks:schema}
 \\
 & the correctly and incorrectly reconstructed pairs, respectively.
 \\
 E & expectation value & Sect. \ref{sec:MC:4vectors} \\
 $E$ & photon energy \\
$E_c$ & critical energy of an active target & eq. (\ref{eq:critical:energy}) \\
 $e$ & wafer thickness & Tab. \ref{tab:telescopes} \\
 $\epsilon $ & data taking exposure \\
 $\varphi$ & generic azimuthal angle of a conversion event & eq. (\ref{eq:1D}) \\
 $\varphi_0$ & polarisation angle of the incoming radiation & eq. (\ref{eq:1D}) \\
 $\phi$ & one way to define the azimuthal angle $\varphi$ & \cite{Gros:2016dmp} \\
 $\Gamma$ & source spectral index \\
 $j$ & imaginary unit \\
 $\ell$ & distance between layers \\
 $\lambda$ & detector scattering length & \cite{Innes:1992ge} \\
 $n_{\gamma}$ & number of photon candidates in an event \\
 $n$ & number of subsamples & Sect. \ref{sec:full:spec} \\
 $n$ & high-order index & Fig. \ref{fig:systematics}\\
$N$ & number of layers & Tab. \ref{tab:telescopes} \\ 
$N$ & number of events in sample & Sect. \ref{sec:MC:4vectors} \\
 $N$ & total number of generated events & Sect. \ref{sec:wrap} \\
$N_d$ & number of pair-converted-and-triggered events & Sect. \ref{sec:discussion:perspectives}\\
 $\ntrack $ & number of track candidates in an event \\ 
 \end{tabular}

% \clearpage

\begin{tabular}{lllllll} 
 p & silicon strip pitch \\
 $p$ & track momentum \\
 $p_0$ & characteristic momentum of multiple scattering &\cite{Patrignani:2016xqp} \\
 $p_1$ & characteristic multiple-scattering momentum of an active target & \cite{Bernard:2013jea}, eq. (\ref{eq:p1})\\ 
 $P$ & (linear) polarisation fraction of the incoming radiation & eq. (\ref{eq:1D}) \\
 $q$ & target recoil momentum upon gamma-ray conversion \\
 $r$ & ratio of the number of correctly to falsely assigned candidates & Fig. \ref{fig:4tracks:assign} \\
 $\rho$ & material density \\
 $\sigma$ & RMS \\
 $\sigma$ & cross section \\
 $\sigma$ & single-(wafer)-measurement space resolution \\
 $\sigma_\theta$ & (polar) angular resolution \\
 $\sigma_\varphi$ & RMS resolution of the measurement of the azimuthal angle \\ 
 $t$ & thickness through which a lepton propagate, $x$, normalised to $X_0$ \\
 $\theta$ & polar angle \\
 $\theta_{+-}$ & pair opening angle of an event \\
 $\hat \theta_{+-}$ & most probable value of $\theta_{+-}$ & \cite{Olsen:1963zz}, eq. (\ref{eq:theta:pm})\\
$\theta_{\LAT}$ & angle between the photon direction and the $z$ axis of the LAT \\
 $V$ & variance \\
 $w$ & weight \\
 $\omega$ & one way to define the azimuthal angle $\varphi$ & \cite{Bogdan:1998}\\
 $W$ & RMS width (of weight distributions) & Sect. \ref{sec:MC:4vectors} \\
 W & tungsten \\
 $X_0$ & material radiation length \\
 $x$ & detector longitudinal sampling normalised to scattering length & Sect. \ref{sec:Tracking} \\
 $x$ & axis name \\
 $y$ & axis name \\
 $z$ & axis name \\
\end{tabular}
%\end{table*}

\clearpage

\appendix

 \section{~}

In Section \ref{sec:A:variation}, I have used a function
$g(\theta)$ in the place of variable $\theta$, 
so as to histogram the distribution with a decent, more or less
homogenous, statistics per bin.
 Function $g(\theta)$ is documented in this Appendix.

~

\begin{eqnarray}
 g(\theta) = \theta / \theta_u , 
\end{eqnarray}
with
\begin{eqnarray}
 \theta_u =
 \theta_0 \left(\gfrac{E}{100\,\mega\electronvolt}\right)^\alpha a^{i}
 \oplus
 \theta_1 b^{\ntrack }
\end{eqnarray}
where 
$\theta_0$, $\alpha$, $a$, $\theta_1$, $b$ are the following parameters:

\begin{tabular}{lllllll} 
 $\theta_0$ & $\alpha$ & $a$ & $\theta_1$ & $b$ \\
 0.2154 & -1.0338 & 0.5260 & -0.0014 & 2.0289 \\
\end{tabular}

$i$ varying from 1 to 4 for W-back to lower-wafer.

\clearpage

\tableofcontents

\end{document}